\documentclass[fleqn,usenatbib]{mnras}

\usepackage{newtxtext,newtxmath}
\usepackage[T1]{fontenc}
\usepackage{ae,aecompl}

\usepackage{graphicx}	% Including figure files
\usepackage{amsmath}	% Advanced maths commands
\usepackage{amssymb}	% Extra maths symbols

\usepackage{verbatim}
\usepackage{xspace}
\usepackage{stackengine}
\usepackage{multirow}
\newcommand{\RX}{RXC\,J1314.4-2515\xspace}

\title[RXC\,J1314.4-2515]{Particle re-acceleration and Faraday-complex structures in the RXC\,J1314.4-2515 galaxy cluster}

\author[Stuardi et al.]
{\parbox{\textwidth}{Stuardi, C.,$^{1,2}$\thanks{E-mail: \texttt{chiara.stuardi2@unibo.it}}
Bonafede, A.,$^{1,2}$
Wittor, D.,$^{1,2,3}$
Vazza, F.,$^{1,2}$
Botteon, A.,$^{1,2,6}$
Locatelli, N.,$^{1,2}$
Dallacasa, D.,$^{1,2}$
Golovich, N.,$^{4}$
Hoeft, M.,$^{5}$
van Weeren, R.J.,$^{6}$
Br\"uggen, M.$^{3}$ and
de Gasperin, F.$^{3}$}\vspace{0.4cm}\\
\parbox{\textwidth}{$^{1}$Dipartimento di Fisica e Astronomia, Universit\`a di Bologna, via Gobetti 93/2, 40122 Bologna, Italy\\
$^{2}$INAF - Istituto di Radioastronomia di Bologna, Via Gobetti 101, I-40129 Bologna, Italy\\
$^{3}$Hamburger Sternwarte, Universit\"at Hamburg, Gojenbergsweg 112, 21029 Hamburg, Germany\\
$^{4}$Lawrence Livermore National Laboratory, 7000 East Avenue, Livermore, CA 94550\\
$^{5}$Th\"uringer Landessternwarte Tautenburg, Sternwarte 5, 07778 Tautenburg, Germany\\
$^{6}$Leiden Observatory, Leiden University, PO Box 9513, 2300 RA Leiden, The Netherlands
}}

\pubyear{2019}

\begin{document}
\label{firstpage}
\pagerange{\pageref{firstpage}--\pageref{lastpage}}
\maketitle

% Abstract of the paper
\begin{abstract}
Radio relics are sites of electron (re)acceleration in merging galaxy clusters but the mechanism of acceleration and the topology of the magnetic field in and near relics are yet to be understood. We are carrying out an observational campaign on double relic galaxy clusters starting with \RX. With {\it Jansky Very Large Array} multi-configuration observations in the frequency range 1-4 GHz, we perform both spectral and polarization analyses, using the Rotation Measure synthesis technique. We use archival {\it XMM-Newton} observations to constrain the properties of the shocked region. We discover a possible connection between the activity of a radio galaxy and the emission of the eastern radio relic. In the northern elongated arc of the western radio relic, we detect polarized emission with an average polarization fraction of $31 \ \%$ at 3 GHz and we derive the Mach number of the underlying X-ray shock. Our observations reveal low levels of fractional polarization and Faraday-complex structures in the southern region of the relic, which point to the presence of thermal gas and filamentary magnetic field morphology inside the radio emitting volume. We measured largely different Rotation Measure dispersion from the two relics. Finally, we use cosmological magneto-hydrodynamical simulations to constrain the magnetic field, viewing angle, and to derive the acceleration efficiency of the shock. We find that the polarization properties of \RX are consistent with a radio relic observed at $70^{\circ}$ with respect to the line of sight and that efficient re-acceleration of fossil electrons has taken place.
\end{abstract}

% Select between one and six entries from the list of approved keywords.
% Don't make up new ones.

%%%%%%%%%%%%%%%%%%%%%%%%%%%%%%%%%%%%%%%%%%%%%%%%%%

%%%%%%%%%%%%%%%%% BODY OF PAPER %%%%%%%%%%%%%%%%%%

\section{Introduction}
\label{sec:intro}

During cluster mergers, the kinetic energy of in-falling materials is injected in the Intra-Cluster Medium (ICM). A fraction of the dissipated energy could amplify the magnetic field and (re)accelerate particles \citep[see, e.g.,][ for a review]{Brunetti14} leading to the formation of diffuse synchrotron emission, observable in the radio band. Extended radio sources on Mpc scale, without any optical counterpart and with low surface brightness (i.e., $\sim0.1-1 \ \mu$Jy arcsec$^{-2}$), have been detected in some merging galaxy clusters \citep[see, e.g.,][and references therein]{vanWeeren19}. They are  termed radio halos and radio relics and they reveal the presence of $\sim \mu$G magnetic fields and $\sim$ GeV relativistic particles in galaxy clusters.

Radio halos permeate the central volume of most dynamically disturbed galaxy clusters. They are typically circular, with sizes $\geq$1 Mpc,  with low/absent polarization, down to a few percent level  \citep[see, e.g.,][for a review]{Feretti12}. Radio halos have a steep spectrum\footnote{Hereafter the spectrum is defined as $S_{\nu}\propto\nu^{-\alpha}$, where $S_{\nu}$ is the radio flux density  at the frequency $\nu$ and $\alpha$ is the spectral index.} with $\alpha > 1$. Their presence in merging galaxy clusters suggests that the energy for particle (re)acceleration could come from the gravitational energy released in the ICM during the process of structure formation (i.e. via turbulence, \citealt{Cassano10a}). The details of the acceleration mechanism are yet to be understood.

Radio relics are arc-shaped sources on Mpc scale, located at the periphery of  some merging galaxy clusters. They are characterized by a steep spectrum (i.e., $\alpha > 1$) and strong polarization ($\sim20-50 \ \%$ at 1.4 GHz). According to simulations, during cluster mergers, shock waves move outwards along the merger axis and form pairs of symmetric radio relics that extend in the direction perpendicular to the merger axis \citep[][]{Bruggen12,Ha18}. Therefore, they are best observed when the merger occurs in the plane of the sky \citep{Golovich18,Golovich19}. Although there  is  evidence that their origin is connected to shock waves generated in the ICM  by merger events \citep{Ensslin98,Roettiger99}, the underlying particle acceleration mechanism is still under debate. The mechanism of Diffusive Shock Acceleration (DSA) has been proposed \citep{Bell78,Jones91}: in this scenario,  cosmic-ray protons and electrons should be accelerated from the thermal pool up to relativistic energies by cluster merger shocks. Although this mechanism can explain the general properties of relic emission, several observational features remain unexplained:

{\it (i) the observation of low Mach number shocks.} Shock waves can be detected in the X-rays as sharp surface brightness discontinuities associated with a  density jump.  To date, a number of shock fronts  have been detected using radio relics as shock tracers \citep[][for some collections]{Akamatsu13,Finoguenov10,Urdampilleta18}.  Typical Mach numbers ($M$) of mergers shocks inferred from X-ray observations are between 1.5 and 3, with some exceptions at $M$>3 \citep{Markevitch02,Botteon16,Dasadia16}. In this regime, the electron acceleration efficiency predicted by the DSA model is  at most of a few per-thousands of the shock injected energy flux;

{\it (ii) the non-detection of $\gamma$-ray emission in merging galaxy clusters.} Protons are also expected to be accelerated by merger shocks and to produce $\gamma$-rays in the interaction with the thermal gas. The most updated $Fermi$ upper limits \citep{Ackermann14} lead to a shock acceleration efficiency for protons even lower than what is normally assumed \citep[i.e., below 10$^{-3}$,][]{Vazza14,Vazza16};

{\it (iii) the high radio power of radio relics.} The shock acceleration efficiency allowed by the DSA mechanism is not enough to match the radio emission observed in most radio relics \citep[e.g.][]{Markevitch05,Botteon16a,Eckert16a,Hoang18};

{\it (iv) the radio spectral index of some relics.} Clear cases in which spectral indices are difficult to reconcile with particle acceleration models are Abell 2256 \citep{vanWeeren12} and the Toothbrush radio relic \citep{vanWeeren16}.

 Additional mechanisms have been proposed, as for example the re-acceleration of pre-existing low-energy relativistic electrons \citep[e.g,][]{Pinzke13,Kang15,Kang16}. The origin of such fossil particles could be either in old remnants of  Active Galactic Nuclei (AGN),  or in an electron population accelerated  by earlier shock waves. In this scenario,  only a fraction of clusters may have adequate seed particle population and, in that case, the acceleration efficiency required to match radio observations would be lower. Furthermore, if the seed population of AGN origin was mostly composed  of electrons, the non-detection of gamma ray emission would also be explained. Unfortunately, up to date, the connection between AGN and radio relic could be established only in few cases \citep[e.g.,][]{Bonafede14,vanWeeren17a}.

Recent models have focused on the role of magnetic fields, that in  particular configurations could allow electrons to  reach supra-thermal energies via the Shock Drift Acceleration mechanism (SDA). Using particle in cell simulations, \citet{Guo14a,Guo14b} have shown that, in a quasi-perpendicular pre-shock magnetic field (i.e., when the magnetic field lines are almost perpendicular to the shock normal,   hence aligned with the shock front), electrons can be pre-accelerated also by shocks with $M$=3. At the same time, \citet{Caprioli14} demonstrated that  for quasi-perpendicular shocks the proton acceleration is quenched also for $M$=5. Although recent studies have confirmed that this might reduce the tension with the upper limits set by the $Fermi$ collaboration \citep{Wittor17}, the role of the magnetic field and its amplification by low Mach number shocks is still poorly constrained.

To this end, obtaining highly resolved information on the polarization of radio emission in relics has the potential of  revealing the local topology of magnetic fields in the electron cooling region. This is possible through the technique of Rotation Measure (RM) synthesis \citep{Brentjens05}. 

We started the first systematic study of magnetic field in radio relics. Our final goal is to constrain the magnetic field strength and structure in a sample of galaxy clusters, where shock waves have been detected in X-ray  and/or radio relics are observed. This will be possible  by comparing values of RM derived from background sources in the  pre-shock and post-shock regions with magnetic field models as in \citet{Bonafede13}. The full sample  is made of 14 galaxy clusters with double radio relics  observed with the {\it Karl G. Jansky Very Large Array} (JVLA) (Stuardi et al. in prep.).

Spectro-polarimetric receivers of the JVLA allow us to simultaneously study continuum and polarized emission with a MHz resolution at  GHz frequencies. We can therefore perform the RM synthesis and study the polarized emission in galaxy clusters in a large range of frequencies and with a high resolution in Faraday space. At the same time JVLA allows us to study the diffuse radio emission in galaxy clusters on a variety of angular scales, and at different frequencies while keeping the same resolution. High-resolution spectral index  images are important to obtain information on the life cycle of the relativistic electron that power radio sources \citep[see, e.g.,][]{vanWeeren17b}.

The combined information on: {\it (i)} magnetic fields at relics through radio polarimetric study, {\it (ii)} discovery of  a possible connection between radio relic emission and AGN activity through high-resolution spectral index imaging, and {\it (iii)} new detection of merger shocks through X-ray data analysis, are the key ingredients to solve the problem of particle acceleration in low Mach number shocks. 

In this paper, we study the radio emission of the galaxy cluster \RX, that belongs to our sample  of double relics systems. We decided to focus on this galaxy cluster since it shows a number of interesting features: although it is a double relic cluster (i.e., the merger axis is expected to be on the plane of the sky) different  works found that a significant component of the merger could lie along the line of sight \citep{Golovich18, Wittman18}, the central radio halo is spatially connected with the western relic making their nature ambiguous, and the eastern radio relic was suspected to host a radio galaxy \citep{Feretti05,Venturi07}.

In Sec.~\ref{sec:clusterinfo}, we briefly review literature information about this galaxy cluster, and in Sec.~\ref{sec:radioobs} we describe the radio observations and data reduction techniques. In Sec.~\ref{sec:continuum}, we analyse the results of continuum radio observations and discuss the spectral properties of the system. polarization and RM synthesis studies are reported in Sec.~\ref{sec:polarization}. We discuss our results and conclude in Sec.~\ref{sec:discuss} and \ref{sec:conclusions}. 

Throughout this paper, we assume a $\Lambda$CDM cosmological model, with $H_0$ = 69.6 km s$^{-1}$ Mpc$^{-1}$, $\Omega_{\rm M}$ = 0.286, $\Omega_{\Lambda}$ = 0.714 \citep{Bennett14}. With this cosmology 1$\arcsec$ corresponds to 3.9 kpc at the cluster redshift, $z$=0.247.

\section{RXC~J1314.4-2515}
\label{sec:clusterinfo}

\begin{figure*}
  \includegraphics[width=1\textwidth]{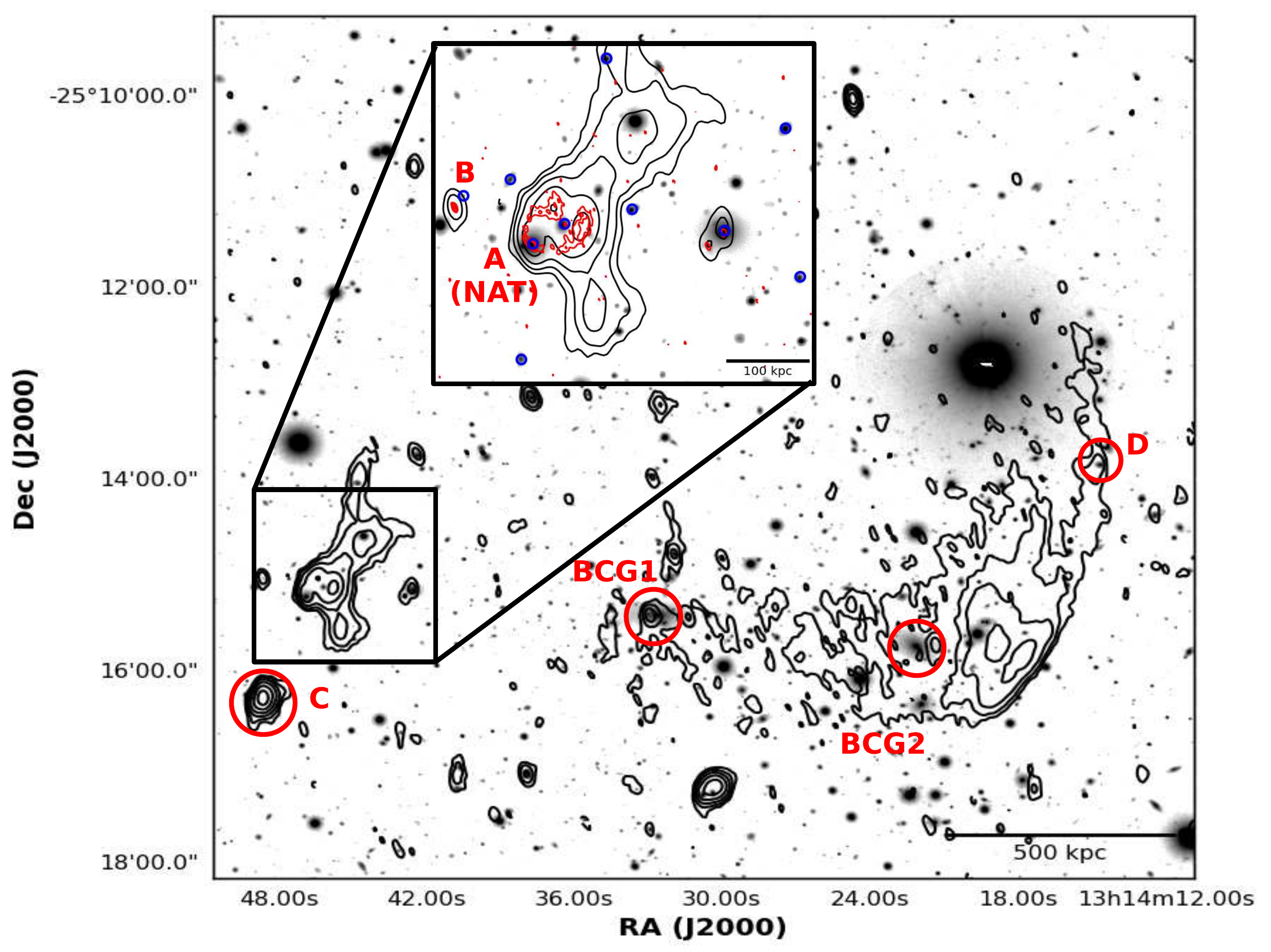}
    \caption{{\it Subaru} r- and g-band image of the cluster \RX with  black radio contours overlaid. Contours are obtained combining B and C configurations and the restoring beam is 9$\arcsec\times5\arcsec$. Black contours start from 3$\sigma$, with $\sigma$=0.015 mJy beam$^{-1}$, and they are spaced by a factor of two. A zoom in the region of the E radio relic at 1.5 GHz is displayed in the top inset panel. Black contours are in B+C configuration, same as above; red contours are from A configuration with a restoring beam of 2$\arcsec\times1\arcsec$, and they start at $3\sigma$, with $\sigma$=0.011 mJy beam$^{-1}$, spaced by a factor of two. Blue circles mark optically identified cluster members. Red letters  and circles mark the sources  with optical counterparts quoted in the paper. $BCG\,1$ is the brightest cluster galaxy of the main sub-cluster;  $BCG\,2$ is the one of the western sub-cluster; $A$ and $B$ are cluster members; $C$ and $D$ do not have redshift estimates \citep{Golovich18,Golovich19} }
    \label{fig:optic}
\end{figure*} 

\begin{figure*}
	\includegraphics[width=1\textwidth]{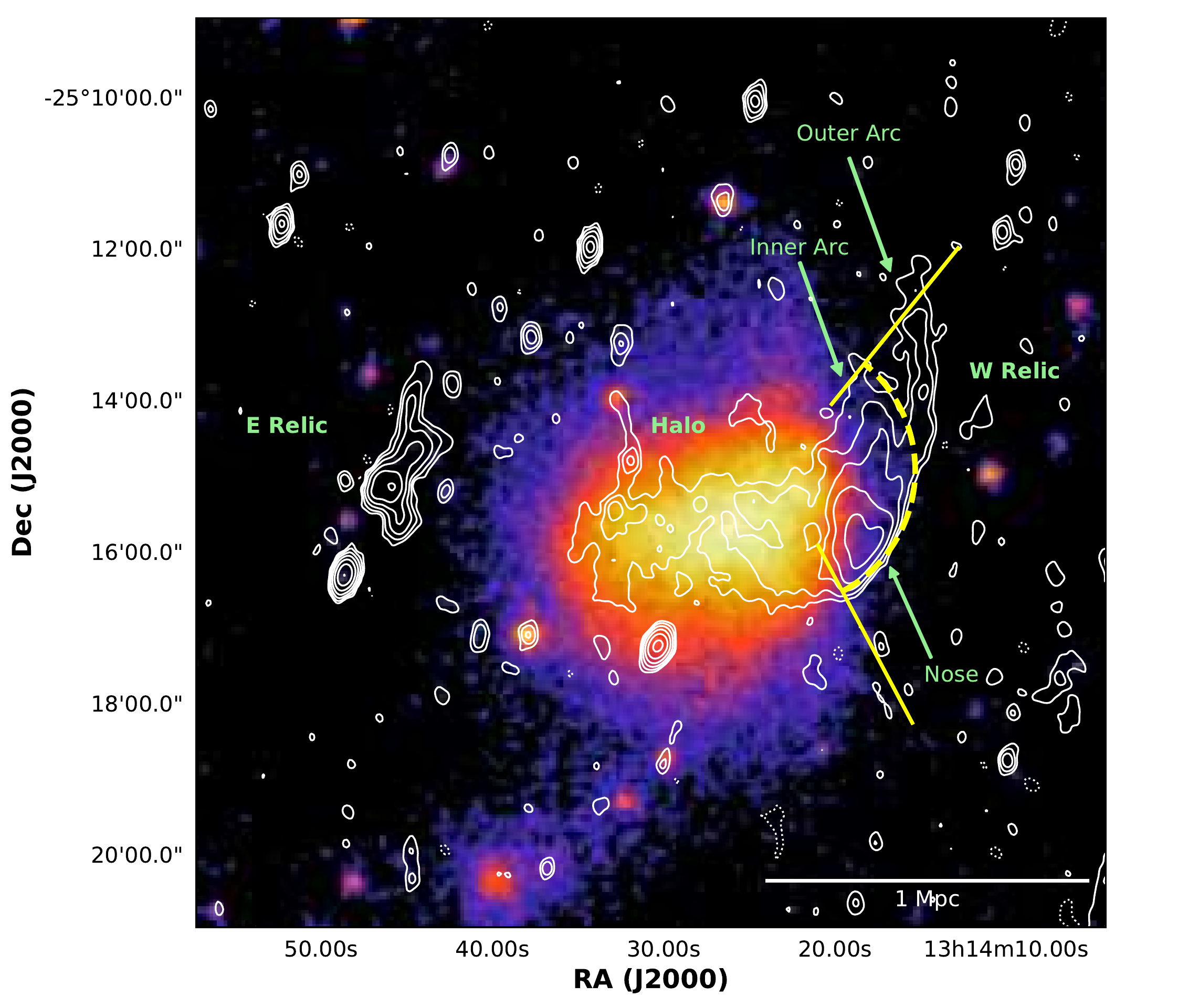}
    \caption{X-ray {\it XMM-Newton} image of the cluster \RX with white radio contours overlaid. Contours are from A+B+C+D configurations at 1.5 GHz and start at $\pm3\times$0.016 mJy beam$^{-1}$. They are spaced by a factor of two and negative contours are dotted. The restoring beam is 15$\arcsec\times8\arcsec$. The X-ray image is smoothed with a Gaussian kernel of 5$\arcsec$. The yellow  dashed line marks the position of the X-ray detected shock and  the yellow sector encloses the region used to extract the surface brightness profile (see Sec.~\ref{sec:discussrelicshock}).}
    \label{fig:X}
\end{figure*}

General information on this cluster  is listed in Tab.~\ref{tab:info}. The radio contours obtained in Sec.~\ref{sec:radioobs} superimposed on optical and X-ray images are shown in Fig.~\ref{fig:optic} and Fig.~\ref{fig:X}, respectively. \RX shows two symmetric radio relics, east and west of the cluster.  They were observed with the {\it Very Large Array} (VLA) at 1.4 GHz \citep{Feretti05}, and with the {\it Giant Metrewave Radio Telescope} (GMRT) at 610 MHz \citep{Venturi07} and at 325 MHz \citep{Venturi13}. The western relic is more extended than the eastern one, and it is connected to a central radio halo. Recently, the galaxy cluster was observed also with the {\it Murchison Widefield Array} (MWA) from 88 to 215 MHz \citep{George17}, leading to an estimate of the integrated spectral index of eastern and western relics: $\alpha^{\rm 118MHz}_{\rm 1.4GHz}$=1.03$\pm$0.12 and $\alpha^{\rm 118MHz}_{\rm 1.4GHz}$=1.23$\pm$0.09, respectively.

\RX has a disturbed morphology in the  X-rays: it is elongated in the east-west direction, suggesting an ongoing merger activity along this axis \citep{Valtchanov02}. In particular, \citet{Mazzotta11} found that the western relic is coincident with a shock front, detected through {\it XMM-Newton} observations, with Mach 2.1$\pm$0.1. They noticed that this shock front is M-shaped, with the nose of the front tilted inward, which they proposed may be produced by the material in-falling along a filament. In the X-ray image, a sub-cluster in the south direction is also visible, with a stream of gas suggesting accretion by the northern main cluster (see Fig.~\ref{fig:X}).

\citet{Valtchanov02} found a bi-modal distribution of the galaxies in this cluster both in velocity space ($\sim$1700 km s$^{-1}$ separation) and in projected space. This was recently confirmed by \citet{Golovich18}, who found also that the two merging sub-clusters have $\sim$1500 km s$^{-1}$ line of sight velocity difference, suggesting that the merger axis has a substantial component along the line of sight. Recently, matching the observed projected separation and relative radial velocities between sub-clusters with cosmological $N$-body
simulations, \citet{Wittman18} constrained the angle between the sub-cluster separation vector and the line of sight.  While in other double relics clusters the merger axis is found on the plane of the sky, for \RX they obtained a maximum likelihood at 42$^{\circ}$, although angles  up to 90$^{\circ}$ cannot be ruled out.

The median Galactic RM in the region of \RX measured with an angular resolution of 8$^{\circ}$ is -30$\pm$2 rad m$^{-2}$ \citep{Taylor09}. We used this value throughout the polarization analysis because we found the same median value outside the galaxy cluster in our field. This value is also consistent with the most updated estimate by \citet{Oppermann12} and \citet{Huts19}.

\begin{table}
	\centering
	\caption{Properties of \RX. Row 1,2: J2000 celestial coordinates  of the X-ray cluster centroid; Row 3: redshift, $z$; Row 4: X-ray luminosity in the energy band 0.1-2.4 keV; Row 5: estimate of the hydrostatic mass. References: (1) \citet{Piffaretti11}, (2) \citet{Valtchanov02}, (3) \citet{Planck16}.}
	\label{tab:info}
	\begin{tabular}{lcc}
	    \hline
		\hline
		R.A. (J2000) & 13$^{\rm h}$14$^{\rm m}$28$^{\rm s}$.0 & (1) \\
		Dec. (J2000) & -25$^{\circ}$15$\arcmin$41$\arcsec$ & (1) \\
		$z$ & 0.247 & (2) \\
		$L_{\rm X(0.1-2.4 keV)}$ & 9.9$ \cdot10^{44}$ erg s$^{-1}$ & (1) \\
		$M^{\rm SZ}_{500}$ & 6.7$ \cdot10^{14}{\rm  M_\odot}$ & (3) \\
		\hline
	\end{tabular}
\end{table}

\section{Data analysis}
\label{sec:radioobs}

\subsection{Radio observations}

The cluster  has been observed with the JVLA in the L-band (1-2 GHz) in A, B, C and D configurations. These observations have a total bandwidth of 1024 MHz, subdivided  into 16 spectral windows of 64 MHz each (with 64 channels at frequency resolution of 1 MHz). We also reduced and analysed archival data in S-band (2-4 GHz) in DnC configuration,  covering a total of 2048 MHz in 16 spectral windows of 128 MHz each (64 channels of 2 MHz channel$^{-1}$ frequency resolution). Both  data sets have full polarization products. Observing date, time, rms noise ($\sigma$) and restoring beam of radio observations are  listed in Tab.~\ref{tab:radio}.

\begin{table*}
    \centering
	\caption{Details of radio observations. Column 1: central observing frequency; Column 2: name of the frequency band; Column 3: array configuration; Column 4: date of the observation; Column 5: observing time; Column 6: \texttt{robust} parameter used for the Briggs weighting scheme \citep{Briggs95} during imaging process; Column 7: size of the Gaussian taper used in the imaging process. If \textquotedbl--\textquotedbl, no taper has been used; Column 8:  Full Width Half Maximum (FWHM) of the major and minor axes of the restoring beam of the final image; Column 9: 1$\sigma$ rms noise of the total intensity image; Column 10: reference of the figures in this paper. Rms noise and beam shape of the images obtained with a combination of different configurations are reported under the horizontal line.}
	\label{tab:radio}
	\begin{tabular}{lccccccccc} 
	    \hline
		\hline
		Freq.  & Band & Array Conf. & Obs. Date & Obs. Time & Robust & Taper & Beam & rms Noise ($\sigma$) & Fig. \\
		(GHz)      &      &               &           &    (hr)   &        &       &
		    & (mJy/beam) & \\
		\hline
		1.5     &   L   &    A       & 2018 Mar. - 2018 Apr.&  5.5  & 0.5 & -- &   2$\arcsec\times$1$\arcsec$ & 0.011 & \ref{fig:optic} \\
		1.5     &   L   &    B       & 2017 Oct.  &  2.0  & 0.5 & -- &  7$\arcsec\times$4$\arcsec$ & 0.018 & \\
		1.5     &   L   &    C       & 2017 Jun.  &  2.0  & 0.5 & -- & 24$\arcsec\times$11$\arcsec$ & 0.035  & \ref{fig:pol_CS_E},\ref{fig:pol_CS_W} \\
		1.5     &   L   &    D       & 2017 Feb.  &  0.5  & 0.5  & -- &  74$\arcsec\times$33$\arcsec$ & 0.3  & \\
		3.0     &   S   &   DnC      & 2014 Sept. &  6.0  & 0.5 & -- & 17$\arcsec\times$13$\arcsec$ & 0.012 & \ref{fig:S}, \ref{fig:pol_CS_E},\ref{fig:pol_CS_W} \\
		\hline
		1.5     &   L   &  B+C     &  &   & 0.5 & -- &  9$\arcsec\times$5$\arcsec$ & 0.015  & \ref{fig:optic} \\
		1.5     &   L   &  C+D     &  &   & 0.5 & -- & 25$\arcsec\times$11$\arcsec$ &  0.014  & \ref{fig:CD} \\
		1.5     &   L   &  B+C+D   &  &   & 0.0 & 15$\arcsec\times$15$\arcsec$  & 17$\arcsec\times$14$\arcsec$ & 0.035 & \ref{fig:Spixmap_E},\ref{fig:Spixmap_W} \\
		1.5     &   L   &  A+B+C+D &  &   & 0.5 & 8$\arcsec\times$8$\arcsec$  & 15$\arcsec\times$8$\arcsec$ & 0.016 & \ref{fig:X} \\
%		2.5     &   S+L &  B+C+DnC &  &   & 0.5 & --  & 25$\arcsec\times$25$\arcsec$ & 0.02 & \ref{fig:polSL}, \ref{fig:phiSL} \\
		\hline
	\end{tabular}
\end{table*}

\subsubsection{Calibration}
\label{sec:reduction}

For calibration and total intensity imaging we used the \texttt{CASA 5.3.0}\footnote{\url{https://casa.nrao.edu/}} package. We started the calibration process from data pre-processed by the VLA \texttt{CASA} calibration pipeline\footnote{\url{https://science.nrao.edu/facilities/vla/data-processing/pipeline}} which performs basic flagging and calibration on Stokes $I$ continuum data. Then, we derived final delay, bandpass, gain/phase, leakage and polarization angle calibrations and applied them to the target. The source 3C\,286 was used as a bandpass, absolute flux density and polarization angle calibrator for all the observations. We used the \citet{Perley13} flux  density scale and we followed the National Radio Astronomy Observatory (NRAO) polarimetry guide for polarization calibration\footnote{\url{https://science.nrao.edu/facilities/vla/docs/manuals/obsguide/modes/pol}}. In particular, we performed a polynomial fit to the values of linear polarization fraction and angle tabulated in \citet{Perley13} for 3C\,286, to obtain a frequency-dependent polarization model. J1248-1959 was used as a phase calibrator for observations in A and B configurations in L-band, and for the S-band observations, while J1311-2216 was used for the observations in C and D configurations (L-band). To correct for the instrumental leakage, an unpolarized source was used: J1407+2827 and 3C\,147 for the L- and S-band observations, respectively.

Radio frequency interference (RFI) was removed by statistical flagging algorithms also from the cross correlation products. At the end of the flagging process, some spectral windows  seriously affected by RFI were entirely removed. We flagged the frequency ranges: 2116-2244 MHz in the S-band observations; from 1520 MHz to 1584 MHz in A configuration; from 1072 MHz to 1136 MHz and from 1520 MHz to 1648 MHz in B configuration; 1136-1264 MHz and 1520-1584 MHz in C configuration; from 1136 MHz to 1328 MHz and from 1520 MHz to 1648 MHz in D configuration. After RFI removal, we averaged the data sets in time down to 10 seconds to speed up the imaging process and we re-weighted the visibilities according to their scatter.

\subsubsection{Imaging and self-calibration}
\label{sec:imaging}

We used the multi-scale multi-frequency de-convolution algorithm of the \texttt{CASA} clean \citep{Rau11} for wide-band synthesis-imaging. We set two terms for the Taylor expansion  (\texttt{nterms} = 2) in order to take into account both the source spectral index (likely a power-law) and the primary beam response. We also used a $w$-projection algorithm to correct for the wide-field non-coplanar baseline effect \citep{Cornwell08} with an appropriate number of $w$-projection planes for each data set. We generally used the Briggs weighting scheme with the \texttt{robust} parameter set to 0.5.  We highlight in Tab.~\ref{tab:radio} the cases in which a different weighting scheme has been used.

There  are two bright sources in the target field, one south-west of the cluster and the other to the north-east. The latter falls at the edge of the primary beam in L-band observations, causing problems for the self-calibration procedure. We set  \texttt{nterms} = 3 to individually image these sources. Then, we used the peeling technique to subtract them  out of the images with direction-dependent gain solutions derived for each one. Some artifacts  around the brightest source in the south are still present in the final images but their effect on the cluster emission is negligible. We used the peeling technique to subtract two  variable sources before combining data at various configurations observed in different dates. 

The combination of various antenna configurations differently flagged in frequency could in principle cause imaging artifacts due to uneven $uv$-coverage. To exclude the presence of strong artifacts we imaged each configuration individually before combining the different configurations.

Finally, cycles of self-calibration were performed to refine the antenna-based phase gain variations on the target field. The residual amplitude errors due to the calibration are estimated to be $\sim5 \ \%$. The local rms noise of the images is reported in Tab.~\ref{tab:radio}. The final images were corrected for the primary beam attenuation using the  \texttt{widebandpbcor} task in \texttt{CASA}. 

The S-band observations were performed  on two pointings roughly centred on the east (E) and west (W) relic. We separately performed data reduction, peeling and imaging of the two fields. Then, we  joined the two final images correcting for the primary beam attenuation of both pointings.

\begin{figure}
	\includegraphics[width=\columnwidth]{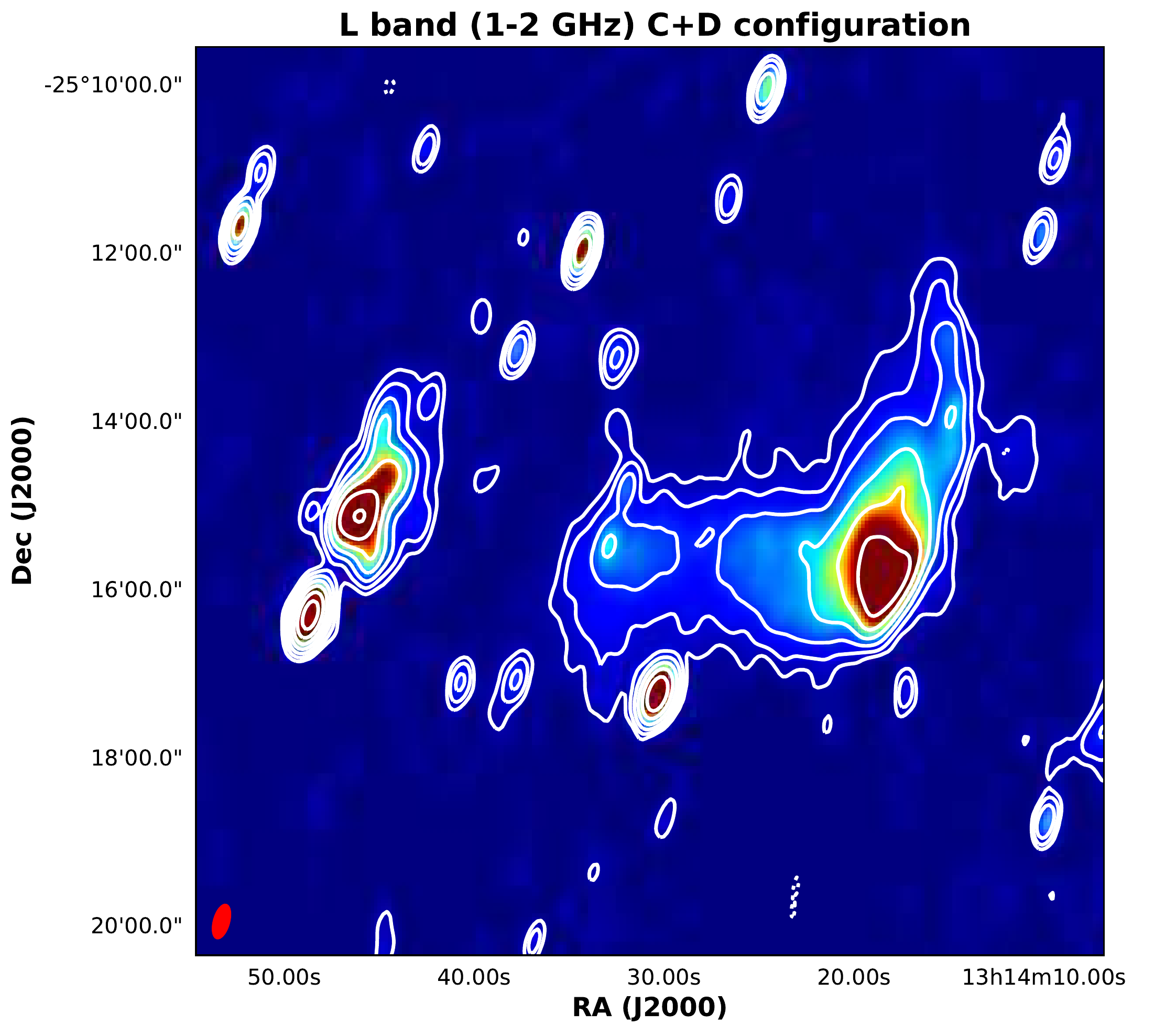}
    \caption{Lowest resolution image of the cluster in C+D configuration at 1.5 GHz. White contours are overlaid, starting from $\pm3\sigma$, with $\sigma$=0.014 mJy beam$^{-1}$, and they are spaced by a factor of two. Negative contours are dotted. The restoring beam of 25$\arcsec\times11\arcsec$ is shown in red in the left-hand corner and has a physical size of $\sim$70 kpc.}
    \label{fig:CD}
\end{figure}

In Fig.~\ref{fig:optic}, radio contours at 1.5 GHz in the combined B and C (B+C) configurations are overlaid  to the optical image of \RX composite  of {\it Subaru} r- and g-band.  A zoom of the E relic with A configuration high-resolution contours is also shown in Fig.~\ref{fig:optic}. In Fig.~\ref{fig:X} the radio contours obtained combining all the L-band observations (A+B+C+D) are overlaid on the X-ray {\it XMM-Newton} image of the cluster. The L-band image obtained combining C and D (C+D) configurations is shown in Fig.~\ref{fig:CD}. The S-band image in DnC configuration is shown in Fig.~\ref{fig:S}. 

\subsection{X-ray observations}
\label{sec:xraysobs}

 We retrieved from the {\it XMM-Newton} Science Archive two observations on \RX (ObsID: 0501730101 and 0551040101), accounting for a total exposure time of $\sim110$ ks. The data sets were processed using the {\it XMM-Newton} Scientific Analysis System (\texttt{SAS v16.1.0}) and the Extended Source Analysis Software (\texttt{ESAS}) data reduction scheme \citep{Snowden08} following the working flow described by \citet{Ghirardini19}. We combined the count images and corresponding background and exposure maps of each ObsID to produce a single background-subtracted image also corrected for the effects of vignetting and exposure time fluctuations. The image in the $0.5-2.0$ keV band is shown in Fig.~\ref{fig:X}.

After the excision of contaminating point sources, we performed surface brightness and spectral analyses in the region of the western radio relic. The cluster emission was described with a thermal model with  fix metallicity of 0.3 $Z_\odot$ \citep[e.g.][]{Werner13} and taking into account the Galactic absorption in the direction of the cluster as reported in \citet{Kalberla05}. The background was carefully treated by modelling both the astrophysical and instrumental components \citep[see][for more details  on the procedure]{Ghirardini19}.

\section{Study of the continuum emission}
\label{sec:continuum}

%\begin{commen}

\begin{figure}
	\includegraphics[width=\columnwidth]{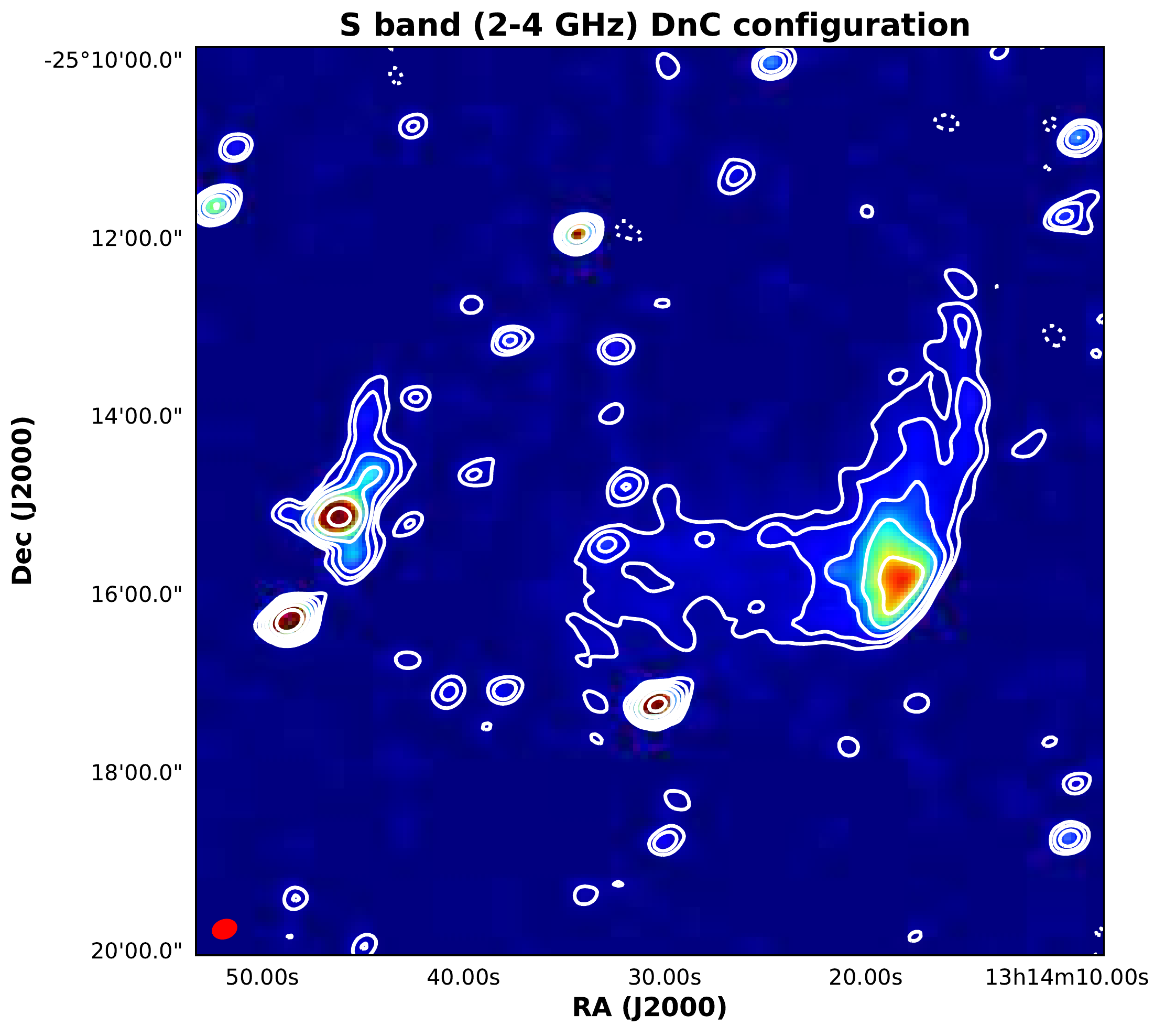}
    \caption{DnC configuration image at 3 GHz. White contours start by $\pm3\sigma$, with $\sigma$=0.012 mJy beam$^{-1}$, and they are spaced by a factor of two. Negative contours are dotted. The restoring beam is 17$\arcsec\times13\arcsec$ and it shown in red at the left-hand corner. It has a physical size of $\sim$58 kpc.}
    \label{fig:S}
\end{figure}

%\end{commen}

\subsection{Description of Radio Sources}

The eastern and western relics have a different shape but they are at  about the same  projected distance of $\sim$750 kpc from the central brightest cluster galaxy of the main sub-cluster (BCG\,1, at redshift $z$=0.246, see Fig.~\ref{fig:optic}). A comparison between radio and X-ray  images (Fig.~\ref{fig:X}) shows that the two relics are  on the opposite sides of the cluster while the radio halo overlaps with the X-ray emission. There is a shift between the peak of X-ray surface brightness and the position of the BCG\,1, as  may be expected from an interacting system \citep{Rossetti16}. The X-ray emission is elongated along the east-west merger axis and it is brighter on the western side of the cluster. At the position of the nose of the W relic, the X-ray emission has a sharp drop  where a shock was first detected by \citet{Mazzotta11}. We confirm and discuss the shock detection in Sec.~\ref{sec:discussrelicshock}. A stream of gas that follows the profile seems to connect the main cluster with a southern sub-cluster but in this region we did not detect any diffuse radio emission.

\subsubsection{The Eastern Relic}
\label{sec:imagingE}

The eastern relic has a largest linear size of $\sim$ 500 kpc. Its morphology and a plausible association with optical sources made \citet{Feretti05} cautious about its identification with a radio relic. With high-resolution A configuration imaging we discovered that a narrow angle tail radio galaxy \citep[NAT, e.g.][]{Miley80} is embedded in the  diffuse emission (marked with $A$ in Fig.~\ref{fig:optic}). This radio galaxy is a cluster member at redshift $z$=0.242 \citep{Golovich18,Golovich19}. The diffuse emission is clearly related to the NAT but it extends well beyond radio galaxy lobes in the N-S direction and its largest size is perpendicular to the E-W merger axis. In Sec.~\ref{sec:discussAGNrelic} we discuss the relic-AGN connection using spectral index and polarization analyses. Another radio galaxy without redshift estimate ($C$ in Fig.~\ref{fig:optic}) lies  to the south of the E radio relic, at a  projected distance of $\sim$200 kpc.

The E relic lies in a region of low X-ray surface brightness that prevents the possible detection of a shock related to the relic emission (see Fig.~\ref{fig:X}.)

\subsubsection{The Western Relic and the Radio Halo}
\label{sec:imagingW}

The dominant radio feature of \RX is in the  western part of the cluster. The faint diffuse emission  of the halo with a roundish shape of radius $\sim65\arcsec$ (i.e., 250 kpc) is visible at the centre of the cluster in Fig.~\ref{fig:X}. The emission broadens and brightens to the west. Then it  bends to the north and two arcs  detach from the brightest region of the W relic along the N-S direction. The innermost one extends for approximately 140$\arcsec$, corresponding to $\sim$550 kpc, and  it is the brightest one. The outermost arc is more extended, reaching a largest linear size of 970 kpc and a transverse size in the thinnest part of $\sim$80 kpc. The inner arc seems to follow the sharp X-ray profile in the west side of the cluster, while the longest one lies outside the region where the X-ray shock was detected. No clear optical counterpart could be associated  with this radio emission. A point-like radio source without redshift estimate lies along the outermost arc  (labelled with $D$ in Fig.~\ref{fig:optic}). All these features are observed also in the S-band image in Fig.~\ref{fig:S}.

\subsection{Spectral index study}
\label{sec:spectralstudy}

Using archival S-band data, we performed  the spectral analysis of the extended emission to locate the site of particle acceleration.

We computed the spectral index between 1.5 and 3 GHz based on the combined L-band (B+C+D in Tab.~\ref{tab:radio}) and S-band observations. We imaged the L-band observations with the same  $uv$-range as the S-band data (0.19-23.7 k$\lambda$). We used the same pixel size and baseline interval, and set data weights in order to reach a similar beam size as in the S-band. After cleaning, we convolved the two images  to the same Gaussian beam with FWHM of 18.5$\arcsec$ and we corrected them for the primary beam response. We have checked the position of  a number of point-like sources  in the two data sets to exclude any  significant astrometric offset between them.

\begin{figure}
	\includegraphics[width=\columnwidth]{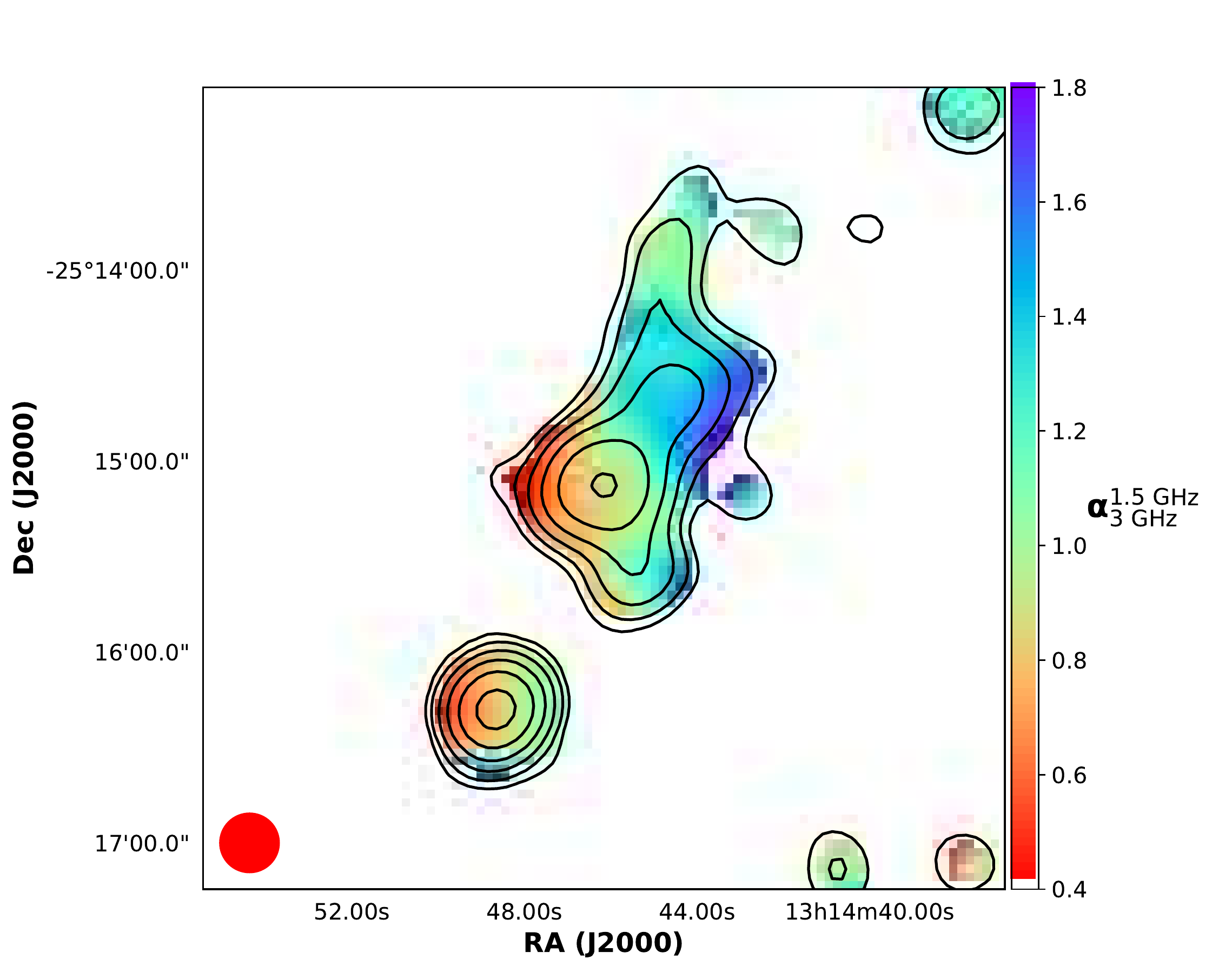}
	\hfill
	\includegraphics[width=\columnwidth]{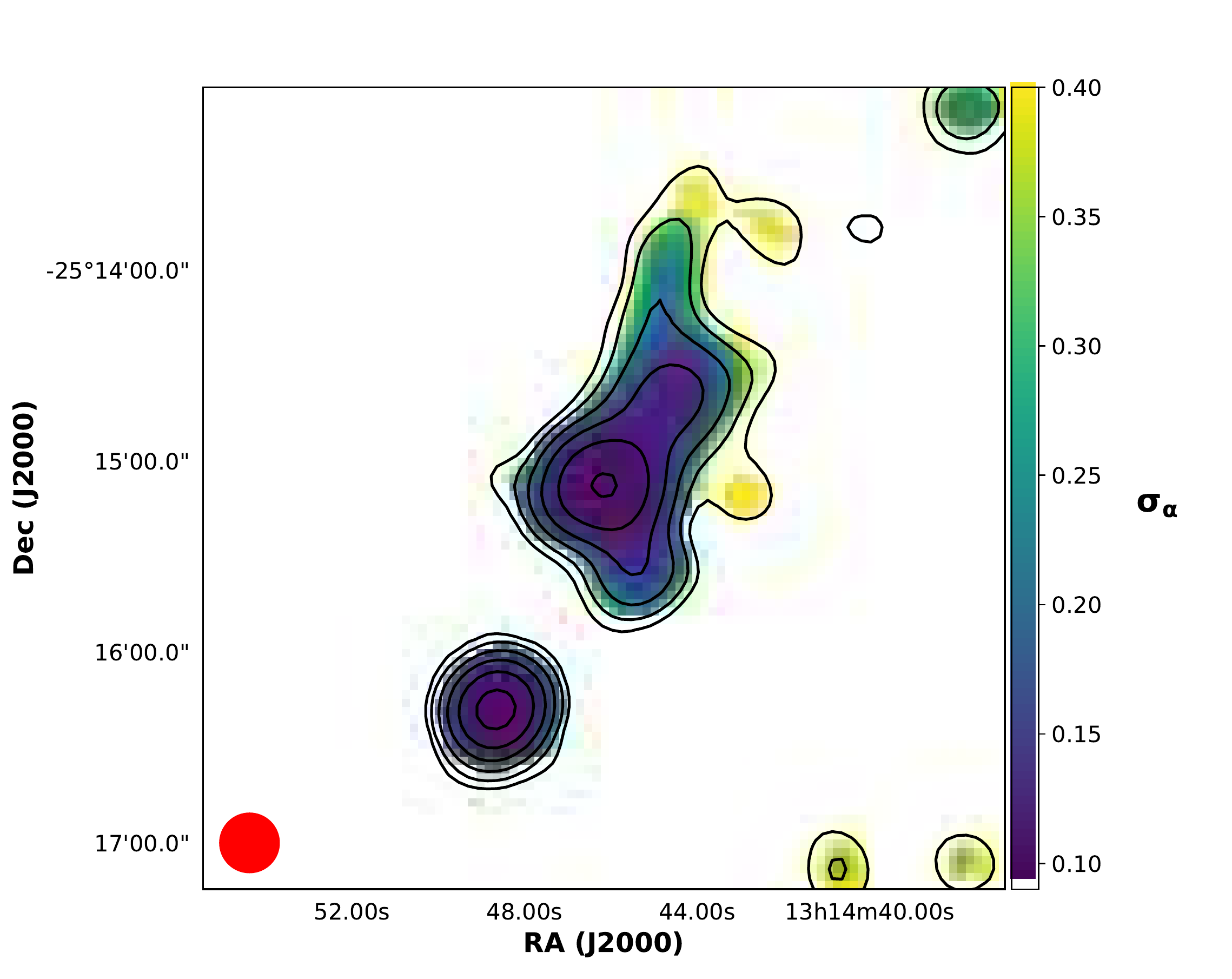}
    \caption{Top panel: spectral index image of the E relic region between 1.5 and 3 GHz. The 1.5 GHz total intensity contours (B+C+D configuration) are overlaid in black: the levels start at 3$\sigma$, with $\sigma$=0.04 mJy beam$^{-1}$,  and are separated by a factor of two. Bottom panel: error map of the spectral index image with the same contours overlaid. The restoring beam of the two images is shown in the left-hand corner and its size is 18.5$\arcsec\times18.5\arcsec$.}
    \label{fig:Spixmap_E}
\end{figure}

\begin{figure}
	\includegraphics[width=\columnwidth]{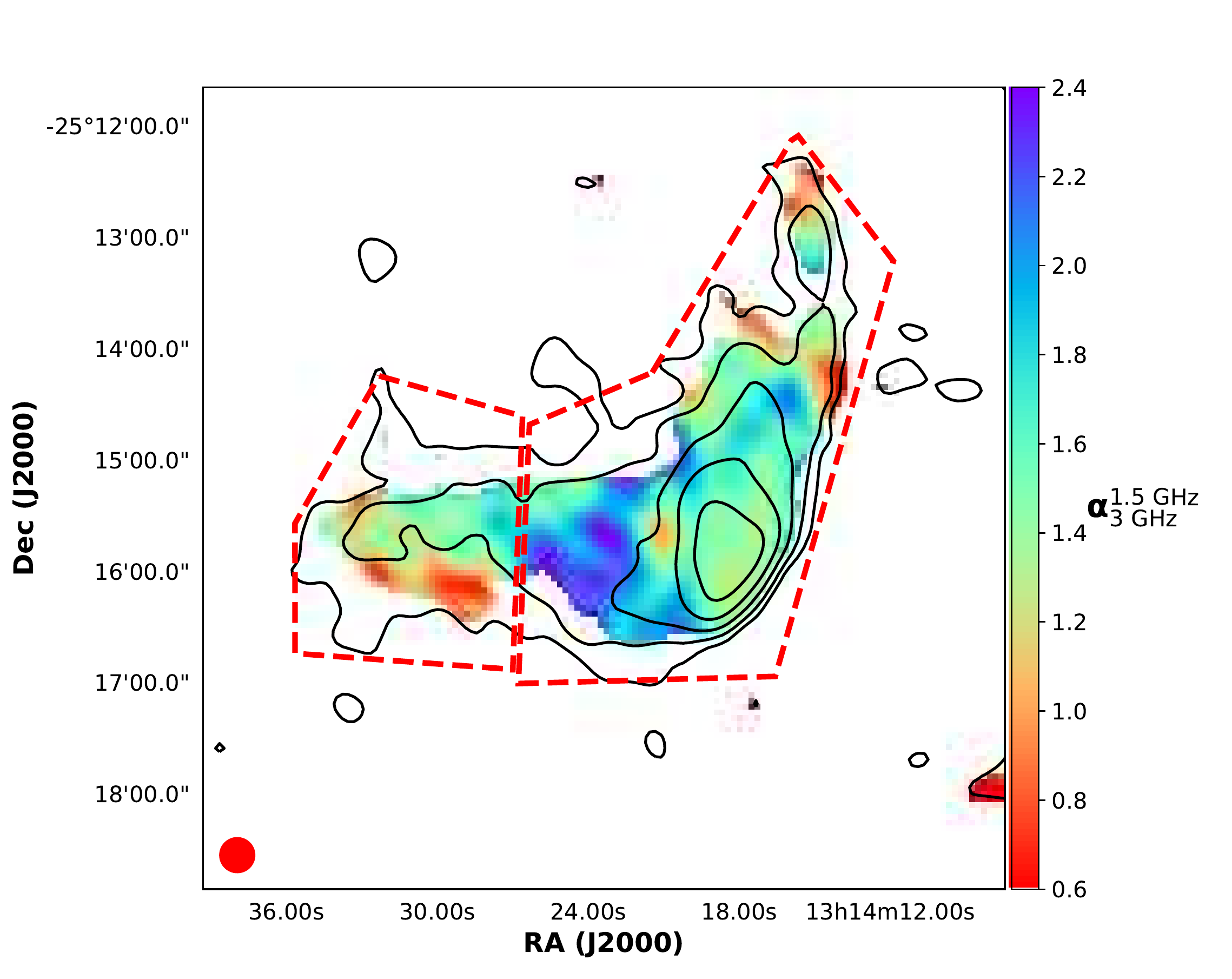}
	\hfill
	\includegraphics[width=\columnwidth]{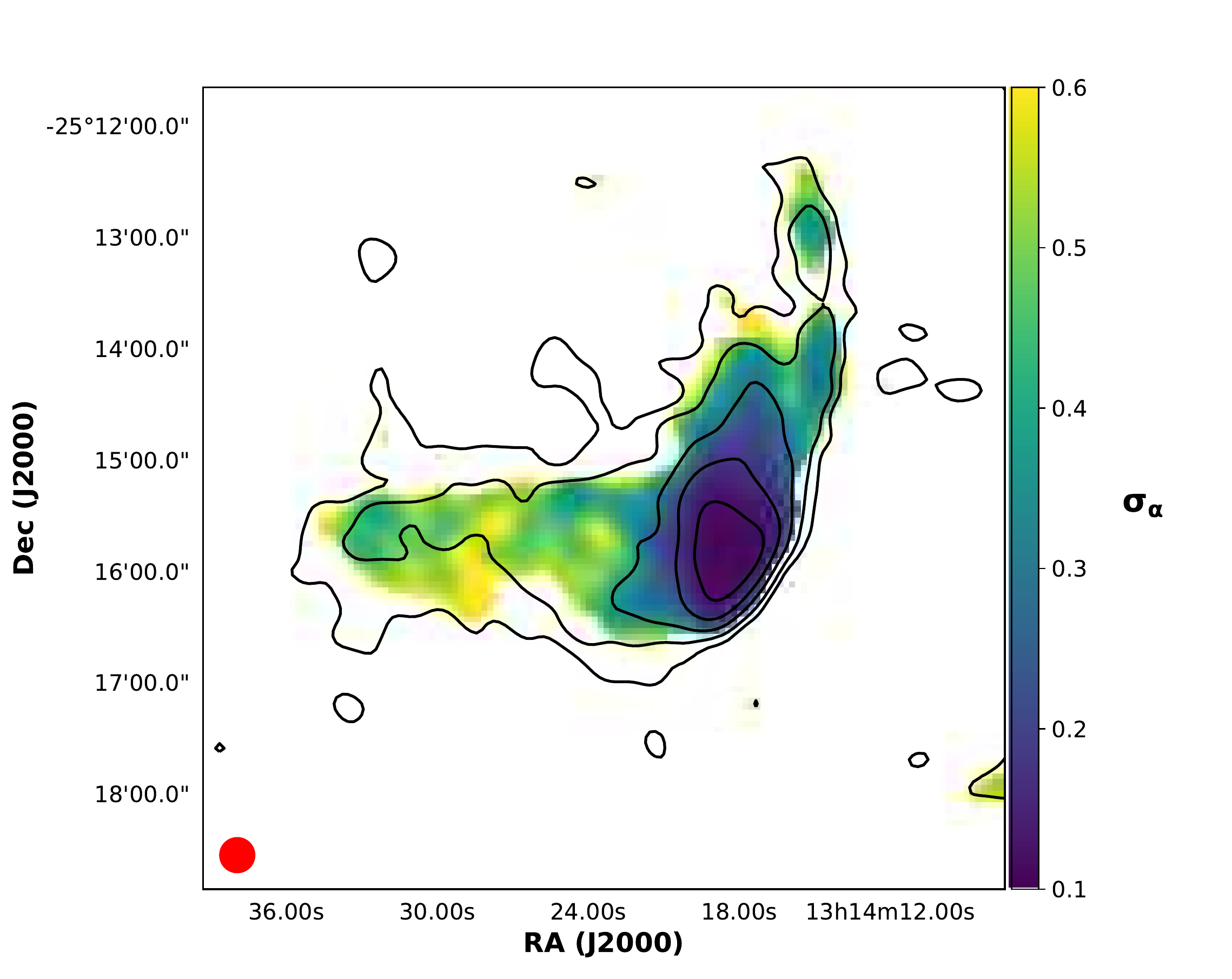}
    \caption{Spectral index image of the W relic and halo region computed between 1.5 and 3 GHz.  Point-like sources were subtracted. The two dashed red polygons mark the flux density extraction regions for the halo and the West relic. Bottom panel: error map of the spectral index image. Contours and restoring beam are as in Fig.~\ref{fig:Spixmap_E}.}
    \label{fig:Spixmap_W}
\end{figure}

For the W relic and the halo, we computed the spectral index excluding point-like sources. We first imaged both S- and L-band data sets excluding short baselines  (i.e., < 3.5 k$\lambda$) which are sensitive to extended emission (i.e.,  larger than 90$\arcsec$ corresponding to 350 kpc). Then, we subtracted from the original visibilities  the corresponding model components and we made new images  using all the baselines  at the resolution of 18.5$\arcsec$. This procedure was also applied to the E relic, but since the emission of the NAT is extended, it is impossible to  properly separate the contribution of the tail from the relic. Hence, we decided not to subtract the radio galaxy. This choice allows us to study the spectral index behaviour from the core of the NAT to the lobes, and to investigate  its connection with the diffuse source.

We computed the average flux density and spectral index of the E relic  (and of the NAT radio galaxy) and of the W relic and the radio halo emission (sources subtracted) between 1.5 and 3 GHz. The values are reported in Tab.~\ref{tab:radiopower}. The uncertainties on the flux  density measurements are computed as:

\begin{equation}
    \sigma_S=\sqrt{(\delta S \times S)^2+(\sigma\times\sqrt{n_{\rm beam}})^2}~,
\end{equation}

where $\delta S=5 \ \%$ is the calibration error, $\sigma$ is the rms noise listed in Tab.~\ref{tab:radio} and $n_{\rm beam}$ is the number of beams in the sampled region. These uncertainties were then propagated  to the spectral index.

We extrapolated the radio power at 1.4 GHz from the 1.5 GHz flux  density measurement, considering a luminosity distance $D_{L}=1253.3$ Mpc \citep{Wright06} and using the spectral  indices for the $k$-correction. The uncertainties  on flux densities were propagated to the radio power.  The sum of the radio power of the two relics (sources subtracted) is consistent with the relation found by \citet{deGasperin14} between the radio power of double radio relics and the cluster mass.

Finally, we computed the spectral index for each pixel with value $>3\sigma$ in both frequency bands. We show the spectral index map of the E relic in Fig.~\ref{fig:Spixmap_E}. We propagated the uncertainties on the flux densities pixel by pixel on the spectral index. The error map of the spectral index for the E relic is shown in the bottom panel of Fig.~\ref{fig:Spixmap_E}, while the spectral index image of the western region and its error map are shown in Fig.~\ref{fig:Spixmap_W}.

\begin{figure}
    \begin{picture}(100,100)
    \put(0,0){\includegraphics[width=\columnwidth,height=6cm]{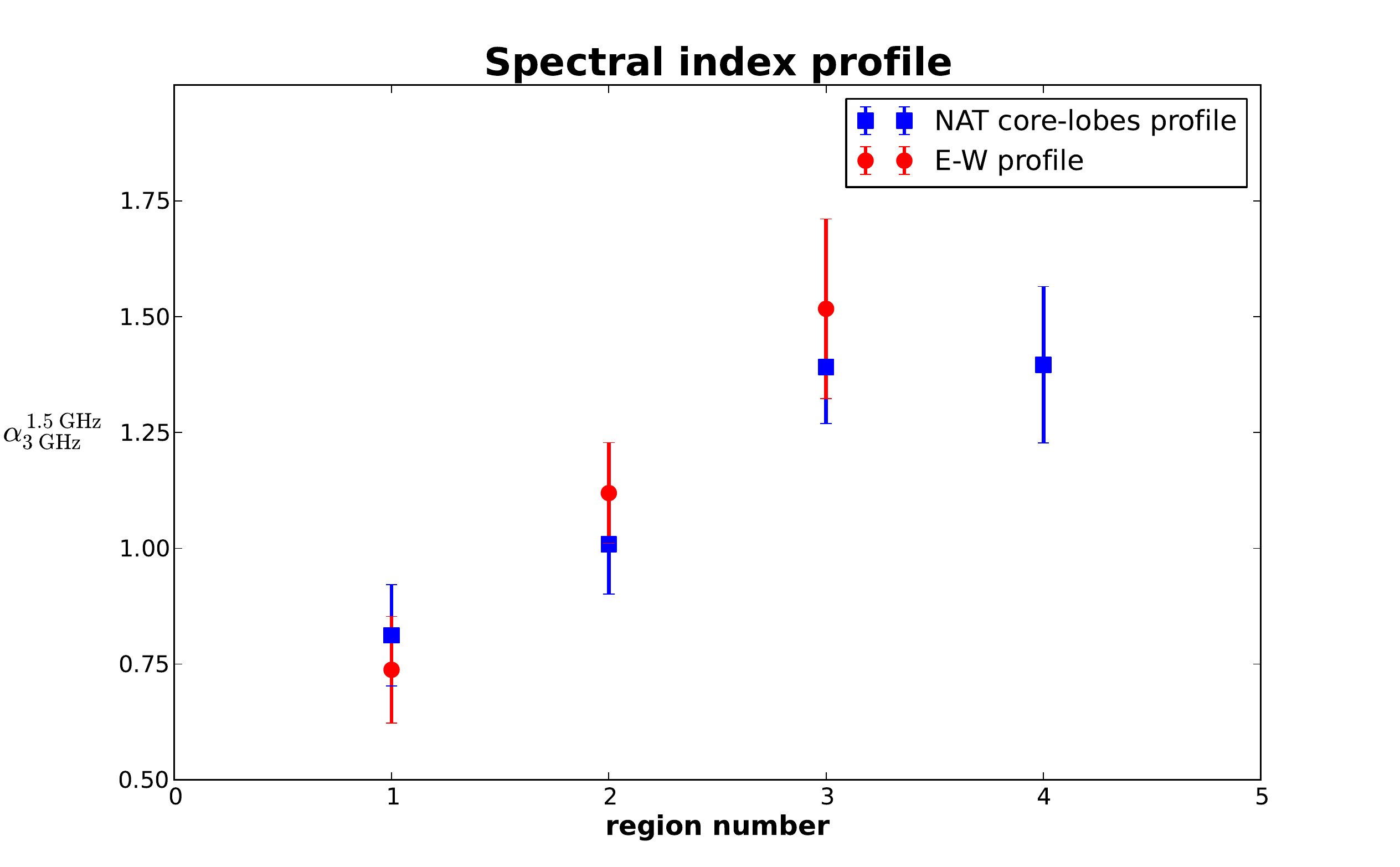}}
    \put(35,90){\includegraphics[width=2cm,height=2.2cm]{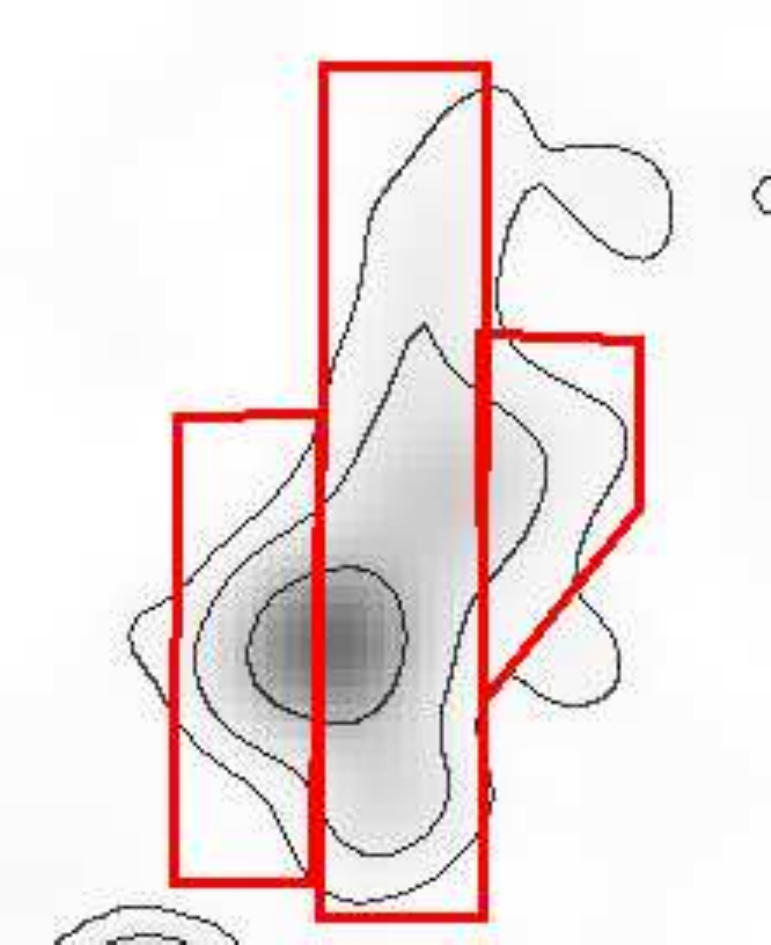}}
    \put(155,20){\includegraphics[width=2.1cm,height=2.2cm]{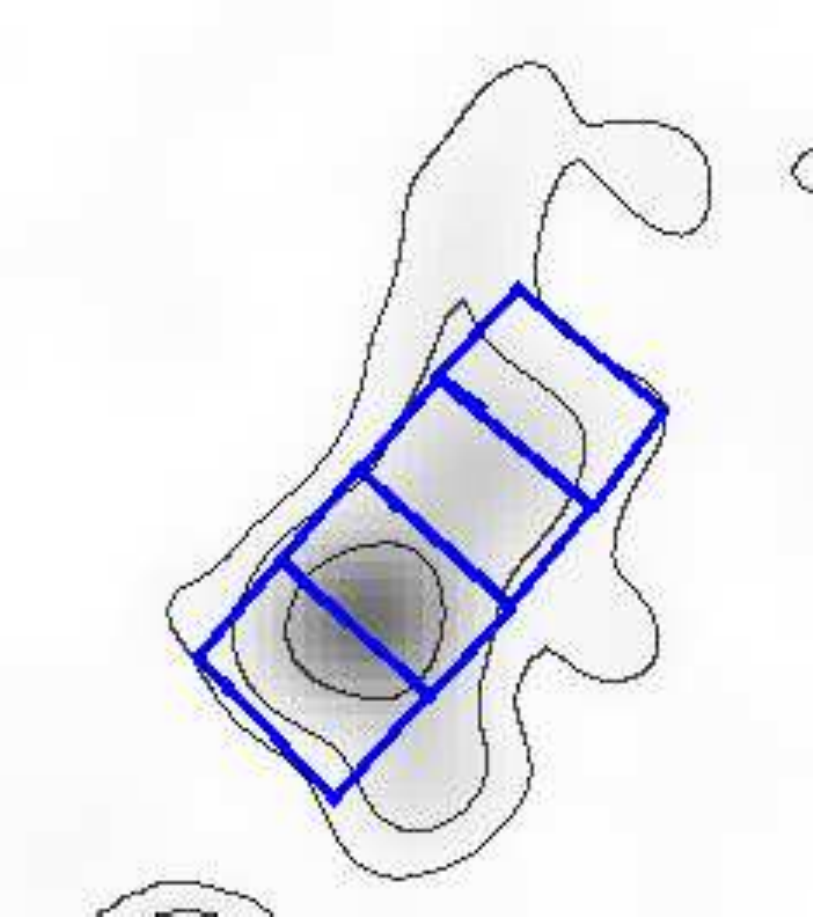}}
    \end{picture}
    \caption{ Spectral index profiles in the E relic. The  spectral index is computed in the regions shown in the inset panels and numbered following the profiles. The regions used for the red profile (plotted with round marks) are shown in the top-left inset panel and are numbered from E to W. The regions used for the blue profile (plotted with square marks) are shown in the bottom-right inset panel and are numbered from the core in the SE to the lobes in the NW. We drew the regions in order to avoid point-like sources surrounding the extended emission.}
    \label{fig:spixprof_E}
\end{figure}

\begin{figure}
    \topinset{\includegraphics[height=2.3cm]{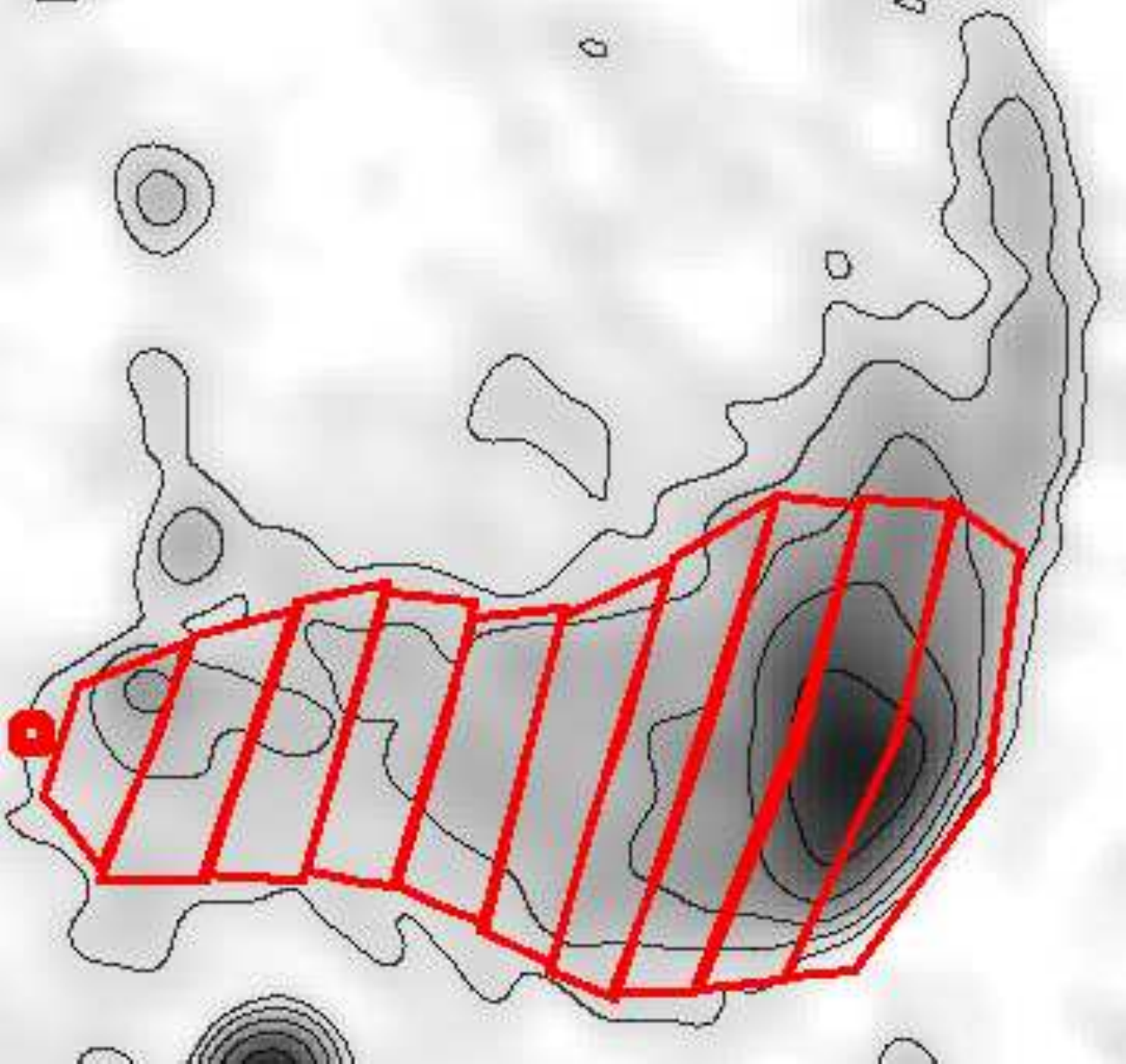}}{\includegraphics[width=\columnwidth,height=6.1cm]{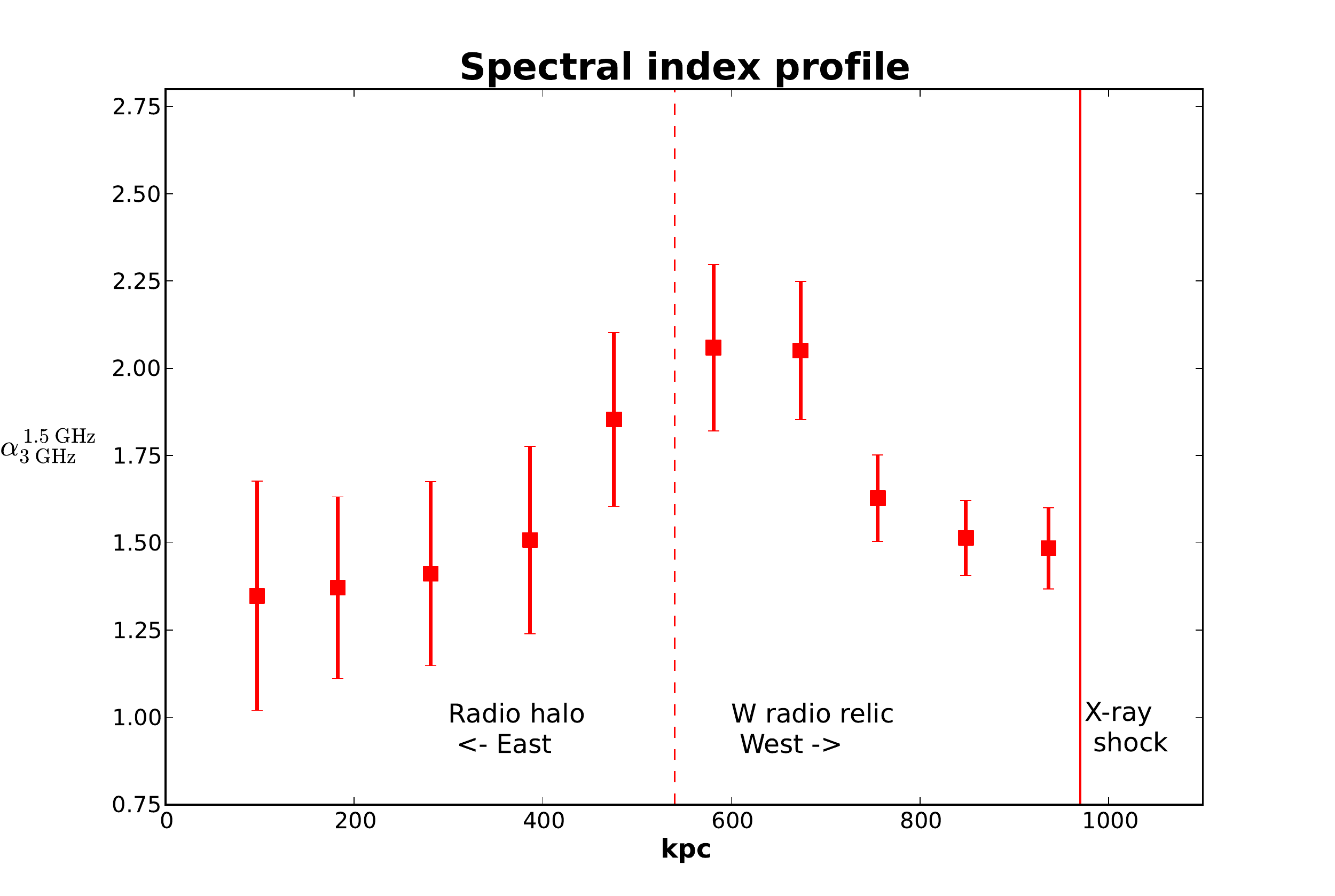}}{20pt}{-50pt}
    \caption{Spectral index profile in the W relic and halo region where point-like sources were  subtracted. The origin of the x axis coincides with the position marked with the red point in the inset panel. The solid red line shows the position of shock front detected in X-ray  while the dashed line separates the W relic and halo region.}
    \label{fig:spixprof_W}
\end{figure}

\begin{table}
    \centering
	\caption{Flux density, spectral index between 1.5 and 3 GHz and radio power of extended sources measured from the images used for the spectral index study. We separated the region of the W relic and the halo on the basis of the spectral index profile as described in Sec.~\ref{sec:spectralstudyW}.  The ``$^{*}$'' stands for point-like sources subtracted.}
	\label{tab:radiopower}
	\begin{tabular}{lcccccc} 
	    \hline
		\hline
		Source & Freq. & Flux Density & $\alpha^{\rm 1.5 \ GHz}_{\rm 3 \ GHz}$ & P$_{\rm 1.4 \ GHz}$ \\
		       &   (GHz)   &  (mJy)       &          & (10$^{24}$W/Hz)    \\
		\hline
		\multirow{ 2}{*}{E relic}  &  1.5  &  11.3$\pm$0.6 & \multirow{ 2}{*}{}{1.0$\pm$0.1} & \multirow{ 2}{*}{2.3$\pm$0.1}   \\
		                                     &  3    &   5.6$\pm$0.3 &   &     \\
	    %\multirow{ 2}{*}{E relic$^{*}$} &  1.5  &  8.0$\pm$0.4 & \multirow{ 2}{*} 2.1$\pm$0.1  &  \multirow{ 2}{*}{2.0$\pm$0.1}  \\
		                                   %  &  3    &   1.8$\pm$0.1 &   &    \\
		\multirow{ 2}{*}{W relic$^{*}$}            &  1.5  &  33$\pm$2  & \multirow{ 2}{*}{1.6$\pm$0.1}    &  \multirow{ 2}{*}{7.9$\pm$0.5}  \\
		                                     &  3    &   10.9$\pm$0.8 &  &  \\
		\multirow{ 2}{*}{Halo$^{*}$}               &  1.5  &  5.3$\pm$0.3 & \multirow{ 2}{*}{1.3$\pm$0.2}  &  \multirow{ 2}{*}{1.16$\pm$0.07}  \\
		                                     &  3    &   2.1$\pm$0.2 &  &   \\
		\hline
	\end{tabular}
\end{table}

Spectral index  images obtained from interferometric data should be treated with caution, since the value of $\alpha$ that we computed  in each pixel is not independent  of the neighbouring pixels. Furthermore, some artifacts  (due to  slightly different sampling of the baselines and to imaging artifacts in the two original images) could affect spectral index images, in particular at the edge of extended emission or where point-like sources were subtracted out. We can recover the spectral index variations with higher significance by integrating the flux over regions larger than the beam size. Regions were chosen by following the features observed in the spectral index maps and  with the goal of obtaining a signal-to-noise ratio higher than 3 in each region. Spectral index profiles of the E relic and of the W relic and the halo emission are shown in Fig.~\ref{fig:spixprof_E} and Fig.~\ref{fig:spixprof_W}, respectively.

\subsubsection{The Eastern Relic}
\label{sec:spectralstudyE}

The  integrated spectral index found in the E relic, including the NAT, is in agreement with the one obtained by \citet{George17} with a  power-law fit of the spectrum between 118 MHz and 1.4 GHz, $\alpha^{\rm 118MHz}_{\rm 1.4GHz}$=1.03$\pm$0.12. No spectral curvature is thus observed for this relic up to 4 GHz.

The spectral index image is shown in Fig.~\ref{fig:Spixmap_E}. It is not trivial to identify a single spectral gradient of the whole extended emission. A steepening trend is visible from the core of the radio galaxy to its lobes but it is possible to draw a steepening trend also from E to W. The first trend is expected if the only radio galaxy originates the emission while the second is expected from the acceleration of particles from a shock wave propagating outwards from W to E. If these two trends were both present they would be mixed in the spectral index map, due to the physical superimposition and of the resolution of our image.

In Fig.~\ref{fig:spixprof_E} we trace both the expected trends and the spectral index profiles are displayed in different colours. The blue profile follows the core-lobes direction as can be derived from the high resolution A configuration image at a position angle of -45$^{\circ}$ (see Fig.~\ref{fig:optic}). The flattest spectral index corresponds to  the core of the radio galaxy in the first blue region, with $\alpha$=0.8$\pm$0.1. The spectral index steepens toward the tails of the radio galaxy and in the fourth region, furthest away from the core, it remains constant. The fact that the northern and southern edges of the emission do not follow the core-lobes direction and show a flatter spectral index (Fig.~\ref{fig:Spixmap_E}) than the fourth region of the profile, indicates that the propagation of jets has been perturbed and/or that another mechanism is possibly accelerating particles.

With the red boxes we traced the E-W profile (avoiding nearby point-like sources in the W and E of the extended emission). A clear steepening trend is observed also in this direction. This is suggestive of an acceleration process that is active along the whole length of the E relic and not only originating in the core of the radio galaxy. This scenario is further discussed in Sec.~\ref{sec:discussAGNrelic}.

\subsubsection{The Western Relic and the Radio Halo}
\label{sec:spectralstudyW}

We computed the spectral index profile of the western  emission region using ten regions between the centre of the cluster to the region of the shock detected in the X-ray  (see Fig.~\ref{fig:spixprof_W}). The emission of point-like sources was subtracted out of the images as explained in Sec.~\ref{sec:spectralstudy}.

In the region close to the shock front, the spectral index is  $\alpha$=1.5$\pm$0.1. In Sec.~\ref{sec:discussrelicshock} we derive the Mach number of the underlying shock wave assuming the DSA mechanism and compare it to the one derived from X-rays.

The spectral index steepens towards the cluster centre, reaching a peak value of  $\alpha$=2.1$\pm$0.2 at $\sim$400 kpc from the shock front. A steepening trend is expected in the downstream of a shock wave where energetic particles cool down, and it is often observed in radio relics \citep[e.g.][]{Hoang17}.

Toward the halo region, the spectral index flattens again. The trend is clear also from Fig.~\ref{fig:Spixmap_W}. The spectral profile strongly resembles the one observed in the Toothbrush radio relic \citep{vanWeeren16}. The flattening of the spectrum can be explained by the presence of another mechanism  re-accelerating particles in the central region of the cluster.  This is further discussed in Sec.~\ref{sec:discusshalo}. 

The profile flattens and remains almost constant in the four central regions. According to the spectral index profile, we disentangled the region of the relic from that of the radio halo. We assumed that the steepening of the spectrum  up to $\alpha\sim$2 is due to the aging of particles in the  region downstream the shock. We considered as radio halo the region where the spectral index flattens again.  In Fig.~\ref{fig:Spixmap_W} we show the two  approximate regions with red dashed polygons. In Tab.~\ref{tab:radiopower} we reported average spectral index and radio power of W relic and radio halo.

\citet{George17} derived a spectral index of $\alpha^{\rm 118MHz}_{\rm 1.4GHz}=1.23\pm0.09$ for the western relic subtracting the flux density of the radio halo extrapolated from the 610 MHz measurement. We instead derived two distinct integrated values: in the halo  $\alpha=1.3\pm0.2$ and in the relic  $\alpha=1.6\pm0.1$. We can also estimate the spectral index of the halo between 610 MHz and 1.5 GHz using the work of \citet{Venturi07} and integrating  over the same physical region: $\alpha^{\rm 610MHz}_{\rm 1.5GHz}=1.1\pm0.1$. Our spectral index estimates within the uncertainties are consistent with those measured by \citet{Venturi07} at 610 MHz. We did not detect any signs of curvature in the integrated spectrum within the sampled frequency range.

\section{Polarized intensity study}
\label{sec:polarization}

\subsection{Theoretical background}

Following \citet{Burn66} we define the  linear polarization vector $P$ as a complex quantity:

\begin{equation}
    P=Q+iU=|P|e^{2i\chi}~,
\end{equation}

where $\chi$ is the polarization angle of the radiation and $Q$ and $U$ are the Stokes parameters.

The high degree of fractional polarization $p=P/I$ observed from radio relics unveils the presence of ordered magnetic field components lying in the plane of the sky ($B_{\perp}$). Faraday depolarization may decrease the observed fractional polarization at large wavelength depending on the physical properties of the magneto-ionic medium between the source and the observer \citep{Sokoloff98}.

The  net component of the magnetic field along the line of sight ($B_{\parallel}$) is responsible for the Faraday rotation on the radiation passing through the magneto-ionized ICM. The rotation of the  observed polarization angle depends on $\lambda^2$, being $\lambda$ the observing frequency:

\begin{equation}
\label{eq:rotation}
    \chi(\lambda^2)=\chi_0+\phi\lambda^2~,
\end{equation}

where $\chi_0$ is the intrinsic polarization angle. The Faraday depth $\phi$ is defined as:

\begin{equation}
    \phi=0.81\int_{\rm{source}}^{\rm{observer}} {n_e B_{\parallel} {\rm d}l} \quad \rm{[rad \ m^{-2}]}~,
\end{equation}

where $n_e$ is the thermal electron density in cm$^{-3}$, $B_{\parallel}$ is  is the magnetic field component parallel to the line-of-sight in $\mu$G and ${\rm d}l$ is the infinitesimal path length in parsecs. 

The Rotation Measure (RM) and the Faraday depth coincide at all wavelengths only when one or several (not emitting) screens lie in between the source and the observer and in the absence of beam depolarization. In this case we term the source Faraday-simple since the RM -- or $\phi$ -- can be recovered from Eq.~\ref{eq:rotation}.

In a general case polarized synchrotron radiation may originate in the same volume that causes Faraday rotation. In particular, in galaxy cluster we expect the magnetic field to be filamentary and the emitting volume of radio relic to be filled with turbulent thermal gas. Hence, polarized emission can be spread over a range of $\phi$ determining a Faraday-complex source.

The RM synthesis technique developed by \citet{Brentjens05} introduces the Faraday dispersion function, hereafter also called Faraday spectrum, $F(\phi)$, which describes the complex polarization vector as a function of the Faraday depth. The reconstructed Faraday spectrum, $\widetilde{F}(\phi)$, can be recovered by Fourier transform of the observed polarization as a function of wavelength squared. The Rotation Measure Sampling Function (RMSF) describes the instrumental response to the polarized signal in the Faraday space based on the wavelength coverage of the observation. We refer to \citet{Brentjens05} for details on this technique.

The amplitude of the reconstructed $\widetilde{F}(\phi)$ peaks at the Faraday depth $\phi_{\rm peak}$, which is the  Faraday depth along the path between the observer and the source contributing the most to the polarized emission. From the value of the reconstructed $|\widetilde{F}(\phi_{\rm peak})|$ it is possible to recover the polarization fraction of the emission, while from the reconstructed Stokes parameters, $\widetilde{Q}(\phi_{\rm peak})$ and $\widetilde{U}(\phi_{\rm peak})$, we can recover the polarization angle at $\lambda_0^2$ (i.e., the weighted average of the observed bandwidth):

\begin{equation}
    \chi(\lambda_0^2)=\frac{1}{2}\arctan{\frac{\widetilde{U}(\phi_{\rm peak})}{\widetilde{Q}(\phi_{\rm peak})}}~.
\end{equation}

\subsection{Polarized intensity imaging}
\label{sec:polimaging}

We made use of \texttt{WSCLEAN 2.6}\footnote{\url{https://sourceforge.net/p/wsclean/wiki/Home/}} \citep{Offringa14} for the polarization intensity imaging. This imager exploits a $w$-stacking algorithm as a faster alternative to $w$-projection and allows multi-scale, multi-frequency and auto-masking algorithms \citep{Offringa17}.

We imaged 1-2 GHz A, B, C configurations data and 2-4 GHz data  in DnC configuration to sample different spatial scales and frequencies. We also  used together L-band B+C and S-band DnC data to cover the whole frequency band 1-4 GHz. The images were cleaned down to 3$\sigma$ level using the \texttt{auto-masking} option. We  used 64 frequency  sub-bands of 16 MHz each for the data in L-band, and 16  sub-bands of 128 MHz for the S-band data. The different sub-band width is required to avoid bandwidth depolarization at the lowest frequency. For the whole 1-4 GHz band we used 96  frequency sub-bands of 32 MHz each. The  restoring beam was forced to be the same in each frequency sub-band, matching the lowest resolution one. The parameters used for each image are listed in Tab.~\ref{tab:pol}.

\begin{table*}
    \centering
	\caption{Details of polarized intensity images. Column 1: observing frequency range, i.e. first and last frequency sub-bands; Column 2: name of the frequency band; Column 2: array configuration; Column 3: $\delta\nu$ is the frequency sub-band width; Column 4: $\sigma_{QU}$ is the best estimate for the rms noise in polarization obtained as $(\sigma_Q+\sigma_U)/2$;  Column 5: FWHM of the major and minor axes of the restoring beam of the image cubes; Column 6: resolution in the Faraday space; Column 7: maximum observable Faraday depth; Column 8: largest observable Faraday-scale; Column 9: reference of the figures in this paper.}
	\label{tab:pol}
	\begin{tabular}{lccccccccc} 
	    \hline
		\hline
		Freq. Range  & Band & Array Config. & $\delta\nu$ & Beam & $\sigma_{QU}$ & $\delta\phi$ & $|\phi_{\rm max}|$ & $\Delta\phi_{\rm max}$ & Figure \\
		(GHz)        &      &               &     (MHz)   &  & (mJy/beam)  & (rad m$^{-2}$)  & (rad m$^{-2}$)    & (rad m$^{-2}$)        &        \\
		\hline
		1.015-2.023  & L   &   A       & 16 & 2.5$\arcsec\times$2.5$\arcsec$ & 0.004  & 45 & 535 & 143 &  \\
		1.015-2.023  & L   &   B       & 16 & 11$\arcsec\times$5$\arcsec$ & 0.005 & 45 & 535 & 143 & \\
		1.015-2.023  & L   &   C       & 16 & 25$\arcsec\times$ 25$\arcsec$ & 0.007 & 45 & 535 & 143 & \ref{fig:pol_CS_E}, \ref{fig:pol_CS_W} \\
	    2.050-3.947  & S   &   DnC     & 128 & 25$\arcsec\times$ 25$\arcsec$ & 0.004 & 188 & 598 & 543 & \ref{fig:pol_CS_E}, \ref{fig:pol_CS_W}\\
		1.022-3.995  & S+L &   B+C+DnC & 32 & 25$\arcsec\times$25$\arcsec$  & 0.004 & 37 & 288 & 558 & \ref{fig:polSL}, \ref{fig:phiSL}\\
		\hline
	\end{tabular}
\end{table*}

 We used \texttt{join-channels} and \texttt{join-polarizations} options to make Stokes $I$, $Q$, $U$ image cubes and full-bandwidth images, as recommended in the \texttt{WSCLEAN} documentation. Each image was corrected for the primary beam calculated for the central frequency of the sub-band. We restricted our analysis to a circular region of radius $\sim$6$\arcmin$ (i.e., $\sim$1.4 Mpc) around the  L-band pointing centre so that the effect of direction-dependent gain, polarization leakage in $Q$ and $U$ and beam squint are  negligible \citep{Jagannathan17}.  We quantified the leakage from Stokes I to V to be $\leq 2 \ \%$ within the closest 6$\arcmin$ to the image centre. This constrains the leakage to Stokes Q and U to be within 1 $\%$ of I. We caution however about the usage of similar data for sources showing lower fractional polarization (<5 $\%$) and further away from the beam centre.
 
\subsection{RM synthesis}
\label{sec:RMsynth}

We performed RM synthesis on the $Q$ and $U$ image cubes using \texttt{pyrmsynth}\footnote{\url{https://github.com/mrbell/pyrmsynth}} and we obtained the cubes in the Faraday space. We thus recovered the reconstructed Faraday dispersion function, or Faraday spectrum, $\widetilde{F}(\phi)$, in each pixel (i.e. each line of sight).

Faraday cubes were created between $\pm$600 rad m$^{-2}$ and  using bins of 2 rad m$^{-2}$. This range is motivated by our sensitivity to large  values of $\phi$. From \citet{Brentjens05} we can estimate the resolution in Faraday space (i.e., the FWHM of the main peak of the RMSF), $\delta\phi$, the maximum observable Faraday depth, $|\phi_{\rm max}|$, and the largest observable scale in Faraday space, $\Delta\phi_{\rm max}$ (i.e., the depth and the $\phi$-scale at which sensitivity has dropped to $50 \ \%$). The parameters for each observation  are listed in Tab.~\ref{tab:pol}.

Notice the different resolutions and $\Delta\phi_{\rm max}$ values of our measurements. The values of $|\phi_{\rm max}|$ to which we are sensitive are well above the value expected from galaxy clusters \citep[see, e.g.,][]{Bonafede10,Bonafede13,Bohringer16}. In this galaxy cluster, we observed $|\phi_{\rm peak}|$<100 rad m$^{-2}$  both in the L- and in the S-band observations, so that the lower $|\phi_{\rm max}|$ obtained combining L- and S-band does not limit our measurements. In the combined data set, with a central frequency of 2.5 GHz, we reach the highest resolution in Faraday space and we are sensitive to polarized emission spread over large Faraday scales.
 
We first run \texttt{pyrmsynth} on the entire central region cleaning the spectrum down to five times the noise level of full-bandwidth $Q$ and $U$ images \citep[see][for the RM clean technique]{Heald09}. We imposed an average total intensity spectral index $\alpha$=1 on the entire field. We noticed the RM cleaning process was improved by the use of $\alpha$=1 instead of the default $\alpha$=0, although it comes from an average estimate for the entire field and it is assumed to be constant at each Faraday depth. We measured the local rms noises $\sigma_Q$ and $\sigma_U$ in the slices of $\widetilde{Q}(\phi)$ and $\widetilde{U}(\phi)$ at $500 \ {\rm rad \ m}^{-2} \ < \ |\phi| \ < \ 600 \ {\rm rad \ m}^{-2}$ 
i.e. outside the sensitivity range of our observations.  Since $\sigma_Q\sim\sigma_U$, we estimated the noise of polarization observation as $\sigma_{QU}=(\sigma_Q+\sigma_U)/2$ \citep[see also][]{Hales12}. This value is reported in Tab.~\ref{tab:pol} for each measurement set. Then, we selected pixels with a peak in the Faraday spectrum above a threshold of 8$\sigma_{QU}$, following \citet{George12}, which corresponds to a false detection rate of $0.06 \ \%$ and to a Gaussian significance level of about 7$\sigma$ according to \citet{Hales12}. This  conservative choice accounts for the Ricean bias  (i.e., the over-estimation of polarized intensity due to $P$ being positive-definite and governed by the Ricean distribution), the non-Gaussian noise in the $Q$ and $U$ images, and the additional bias due to error in $\phi_{\rm peak}$ estimates. We run again \texttt{pyrmsynth} only on these pixels, cleaning the spectrum down to 8$\sigma_{QU}$ level.
 
We computed polarization intensity images using the peak of the Faraday dispersion function and correcting for the Ricean bias as $P=\sqrt{|\widetilde{F}(\phi_{\rm peak})|^2-2.3\sigma_{QU}^2}$ \citep{George12}. We then obtained fractional polarization images dividing the $P$ images (with the 8$\sigma_{QU}$ threshold) by the full-band Stokes $I$ images  with a cutoff of three times the rms noise. We also obtained $\phi_{\rm peak}$ images with the same cutoff. From the reconstructed values of $Q$ and $U$ at $\phi_{\rm peak}$ we can also recover the intrinsic polarization angle (i.e., corrected for the value of RM determined by $\phi_{\rm peak}$), $\chi_0$, as:
 
 \begin{equation}
 \label{eq:chi0}
     \chi_0=\chi(\lambda_0^2)-\phi_{\rm peak}\lambda_0^2=\frac{1}{2}\arctan{\frac{\widetilde{U}(\phi_{\rm peak})}{\widetilde{Q}(\phi_{\rm peak})}}-\phi_{\rm peak}\lambda_0^2~,
 \end{equation}
 
where $\lambda_0$ is 19.7 cm, 11.9 cm and  10 cm for the L, S+L and S-band, respectively.
 
\begin{figure*}
	\includegraphics[width=\columnwidth]{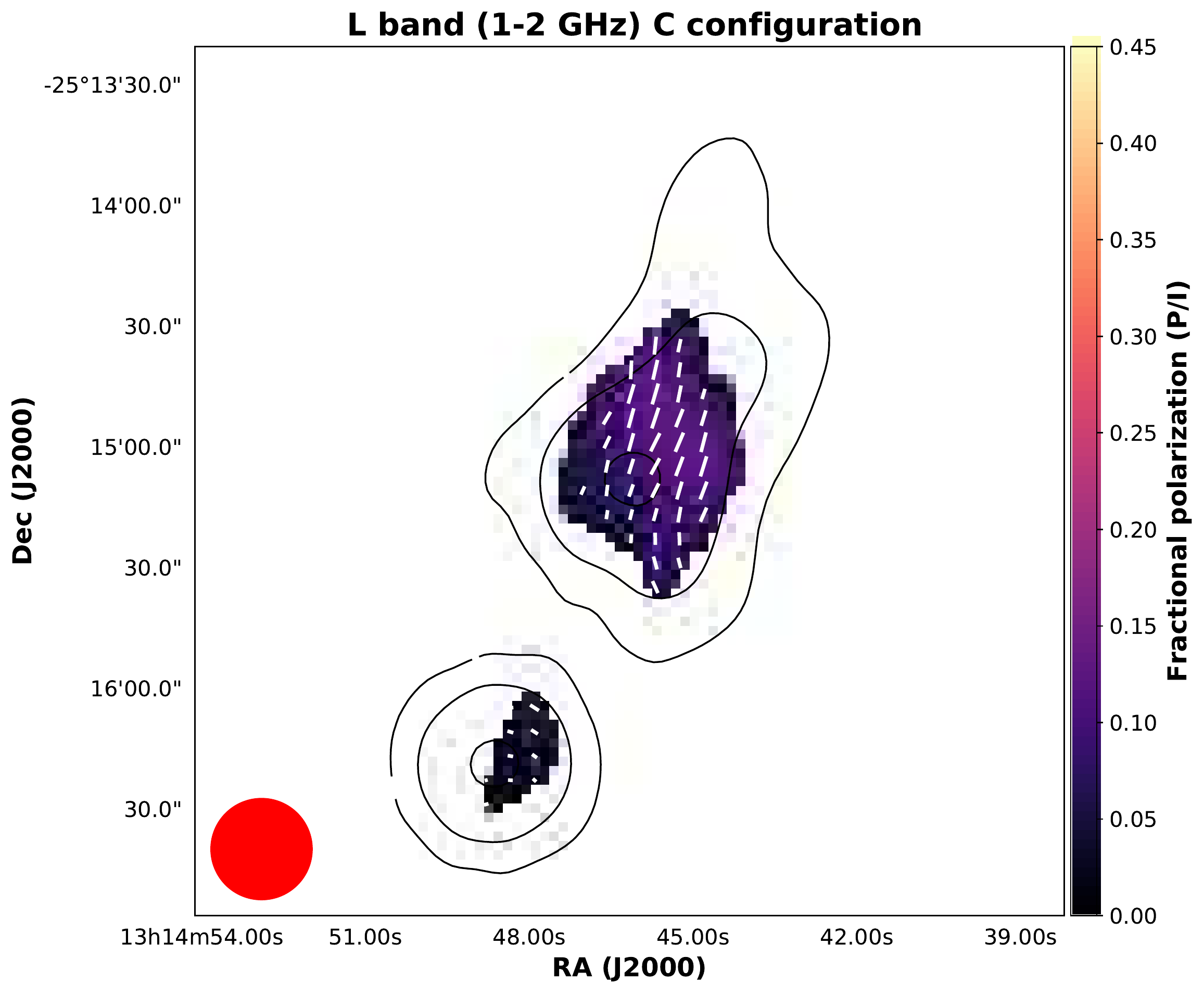}
	\includegraphics[width=\columnwidth]{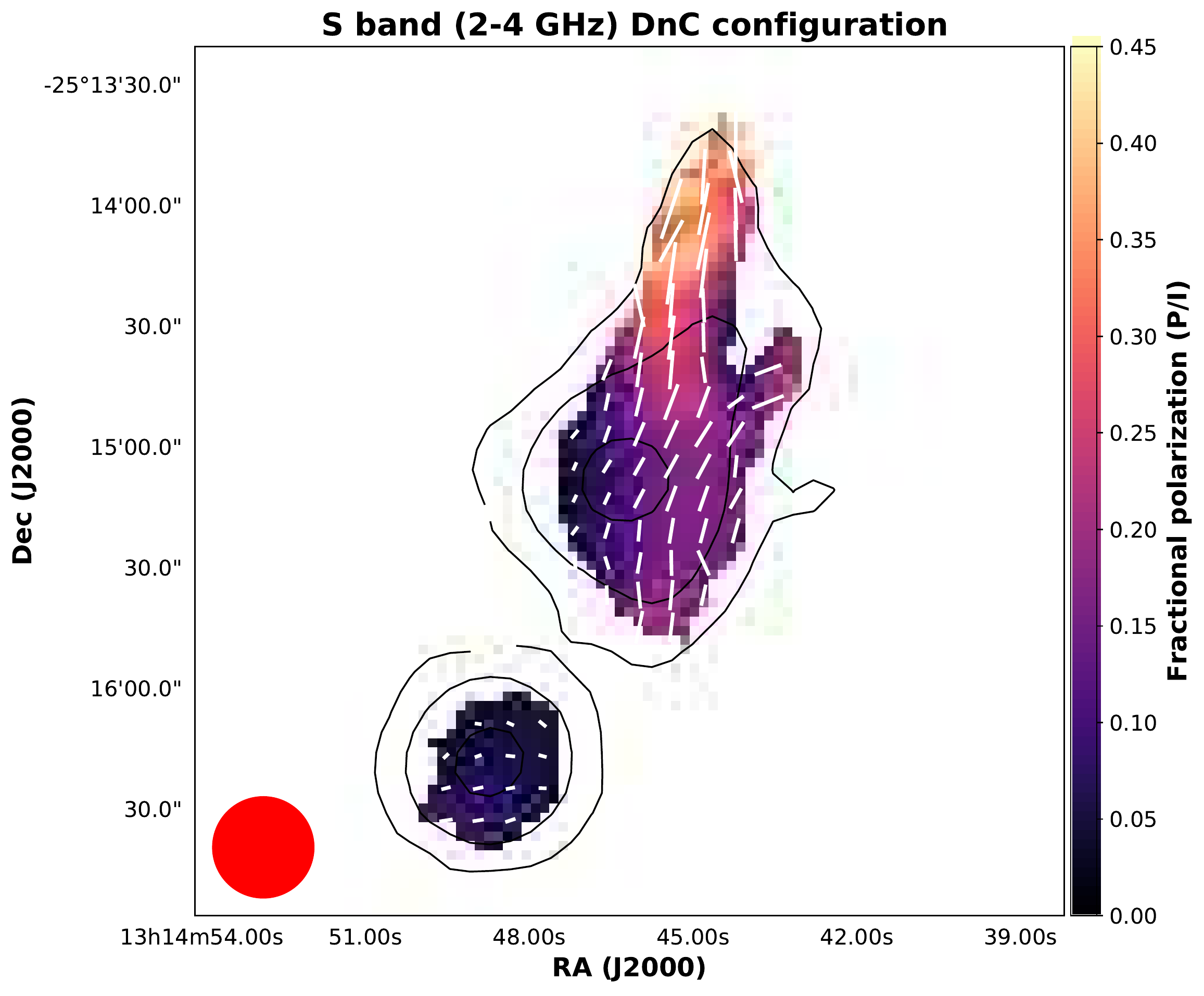}
    \caption{Fractional polarization (color scale) and magnetic field vectors (white lines) of the E radio relic: L-band data in C configuration (right panel) and S-band data (left panel). An 8$\sigma_{QU}$ detection threshold was imposed polarization and values were corrected for the Ricean bias. See Tab.~\ref{tab:pol} for details on the images. The length of white vectors is proportional to the fractional polarization. Black contours are from the total intensity images in the same frequency band and configuration. The restoring beam is 25$\arcsec\times25\arcsec$ in both images. Contour levels start from 3 times the rms noise (0.08 mJy beam$^{-1}$ and 0.03 mJy beam$^{-1}$ for the L- and S-band respectively)} and are spaced by a factor of four.
    \label{fig:pol_CS_E}
\end{figure*}

\begin{figure*}
	\includegraphics[width=\columnwidth]{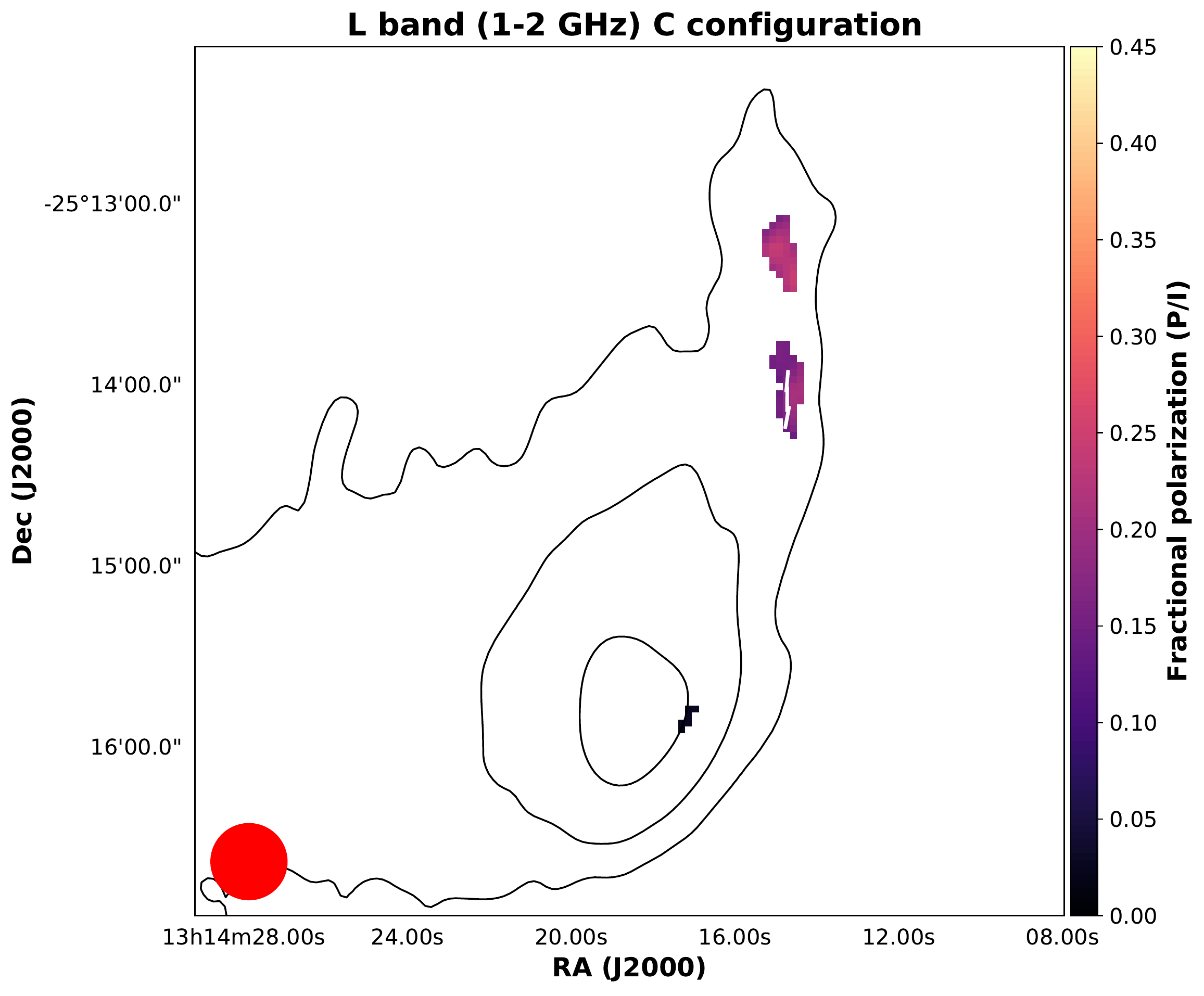}
	\includegraphics[width=\columnwidth]{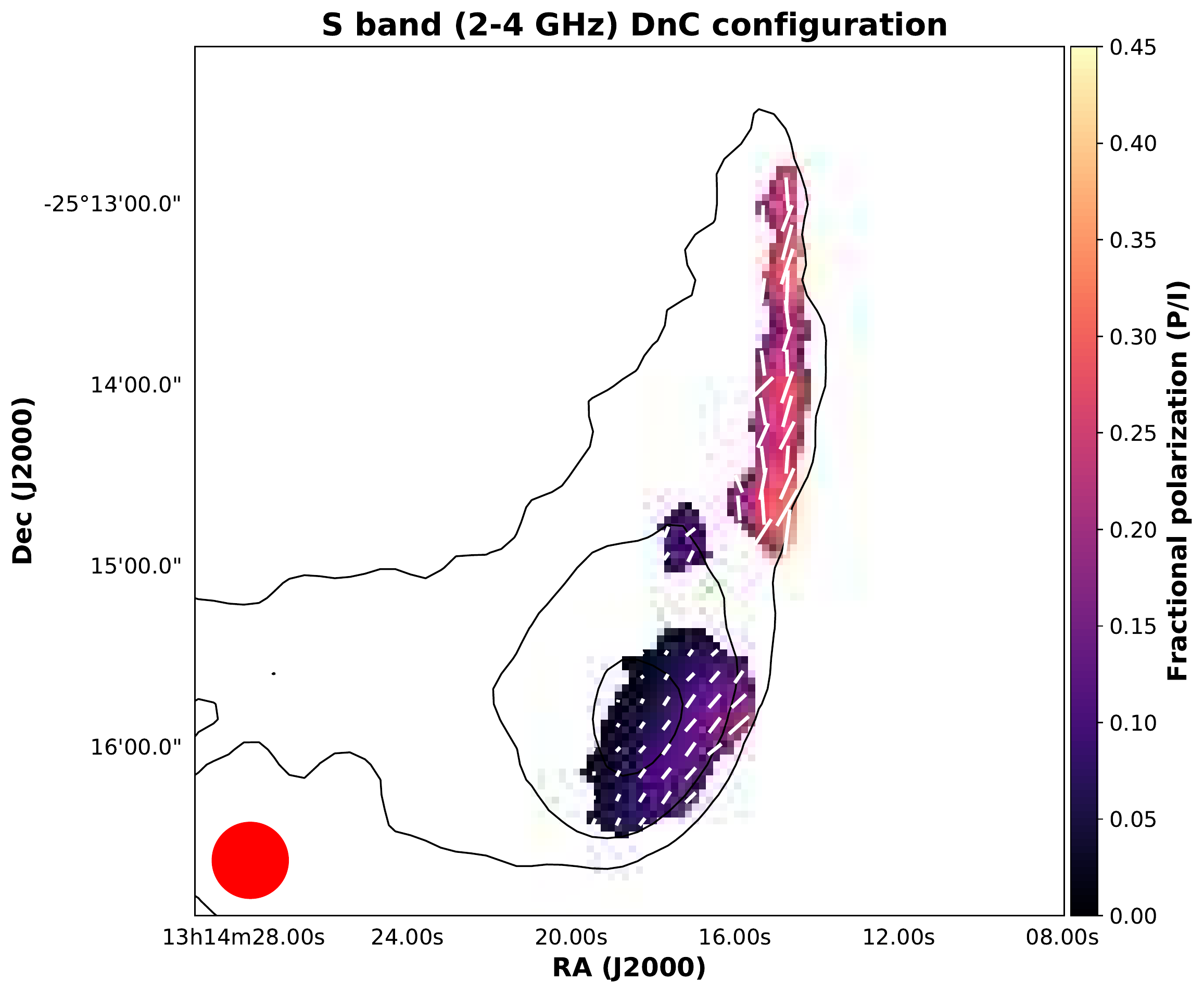}
	    \caption{Same as Fig.~\ref{fig:pol_CS_E} for the W relic.}
    \label{fig:pol_CS_W}
\end{figure*}

Fractional polarization and magnetic field vector images of the L-band C configuration and S-band are shown in Fig.~\ref{fig:pol_CS_E} and Fig.~\ref{fig:pol_CS_W} for the E and W relics, respectively.  We also obtained B configuration L-band images but, for the purpose of this work, we are more interested  in  the polarized diffuse emission of the relics, more visible in the C configuration image. We obtained only few pixels above the 8$\sigma_{QU}$ cutoff with the A configuration L-band image. Fractional polarization and magnetic field vector images obtained from the whole frequency band 1-4 GHz are shown in the upper panel of Fig.~\ref{fig:polSL}. We show  the same quantities obtained at different frequency bands to show the importance of combining multi-band observations for this analysis. 

The Faraday depth image of \RX resulting from the combined S+L-band data set is shown in Fig.~\ref{fig:phiSL}. We show only the S+L-band map since we obtained a good trade-off between resolution in Faraday space and sensitivity. Thanks to the  wide and contiguous frequency coverage of our observations, we can identify some regions of the W relic which clearly show Faraday-complex structures. In this case, the value of $|\widetilde{F}(\phi_{\rm peak})|$ and of $\phi_{\rm peak}$ are not sufficient to describe the polarization and rotation effect experienced by the radiation since the polarized emission would be spread at different values of $\phi$ and affected by Faraday depolarization. For this reason, the results of RM-synthesis in the southern region of the W relic (i.e., the nose) should be regarded with caution. Faraday-complex structures are separately discussed in Sec.~\ref{sec:discussthick}.

The RM value of the Galactic foreground (i.e., -30 rad m$^{-2}$) was subtracted  out: we will refer to  the cluster Faraday depth, $\phi_{cl}$, to indicate foreground subtracted values.

\subsubsection{Uncertainties on the measure of $\phi$}
\label{sec:RMuncertainties}
The method commonly used to estimate the uncertainties on $\phi$ is derived from \citet{Brentjens05} where:

\begin{equation}
    \sigma_{\phi}=\frac{\delta\phi}{2 P/\sigma_{QU}}~,
\end{equation}
 
that is the Half Width Half Maximum of the RMSF divided by the signal-to-noise of the detection. This expression is derived under the assumption $\alpha$=0 and $\sigma_{Q}=\sigma_{U}$. In other cases, it can lead to over- or  under-estimates of the errors \citep{Schnitzeler17}.
However, we computed $\sigma_{\phi}$ since it can be useful to compare the uncertainties pixel by pixel, and we added in quadrature the error  of 2 rad m$^{-2}$ on the estimate of the Galactic foreground by \citet{Taylor09} to obtain $\sigma_{\phi_{cl}}$ . The error map is shown in the bottom panel of Fig.~\ref{fig:phiSL}. From $\sigma_{\phi}$ we derived the error on the polarization angle by propagating the uncertainties on the quantities in Eq.~\ref{eq:chi0} as:

\begin{equation}
    \sigma_{\chi_0}^2=\sigma_{\chi}^2+\sigma_{\phi}^2\lambda_0^4=\frac{\sigma_{QU}^2}{4P^2}+\bigg(\frac{\delta\phi}{2 P/\sigma_{QU}}\bigg)^2\lambda_0^4~.
\end{equation}

\begin{figure}
	\includegraphics[width=\columnwidth]{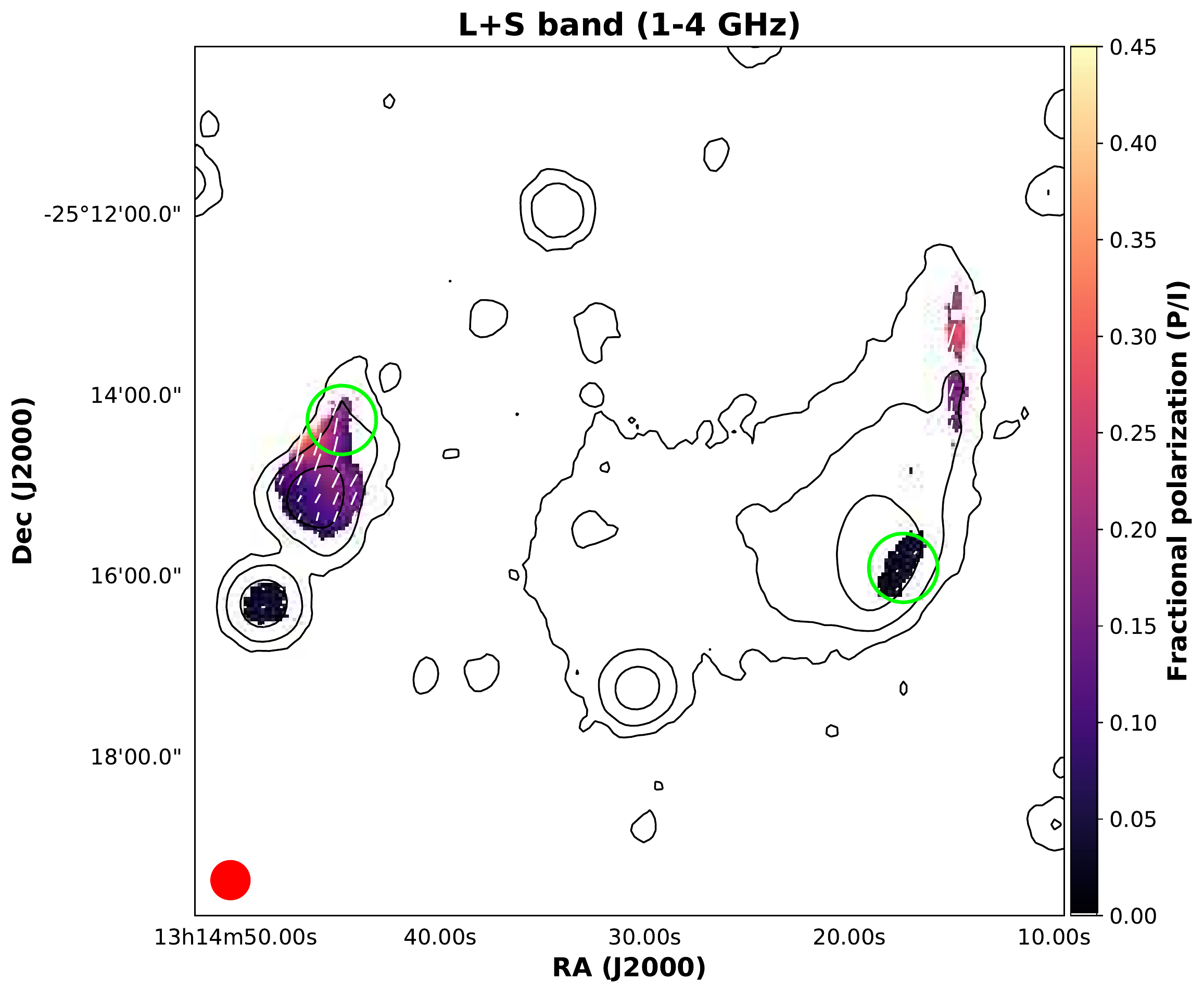}
	\hfill
	\includegraphics[width=\columnwidth]{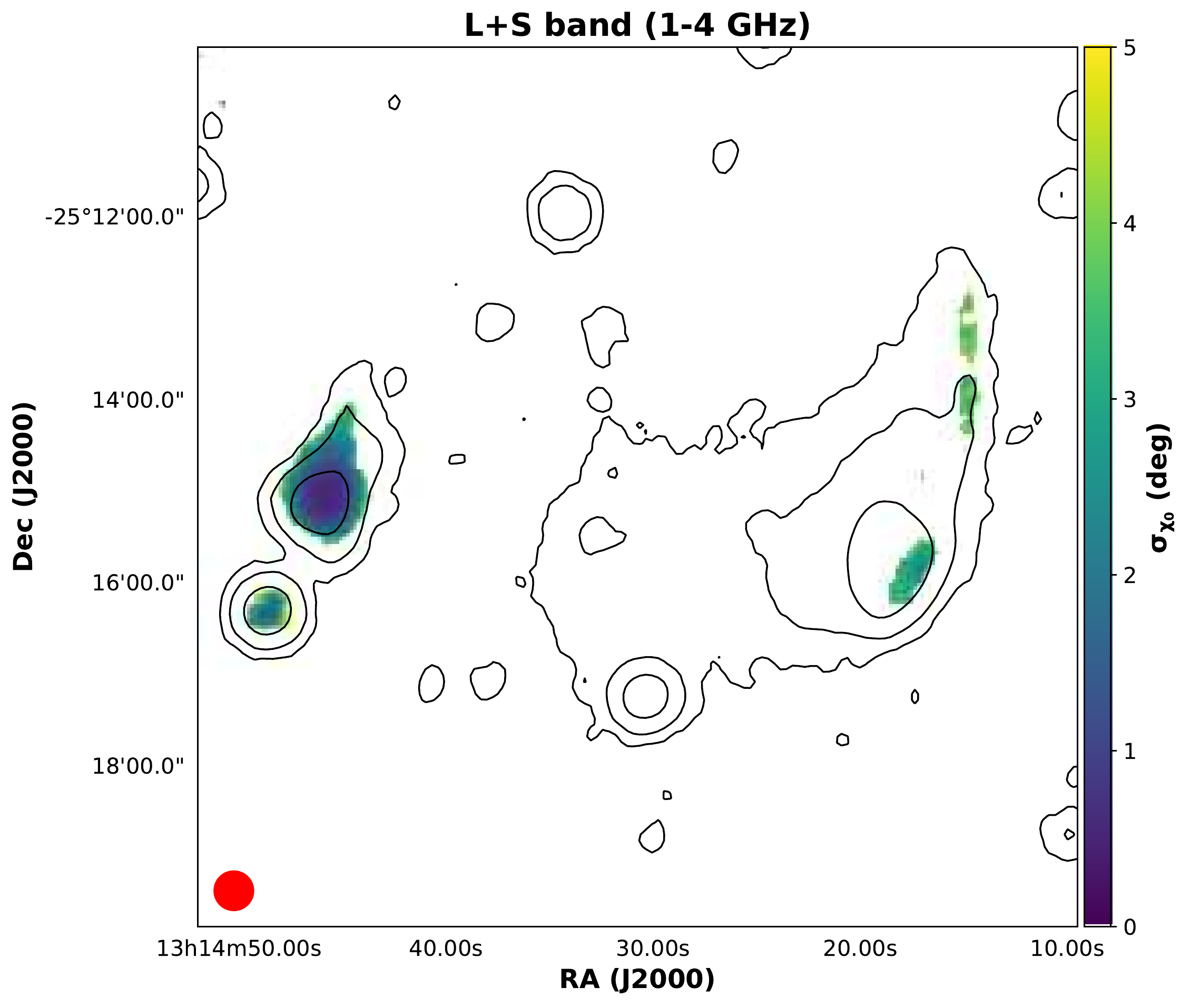}
    \caption{Top panel: fractional polarization (color scale) and magnetic field (white lines) with a 8$\sigma_{QU}$ detection threshold for the 1-4 GHz data in B+C+DnC configuration (see Tab.~\ref{tab:pol} for image details). Fractional polarization values were corrected for the Ricean bias. The green circles mark the regions used for the Faraday depolarization study in Sec.~\ref{sec:discussthick}. The length of white vectors is proportional to the fractional polarization. Bottom panel: error map of the polarization angle. In both panels the black contours are from the total intensity image with a restoring beam of 25$\arcsec\times25\arcsec$, shown in the left-hand corner of the image. They start from 3 times the rms noise of 0.02 mJy beam$^{-1}$ and are spaced by a factor of four.}
    \label{fig:polSL}
\end{figure}

\begin{figure}
	\includegraphics[width=\columnwidth]{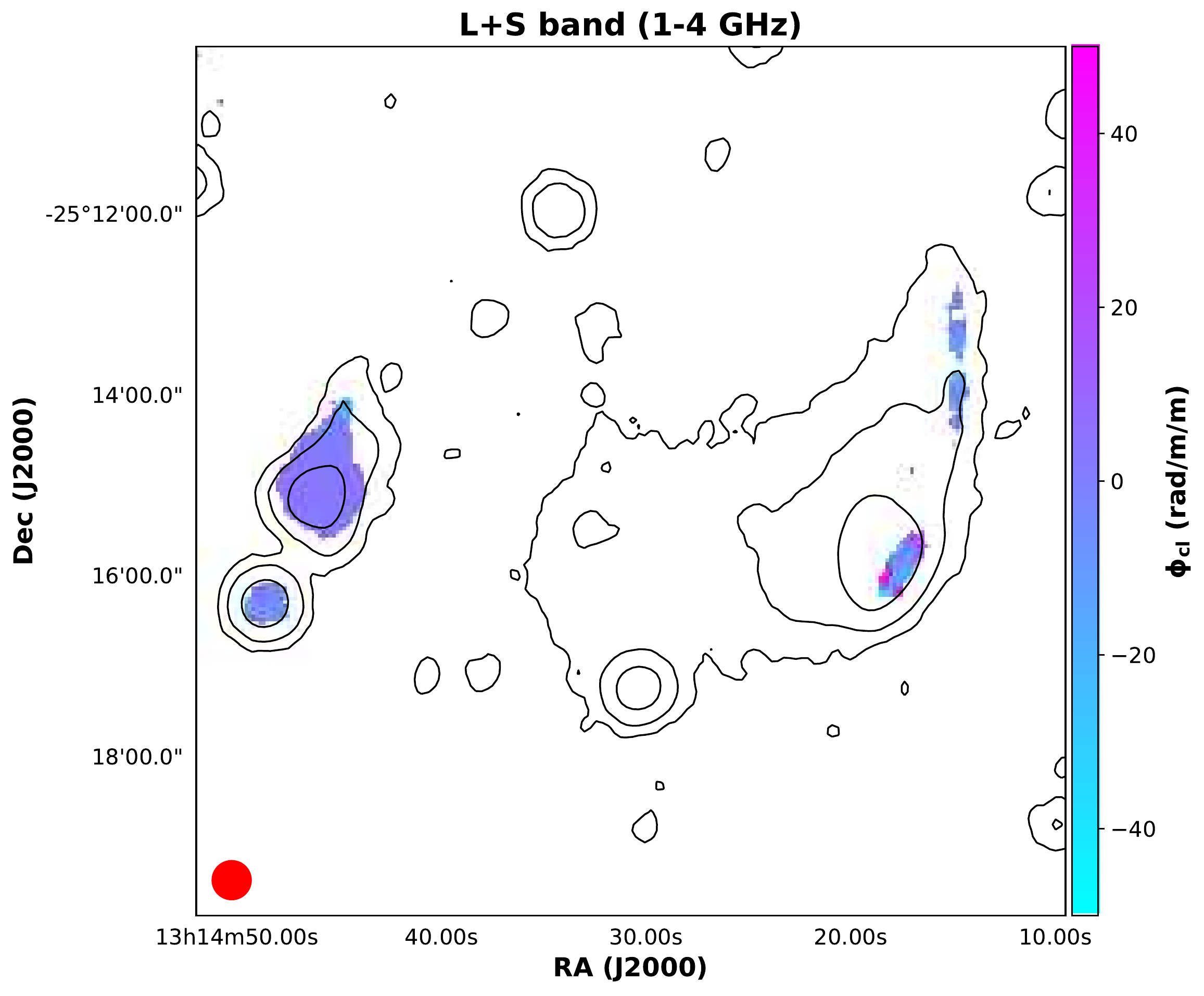}
	\hfill
	\includegraphics[width=\columnwidth]{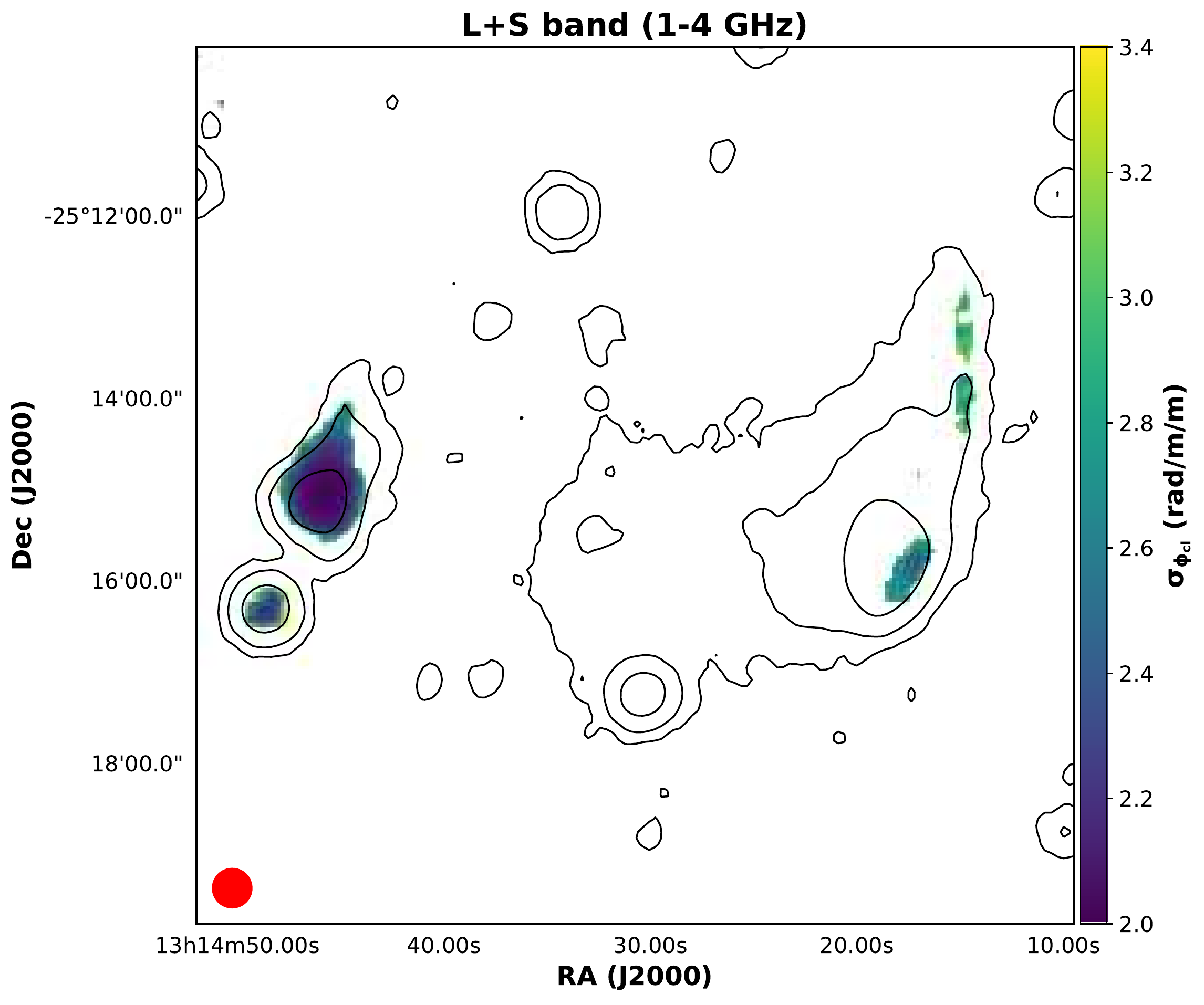}
    \caption{ Top panel: value of $\phi_{cl}$ on a pixel by pixel basis with a 8$\sigma_{QU}$ detection threshold for the 1-4 GHz image in B+C+DnC configuration. Bottom panel: error map of the $\phi_{cl}$ image.  Black contours and restoring beam are the same as Fig.~\ref{fig:polSL}. }
    \label{fig:phiSL}
\end{figure}

These uncertainties for the vector map for the S+L-band data set are shown in the bottom panel of Fig.~\ref{fig:polSL}.

\begin{figure*}
   \includegraphics[width=15cm, height=12cm]{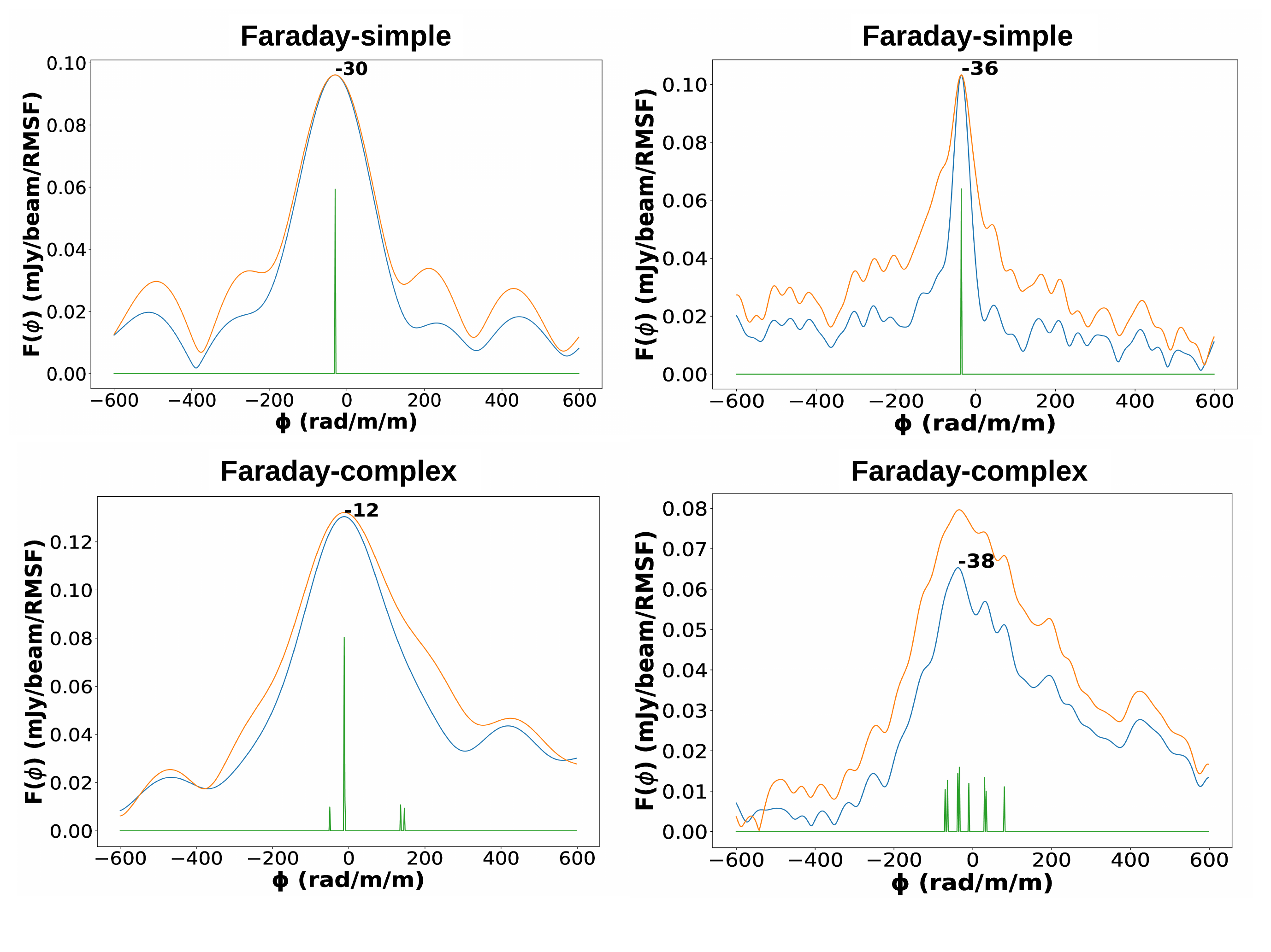}
  \caption{Faraday dispersion function of two representative pixels in the 2-4 GHz band observation (left column) and in the whole 1-4 GHz band (right column).  The orange line is the dirty spectrum. the blue is the clean spectrum and in green the clean components. The pixel in the first row has a Faraday-simple structure while the one in the second row has a Faraday-complex structure, barely resolved in the first column and resolved as a convolution of peaks in the second one.}
  \label{fig:pixels}
\end{figure*}

%\end{comment}

As pixel values are not independent  of each other, in the text we will refer to beam-averaged quantities, which are the average values of $\phi_{cl}$ over beam-size regions, weighted by the signal-to-noise ratio of pixels. To a first approximation, in the case of Faraday-simple sources, the distribution of RM values in each beam has a Gaussian distribution for signal-to-noise ratios higher than 8 \citep{George12} and its variance is not due to a physical variation of the Faraday rotating medium, but to the underling noise in the Faraday spectrum, which shifts the position of the peak. Hence, we calculated also the standard deviation of neighbouring pixels in a region equivalent to the beam area to estimate the RM uncertainties.

 \subsubsection{The Eastern Relic}
 \label{sec:polarizationE}

In the B configuration we detected polarized emission arising only from a barely resolved source in the south-east of the E relic  (source $C$ in Fig.~\ref{fig:optic}) and from two regions close to the lobes of the NAT radio galaxy embedded in the relic.
 
In the S-band and L-band C configuration images (Fig.~\ref{fig:pol_CS_E}), the average polarization fraction of the  source $C$ is 4.3$\pm0.4 \ \%$ and 3.6$\pm0.4 \ \%$, respectively. The polarization fraction is 3.6$\pm0.3 \ \%$ also in the combined S+L-band image (Fig.~\ref{fig:polSL}). The weighted average $\phi_{cl}$ of this source derived from the Faraday depth map of Fig.~\ref{fig:phiSL} is -4 rad m$^{-2}$ while the standard deviation is 9 rad m$^{-2}$.

The E relic shows a resolved polarized emission in the C configuration L-band and in the S-band images (Fig.~\ref{fig:pol_CS_E}). The fractional polarization observed in the L-band is lower than what is detected in other radio relics, while in the S-band it is consistent with the literature \citep{vanWeeren19}. The region of the core of the NAT radio galaxy in the eastern edge of the relic is polarized at an average 9.9$\pm0.9 \ \%$ level in the S-band. The fractional polarization increases in the region of the lobes and in the N-S direction. In particular, in the S-band image we detect polarized emission from the E radio relic, reaching a polarization fraction of $\sim45 \ \%$ in the northern part of the relic. The magnetic field is oriented almost in the direction of the radio lobes of the galaxy and then  bends to be aligned in the N-S direction along the radio relic. The same is observed in Fig.~\ref{fig:polSL}.

The values of average $\phi_{cl}$ within beam-size regions located in the E relic are almost constant ranging between 2 and 4 rad m$^{-2}$ with a standard deviation of $\sim$3 rad m$^{-2}$. Only in the northern side of the relic the average $\phi_{cl}$ changes to -3 rad m$^{-2}$ with a standard deviation  of 11 rad m$^{-2}$.

All the values of $\phi_{cl}$ measured are consistent with zero considering the standard deviation as the uncertainty, meaning that the electron density and the magnetic field in this region of the cluster are not responsible for the Faraday rotation effect which is mainly due to the external screen of our Galaxy. This is confirmed by the moderate Faraday depolarization detected in this region which can be explained by the low thermal electron density (see Fig.~\ref{fig:X}).

\subsubsection{The Western Relic and the Radio Halo}
\label{sec:polarizationW}

The western side of \RX does not show polarization in the B configuration image. Few pixels with a signal-to-noise ratio in polarization higher than 8 appear in the thinnest part of the outer W relic in the L-band C configuration image. Here the degree of polarization reaches the $25 \ \%$ level but it is measured only in few pixels (Fig.~\ref{fig:pol_CS_W}). Hence, the fractional polarization observed in the L-band is lower than what expected from the literature also for the W relic \citep[see][and references therein]{vanWeeren19}.

In the S-band, we found a totally different situation (see Fig.~\ref{fig:pol_CS_W}). Almost all  of the outer arc shows high levels of intrinsic fractional polarization reaching a value of $\sim40 \ \%$ in the northern part. The average fractional polarization  in the northern arc is 24$\pm4 \ \%$. The magnetic field lines are almost aligned along the radio relic arc. In Fig.~\ref{fig:phiSL}, the values of $\phi_{cl}$ in the northern part of the relic are scattered between $-44 \ {\rm rad \ m^{-2}}$ and $42 \ {\rm rad \ m^{-2}}$. A spot of polarized emission, without any optical counterpart, is also detected along the inner arc at 10$\pm3 \ \%$ level of polarization and with values of $\phi_{cl}$ reaching $50 \ {\rm rad \ m^{-2}}$.

Also the southern part of the arc  (i.e., the nose) shows polarization with an average value of 7.6$\pm0.6 \ \%$ in the S-band and of 3.0$\pm0.3 \ \%$ in the S+L combined data set. In the S-band,  the fractional polarization increases, from $\sim2 \ \%$ in the inner region to $\sim25 \ \%$ at the shock front detected in the X-rays  (see right panel of Fig.~\ref{fig:pol_CS_W}). In the whole southern region the $\phi_{cl}$ values in the S+L-band are in the range between -82 and 78 rad m$^{-2}$. 

The  emission coincident with the shock region, detected only in the S- and S+L-band observations, clearly shows a Faraday-complex emission unveiled by  a broadening of the Faraday spectrum larger than the FWHM of the RMSF (i.e., at least larger than 150 rad m$^{-2}$, see Fig.~\ref{fig:pixels}). The strong Faraday depolarization between S- and L-band in this region is probably explained by the higher thermal electron density in respect to the E relic \citep{Kierdorf17}. This is further discussed in Sec.~\ref{sec:discussthick}. 

The emission that extends from the W relic toward the central region of \RX does not show  any polarized emission. The upper limit on the fractional polarization in this area is $17 \ \%$, in agreement with literature results for other radio halos \citep[e.g.][]{Feretti12}.
 
 \section{Discussion}
 \label{sec:discuss}
 
\subsection{AGN-relic connection in the eastern relic}
\label{sec:discussAGNrelic}

The NAT radio galaxy in Fig.~\ref{fig:optic} is a cluster member, and its lobes extend for $\sim90$ kpc fading into the radio relic  which extends in the N-S direction. Although the detection of a shock related to the E relic emission is prevented by the low X-ray surface brightness (see Fig.~\ref{fig:X}) the combined information gained from the spectral index and the polarized intensity study support the idea that a shock wave is at the origin of the extended eastern emission together with the AGN activity. Since pairs of shock waves propagating along the merger axis are generated during cluster mergers, a shock wave moving outwards from W to E is in fact expected to be present at the position of the relic.

Assuming a magnetic field between 0.1 and 10 $\mu$G, as observed in other galaxy clusters \citep[e.g.,][]{Govoni04}, the lifetime of relativistic electrons that emit at 3 GHz at redshift 0.247 due to synchrotron and inverse Compton losses is  $\leq$10$^8$ yr  \citep[][Eq. 3]{vanWeeren19}. Considering a plasma diffusion and/or bulk velocity of $\sim$10$^5$ m s$^{-1}$ (either due to AGN jet activity or to the merger shock) these high-energy electrons are already aged after distances of few tens kpc. The observed emission at 3 GHz spreads over 500 kpc and implies that another mechanism is actively energizing the electrons. We suggest that the re-acceleration originates in the shock front as suggested by the E-W spectral steepening (see Sec.~\ref{sec:spectralstudyE}).

Low-energy electrons with Lorentz factor $\gamma<10^2$ have radiative lifetimes larger than the Hubble time. During a maximum AGN lifetime of $\sim$1 Gyr these electrons can travel distances of hundreds kpc in the ICM. Hence, it is plausible that the AGN  may have supplied the low-energy electrons re-accelerated by the shock.

In the northern region of the E radio relic, we measured an integrated spectral index of $\alpha$=1.2$\pm$0.2, flatter than the value reached in the region of the lobes of the radio galaxy (see Fig.~\ref{fig:spixprof_E}). The flattening of the spectral index at the edge of the E relic could indicate that the particles are first injected by AGN jets in the ICM, where they lose energy due to synchrotron and Inverse Compton losses in the radio galaxy lobes, before being re-accelerated by a shock wave.  A similar scenario was invoked by \citet{Bonafede14} for PLCKG287.0+32.9. Recently, \citet{vanWeeren17a} reported the  clearest connection known to date between an AGN and a relic in Abell 3411 - Abell 3412. 

If spectral index of the pre-existing population is flatter than the one that could be produced by the shock, the spectrum is amplified and retains the spectral index of the pre-existing population \citep[e.g][]{Gabici03,Kang11}. In the opposite case, the post-shock electron population has the spectral index of the DSA and loses the memory of the injection spectrum. Since we observe a flatter spectrum in the northern edge compared to the lobes region, we are in this latter case. The sampled regions have sizes $>$ 50 kpc, thus the electron population should have already reached the equilibrium with the energy losses \citep[continuous injection model,][]{Kardashev62}. Hence, the Mach of the shock can be derived from the integrated spectral index in the region \citep[see, e.g.,][]{Trasatti15} as :

\begin{equation}
\label{eq:mach}
    M^2=\frac{\alpha+1}{\alpha-1}~.
\end{equation}

This implies that,  if a shock wave is re-vitalizing the aged particles from the AGN lobes, its Mach number would be $M=3\pm{1}$.

Furthermore, polarization vectors suggest that the projected magnetic field lines follow the AGN jets and then are bent along the north-south direction (see \ref{fig:pol_CS_W}). A shock wave propagating from W to E along the merger axis (i.e. the line connecting the two sub-clusters) would be able to align the magnetic field lines in the N-S direction. A shock wave with $M$ $\sim$ 3 can in fact amplify the magnetic field components parallel to the shock front of a factor 2.5 \citep{Iapichino12}, so that the resulting magnetic field on the plane of the sky would have a preferential direction aligned with the N-S shock front. As a result, the polarization fraction of the emission is enhanced and reaches values of $\sim50 \ \%$. The fact that the highest polarization is observed in the same northern region where we found a flattening of the spectrum, suggests that in this region the shock wave could be the active acceleration process. Where the emission of the radio galaxy dominates, the fractional polarization is lower, as expected for radio galaxies. This further indicates that the plasma could be a mixture of radio-emitting particles accelerated by the AGN jets and of plasma tracing a shock-wave with highly ordered magnetic fields.

The Faraday depth values of the cluster that we derived from the NAT and the radio relic (i.e., $\phi_{cl}$ in Fig.~\ref{fig:phiSL}) are in agreement and consistent with zero within the uncertainties. This indicates that  in both the sources the Faraday rotation is only caused by the external screen of our Galaxy and thus, in the regions of the cluster where they lie, the ICM has similar properties (i.e. either low thermal electron density and/or low magnetic field).

We can obtain an equipartition estimate of the magnetic field in the region of the E relic. We refer to \citet[e.g.,][]{Govoni04} for the details of this method. We integrated the synchrotron luminosity between 10 MHz and 10 GHz. Assuming that all the relic volume is occupied by magnetic fields and have a width of 500 kpc along the line of sight, the equipartition magnetic field is estimated to range from $0.9$ to $2.7$ $\rm \mu G$, depending on the $k$ value for the ratio between the energy density of relativistic protons and electrons. The upper bound comes from the assumption of $k=10^2$, as is usually inferred for the Milky Way \citep[][]{Schlickeiser02}, while the lower bound is for $k=1$. A similar electron to proton ratio is implied by our modelling of the W Relic, at variance with standard Diffusive Shock Acceleration model (see Sec.~\ref{sec:simulation}). This magnetic field value should be used with caution because it is based on some working and simplified assumptions. Considering our uncertainties on the RM determination, an ordered magnetic field with this strength along the line of sight would have been detected if the electron density was higher than $10^{-5}$ cm$^{-3}$.

Overall, our analysis of the spectral index and polarization properties of the NAT and E relic supports the idea that remnants of radio lobes from AGN are strongly related to the origin of radio relics  providing a fossil, low-energy (i.e., $\gamma<10^2$) electron population \citep{Markevitch05,Bonafede14,Kang16,vanWeeren17a}.

\subsection{Relic-shock connection in the western relic}
\label{sec:discussrelicshock}

 Our  deep JVLA images show intriguing features in the western radio relic, that appear constituted by two arcs in Fig.~\ref{fig:X}. A peculiar morphology is also observed in the X-rays, where a shock front with an unusual "M" shape was reported  \citep{Mazzotta11}. We re-analysed the archival {\it XMM-Newton} observations and used our new radio images to study the connection between the shock and the radio relic.

\begin{figure}
	\includegraphics[width=\columnwidth]{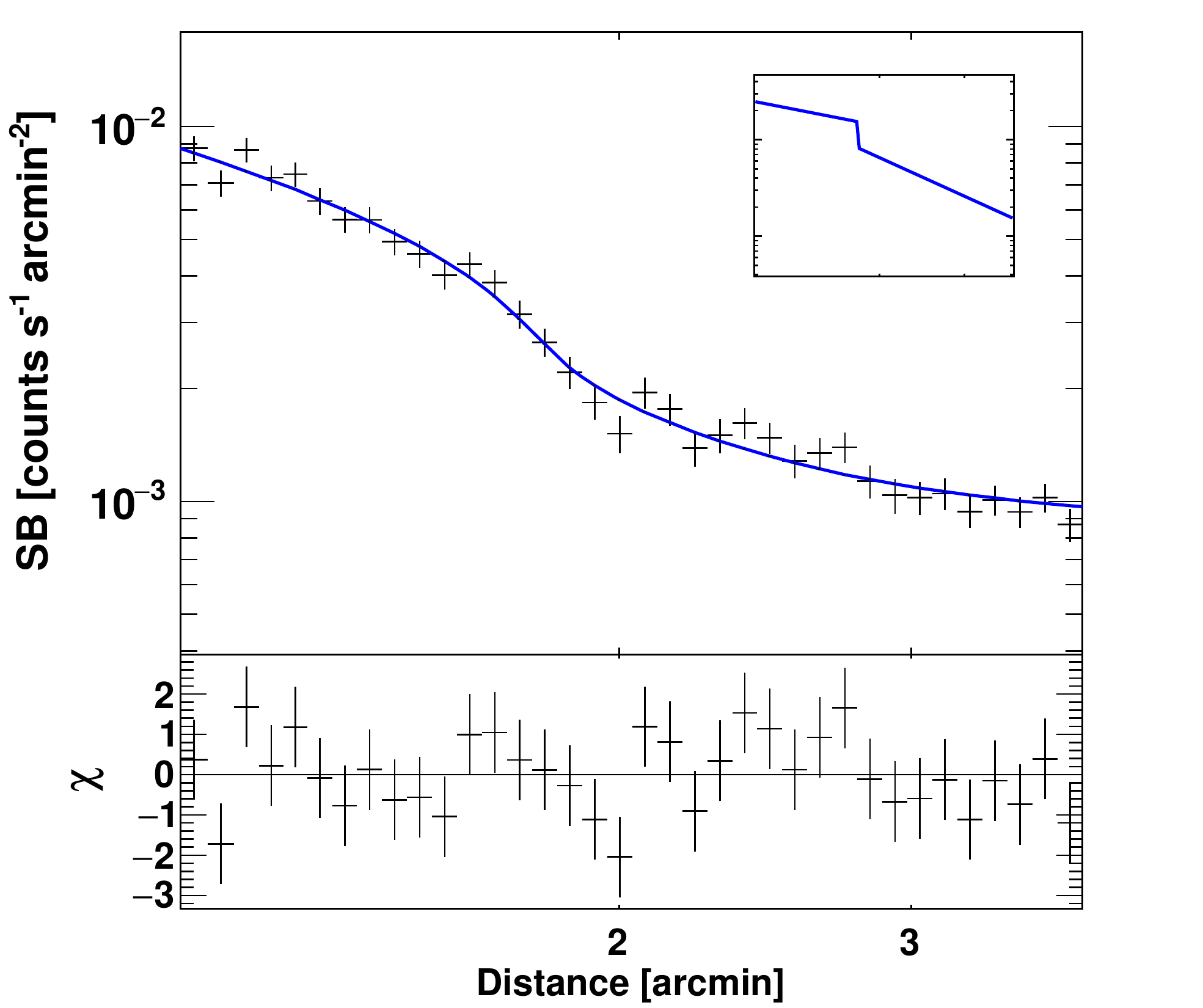}
    \caption{{\it XMM-Newton} surface brightness profile in the $0.5-2.0$ keV band extracted in the yellow sector shown in Fig.~\ref{fig:X}. The best-fitting broken power-law model provides a good description of the data ($\chi^2 = 32$ for 31 d.o.f) and the discontinuity, located at a radial distance of $1.81^{+0.06}_{-0.05}$ arcmin from the centre of the sector, is coincident with the outer edge of the radio emission. The profile is convolved with the {\it XMM-Newton} point spread function. The inset panel shows the corresponding 3D gas density model.}
    \label{fig:shock}
\end{figure}

We used \texttt{PROFFIT v1.5} \citep{Eckert11} to extract a number of surface brightness profiles following the "M" shape of the shock. We then decided to focus our analysis in the yellow sector reported in Fig.~\ref{fig:X} which  better highlights the discontinuity and encloses the brightest part of the radio relic. We adopted a broken power-law model to fit the data in such a region, as this model shape is generally used to describe density discontinuities in the ICM such as shocks and cold fronts \citep[e.g.][]{Markevitch07}. The best-fitting model convolved for the {\it XMM-Newton} point spread function \citep[see][for details]{Eckert16} is reported in Fig.~\ref{fig:shock} and appears in good agreement with the data. The surface brightness jump is observed to be co-spatially located with the outer edge of the nose-inner arc of Fig.~\ref{fig:X} and it would imply a shock with Mach number $M=1.7^{+0.4}_{-0.2}$  as from the Rankine-Hugoniot density jump conditions. To confirm the shock nature of the edge, we extracted and fitted spectra downstream and upstream the front, obtaining temperatures of $kT_d = 8.2^{+2.3}_{-1.3}$ keV and $kT_u = 3.2^{+2.3}_{-1.2}$ keV, respectively. In this case, the Rankine-Hugoniot temperature jump conditions  provide a Mach number of $M=2.4^{+1.1}_{-0.8}$, consistent with that derived from the surface brightness analysis. The Mach number that we measured is in agreement with the value of $2.1\pm0.1$ reported in \citet{Mazzotta11}.

In Sec.~\ref{sec:spectralstudyW} we measured a spectral index  $\alpha$=1.5$\pm$0.1 in the region of the nose close to the X-ray detected shock.
For the consideration of the previous section this value can be considered as the integrated spectral index of particles which are accelerated via DSA by a shock moving outwards, and than reach the equilibrium with energy losses as in the continuous injection model \citep{Kardashev62}. The corresponding Mach number is given by Eq.~\ref{eq:mach} that implies  $M$=2.2$\pm$0.2. This value is consistent with the one derived from the X-rays.

The detection of a shock front coincident with a radio relic strongly supports the relic-shock connection and allows to evaluate the efficiency of particle acceleration in cluster outskirts. Recently, \citet{Botteon19} studied the acceleration efficiency in a sample of relics, including \RX, and concluded that DSA from the thermal pool is severely challenged by the large acceleration efficiency required to reproduce the observed relic luminosity.

The sector chosen for X-ray analysis maximises the surface brightness jump which bounds the radio emission of the inner arc of the relic and is part of the ``M'' shape structure observed by \citet{Mazzotta11}. Conversely, the outer arc of the relic  lies in front of this feature, which is also located in a region where the X-ray brightness is too low for a proper characterization in surface brightness. The presence of this feature and of a possible, still undetected, underlying shock supports the idea of a complex merger dynamic.

\subsection{Particle re-acceleration in the radio halo}
\label{sec:discusshalo}

The radio halo in \RX is not a classical giant (i.e., $\geq$1 Mpc) radio halo since its extension reaches a maximum of 550 kpc in the W-E direction. \citet{Cassano07} found a scaling relation between the size of radio halos and their radio power. This relation is motivated in the framework of the re-acceleration scenario in which relativistic electrons are re-energized {\it in situ} by mechanisms associated with the turbulence originated during cluster merger \citep{Brunetti01,Petrosian01}. Using their relations, we found that, for the halo radio power at 1.4 GHz, listed in Tab.~\ref{tab:radiopower}, the expected halo radius, defined as $\sqrt{R_{min}\times R_{\rm max}}$, $R_{\rm min}$ and $R_{\rm max}$ being the minimum and maximum radii measured on the 3$\sigma$ radio contours, is  $409^{+26}_{-35}$ kpc. The average radius of the halo in \RX is instead 250 kpc. Hence, this halo is slightly smaller than what would be expected. This is not an observational effect due to missing short baselines since with the JVLA D configuration it is possible to recover emission on a scale of several Mpc at the cluster redshift, if present.

The radio emission permeating the central region of \RX is spatially connected to the W radio relic although they show different spectral index  trend and polarization properties. The halo emission shows a smooth spectral index distribution with an average value of  1.3$\pm$0.2 ( Fig.~\ref{fig:Spixmap_W}). Instead, the radio relic shows a typical steepening of the spectrum in the downstream region. The spectral index profile of the whole western extended emission (see Fig.~\ref{fig:spixprof_W}) is similar to the one observed in  the Toothbrush radio relic by \citet{vanWeeren16} and \citet{Rajpurohit18}, with a steepening between the relic and the halo, followed by flattening of the spectrum. This spectral  index distribution suggests a link between different mechanisms of particle acceleration. The flattening of the spectral index  in the halo region may be due to the re-acceleration of aged electrons  previously accelerated by the shock in the W relic. If this is the case, this cluster indicates that the seed electrons necessary for turbulent re-acceleration to work could be supplied by shock waves that also inject turbulence in the cluster core \citep{vanWeeren16}.

It is well-established that the radio power of giant radio halos scales with the cluster X-ray luminosity, and thus with the cluster mass \citep[e.g.][]{Feretti12}. The Sunyaev-Zel'dovich (SZ) effect has also been used as a proxy of the cluster mass to derive a  well-defined scaling relation between the radio power and cluster mass \citep{Cassano13}. The radio halo hosted in \RX follows this relation. If we consider the halo emission as detached from the relic on the basis of the spectral index profile, the radio halo luminosity at 1.4 GHz is 1.16$\pm$0.07$ \cdot10^{24}$ W Hz$^{-1}$ (see Tab.~\ref{tab:radiopower}) while the one expected from the correlation is  $2.2^{+0.7}_{-0.5}\cdot10^{24}$ W Hz$^{-1}$. Despite its small size, the radio power originating from this radio halo is in agreement within the 3$\sigma$ confidence level of the correlation.

\subsection{Projection effects and polarization}
\label{sec:proj}

Although projection effects are expected to be  minimal in double relic clusters, they are probably playing a role in the observed properties of this system. \citet{Golovich18} derived that \RX has a large line of sight velocity difference between the merging clusters, which they found is atypical for double relic systems. The viewing angle of the merger $\theta$ (i.e.  the angle between the line of sight and the merger axis) was constrained by means of a comparison with simulated cluster by \citet{Wittman18}. They found that $\theta>37^{\circ}$ within the $68 \ \%$ confidence interval.

In the absence of optical observations, polarization data can also be used to get an independent constraint on the merger axis. \citet{Ensslin98} developed a simplified relation which links the observed average polarization of a radio relic with its viewing angle, given its integrated spectral index. This relation is derived under a number of assumptions and it does not take into account the effects due to beam and Faraday depolarization.In spite of this, it could be used to derive a lower limit on the viewing angle \citep[e.g.][]{Hoang18}.

Referring to the S-band measurements, in the northern part of the E relic the average polarization fraction is 41$\pm6 \ \%$ and $\alpha=1.2\pm0.2$. This implies a viewing angle of $\sim62^{\circ}$. For the W relic we considered only the northern arc, since the southern nose of the shock is experiencing stronger internal Faraday depolarization as discussed in Sec.~\ref{sec:discussthick}. The arc is polarized on average at 31$\pm3 \ \%$ and we can use a spectral index of $\alpha=$1.5$\pm$0.1 as derived at the edge of the relic: the viewing angle is therefore $\sim$55$^{\circ}$. A value  of around 60$^{\circ}$ can thus be considered as a lower limit for the viewing angle of the merger.

With the fractional polarization value of relics we can also derive the Mach number of the underlying shock as done in \citet{Kierdorf17}. Here the polarization is derived for relics seen perfectly edge-on assuming a weak upstream magnetic field that is tangled and compressed on small scales at the shock front. Under these assumptions, a fractional polarization of 41$\pm6 \ \%$ corresponds to a Mach number of $\sim$1.7 for the E relic while for a fractional polarization of 31$\pm3 \ \%$, as in the W relic arc, $M\sim$1.4. These values are both lower than the Mach numbers that we estimated from the spectral index and X-ray analysis. This supports the idea that projection effects may lower the observed fractional polarization of the relics.

\subsection{Comparison with a simulated radio relic}
\label{sec:simulation}

In the following we compare our results with mock X-ray and radio images of a simulated radio relic. The 3D cubes of simulated quantities can be integrated along different directions, changing the angle $\theta$ and obtaining projected images of the simulated cluster at different viewing angles. This comparison allows us to: {\it (i)} check if the values of RM obtained in the simulation can be compared to the observed ones, with the aim of deriving a rough estimate of the magnetic field in the region of the  diffuse radio emission and to constrain the electron acceleration efficiency at the shock {\it (ii)} test the viewing angle using realistic geometrical conditions and introducing the effects of beam depolarization and Faraday rotation, both external and internal.

We used a high-resolution cosmological magneto-hydrodynamical simulation of a galaxy cluster produced with the adaptive mesh refinement code ENZO \citep[][]{enzo14}, as in \citet{va18mhd}. The cluster has  a virial mass  of $M_{\rm 100} \approx 1.3 \cdot 10^{15} M_{\odot}$ at $z=0.05$.
The maximum spatial resolution in the simulation is $3.95 ~\rm kpc/cell$. However, in the region where a simulated radio relic is used for a comparison with our observations  the effective resolution is $7.9 ~\rm kpc/cell$.

At the beginning of the simulation ($z=30$), we assumed a uniform magnetic field value $B_0=0.1 ~\rm nG$ along each  coordinate axis, which mimics a primordial magnetic field and is below the upper limits from the analysis of the Cosmic Microwave Background \citep[e.g.][]{Subramanian16}. The field is then amplified  in the course of the cluster formation process, both due to gas compression (for $z > 1$) and small-scale dynamo (for $z \leq 1$) and finally reaches a typical magnetic field strength of $\sim 2-5 \ \rm \mu G$ in the cluster core, and of $\sim 0.1-0.3 ~\rm \mu G$ at the cluster virial radius, showing a tangled 3D magnetic field with a typical scale of $\sim 50 ~\rm kpc$ ( see \citealt{Dominguez19} for more details on the simulation).

We followed an approach similar to Wittor et al. (submitted) to compute the polarized emission of the simulated radio relic. Therefore, we used a velocity jump method as in \citet{va09shocks} to detect and flag shock waves in the simulation  and for each flagged cell, we computed the integrated polarized emission following the model by \citet[Eq. 1]{Burn66}. We follow the formalism of \citet[Eq. 29]{hb07} to compute the downstream radio profile of an electron population injected at the shock via DSA and that cools via synchrotron and Inverse Compton. We evaluated the profile at nodes on the simulation grid along the shock normal. For the frequency range probed in this work, the electron cooling length is of a few $\mathrm{kpc}$ and hence most profiles are computed within one simulation cell. Therefore, we assume a homogeneous magnetic field and downstream temperature for each profile. Analogous, we compute the downstream profiles of the perpendicular and parallel component of the radio emission, that determine the intrinsic degree of polarization \citep[see Eq. 63.7 in][]{rybickiandlightman}. 

In order to match the observed radio power of \RX the acceleration efficiency by \citet{hb07} is not enough for this weak shock. Following \citet{Vazza15}, we included also the effect of re-accelerated particles by shocks, which is expected to dominate over the contribution from freshly injected particles for $\mathcal{M} \leq 3$ shocks \citep[e.g.][]{Pinzke13}. We include both internal and external Faraday rotation, directly using the three-dimensional magnetic field from the simulation.

\begin{figure}
	\includegraphics[width=\columnwidth]{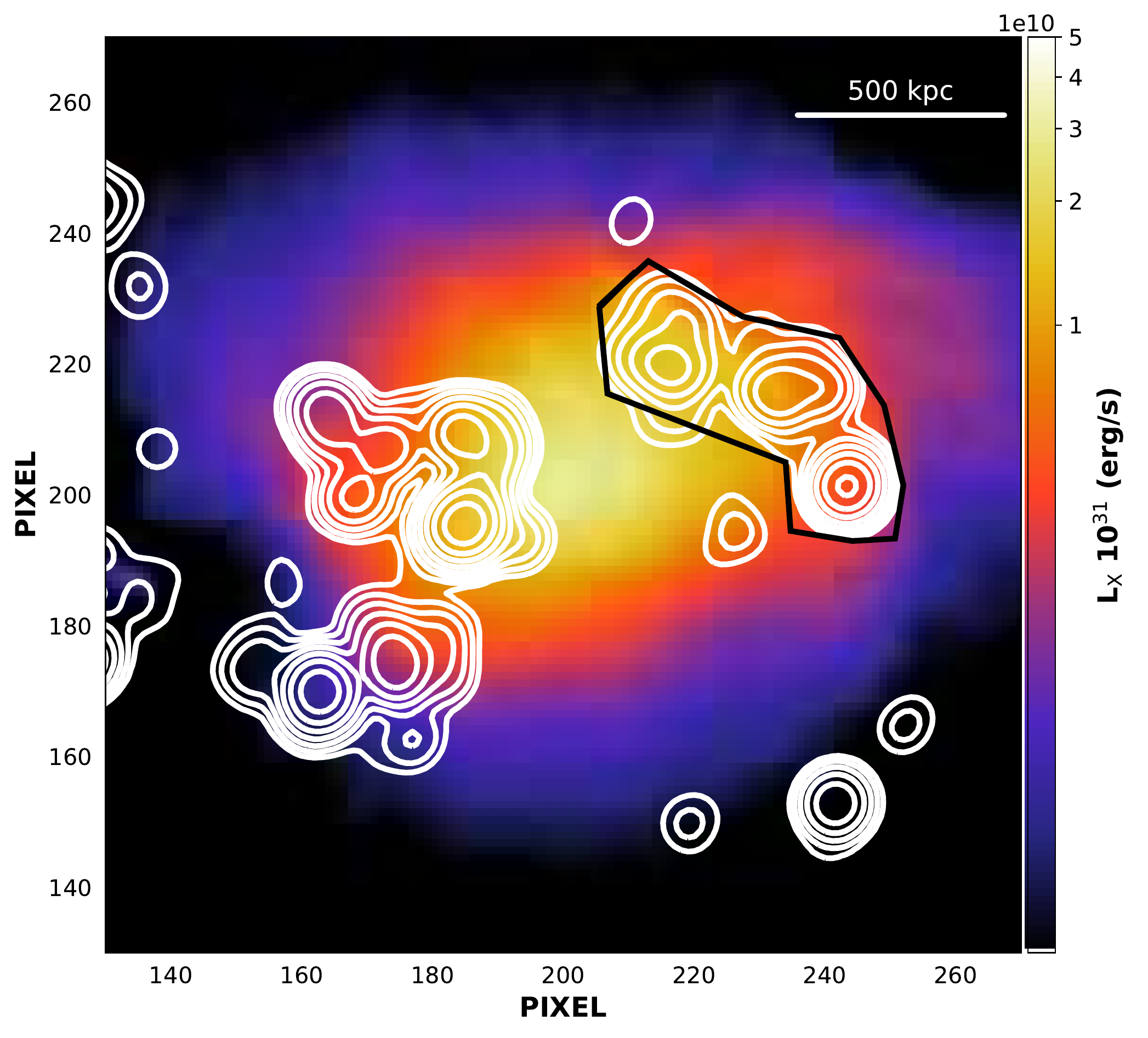}
	\includegraphics[width=\columnwidth]{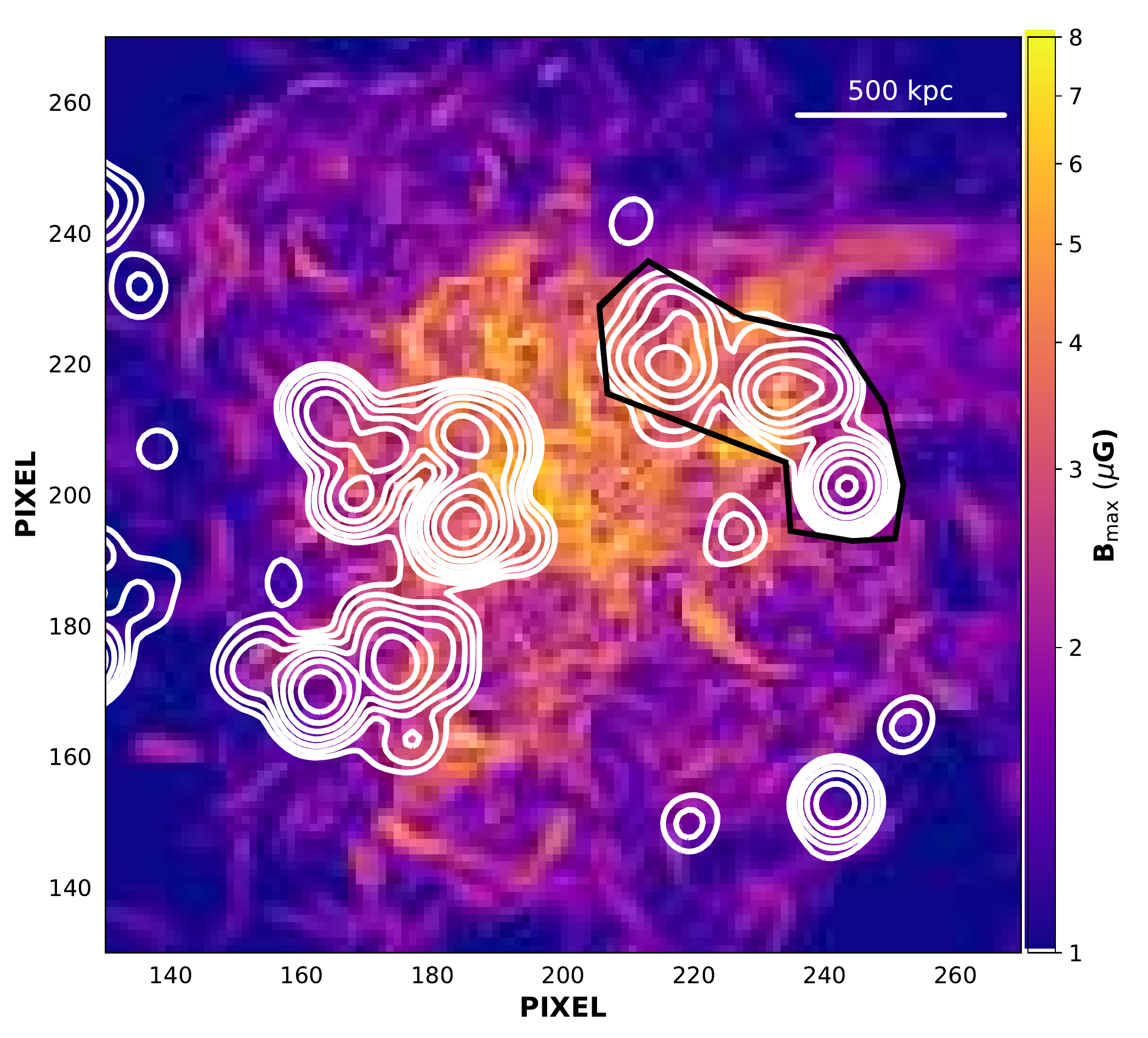}
    \caption{ Top panel: X-ray luminosity map of the simulated cluster (SIM), with 3 GHz radio contours overlaid.  Contours start at $1.8\cdot10^{20}$ W Hz$^{-1}$ and are spaced by a factor of two. The viewing angle is 70$^{\circ}$. The resolution of the image is the same of the simulation (i.e., 15.8 kpc per pixel) while the radio power image was smoothed  with a Gaussian kernel of $\sim$80 kpc. Bottom panel: maximum magnetic field along each line of sight (pixel) of the 3D simulated cluster for a 70$^{\circ}$ viewing angle. Radio contours are the same as in the top panel.  In both panels the black region is the one used to extract the relic radio power.}
    \label{fig:RXSIM}
\end{figure}

We selected a galaxy cluster (SIM for short) in a post-merger state at redshift 0.05. We found that the 2D projection of this cluster at a viewing angle of $70^{\circ}$ matches the observed properties: {\it (i)} the selected cluster shows a shock front in the X-rays, with an ``M'' shaped morphology, similar to the one detected in \RX, {\it (ii)} a radio relic $\sim$900 kpc long traces the shock front at a distance of $\sim$500 kpc from the X-ray centroid, with a region of the relic lying beyond the bulk of the shock, where the X-ray luminosity decreases, {\it (iii)} the average Mach number at the shock is similar to the one detected in \RX. The main characteristics of the simulated cluster are listed in Tab.~\ref{tab:simul}, with the same quantities estimated for \RX for comparison. The 2D projected gas density and maximum magnetic field strength along each line of sight are shown in Fig.~\ref{fig:RXSIM}. The simulated magnetic field shows filament and complex substructures. The average width of magnetic field filaments in Faraday space is few tens rad m$^{-2}$ thus they cannot be resolved at the maximum resolution in Faraday space that we reach combining L- and S- band data.

\begin{table}   
    \centering
	\caption{Quantities of interest in the comparison between the observed \RX (RX1314) and the simulated cluster (SIM). Column 1: cluster name; Column 2: radio power at 3 GHz. The flux density was extracted from a region of the same physical size (i.e., $\sim1.5 \cdot10^5$ kpc$^2$) in both clusters; Column 3: the X-ray luminosity of \RX from \citet{Piffaretti11} (given without uncertainties) was used to compute the luminosity of \RX in the range 0.5-2 keV using WebPIMMS (\url{https://heasarc.gsfc.nasa.gov/cgi-bin/Tools/w3pimms/w3pimms.pl}). The luminosity of SIM was computed as the sum in the whole cube; Column 4: thermal electron density at the shock front. From the density profile derived in \citet{Mazzotta11} we listed the post-shock electron density with an uncertainty derived from their profile. For SIM we computed an average value in the post-shock region and we used the standard deviation as uncertainty. Column 5: Mach number of the shock. We list the value derived in Sec.~\ref{sec:discussrelicshock} for \RX. For SIM, we computed an average value in the cells within the shock region where $M$>2, and we used the standard deviation of the distribution as uncertainty.}
	\label{tab:simul}
	\begin{tabular}{lcccc} 
	    \hline
		\hline
		  Cluster & P$_{\rm 3GHz}$ & L$_{\rm X}$ & n$_{\rm e}$ & $M$ \\
	      &  (10$^{23}$ WHz$^{-1}$) & (10$^{44}$ergs$^{-1}$) & (10$^{-3}$cm$^{-3}$) & \\
		\hline
         RX1314 & 11.5$\pm$0.6 & 8.91 & 0.7$\pm$0.05 & 1.7$^{+0.4}_{-0.2}$ \\
         SIM & 11.1$\pm$0.4 & 5.65 & 3.7$\pm$0.9 & 2.2$\pm$0.2 \\
		\hline
	\end{tabular}
\end{table}

\begin{figure}
	\includegraphics[width=\columnwidth]{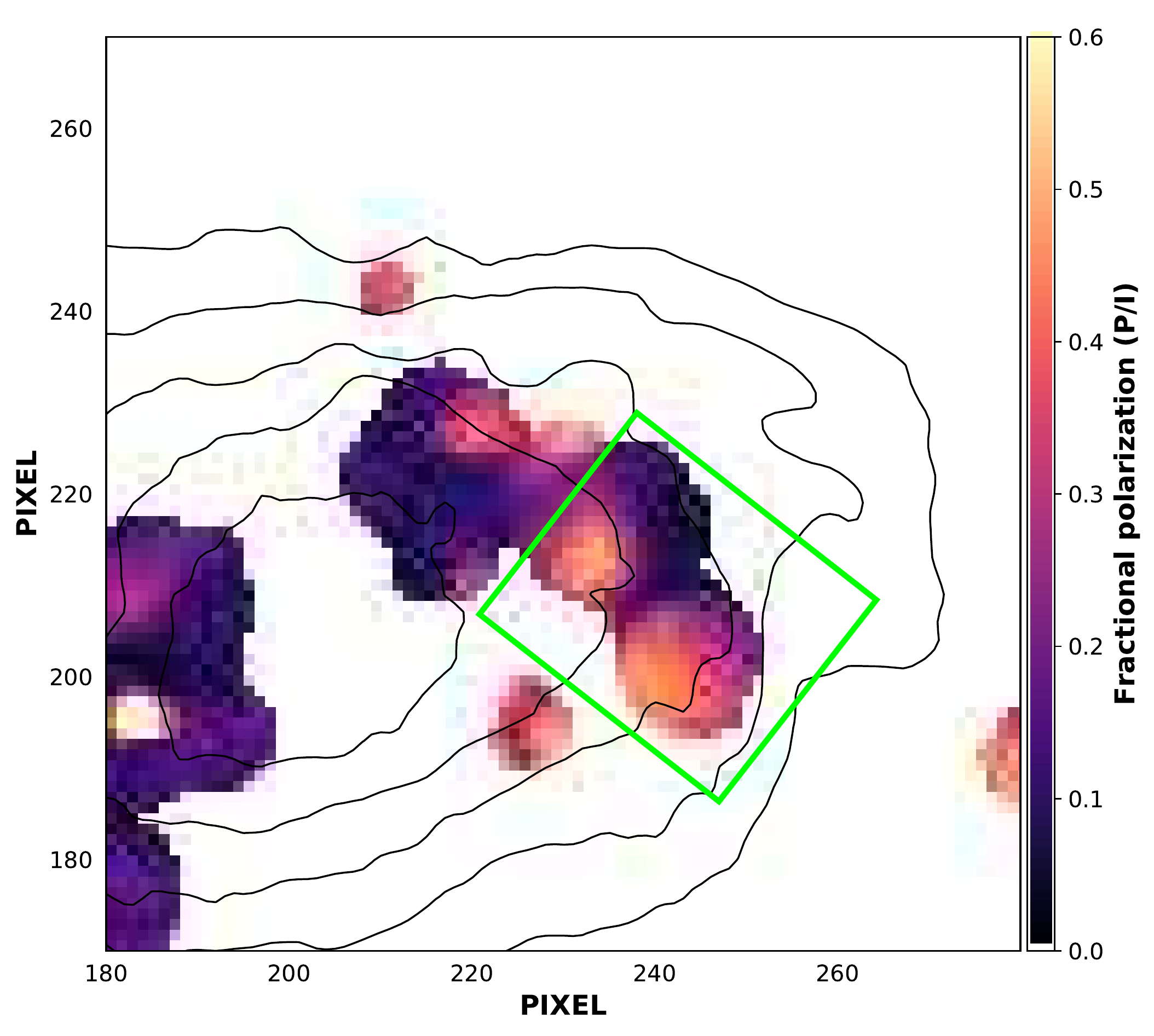}
	\includegraphics[width=\columnwidth]{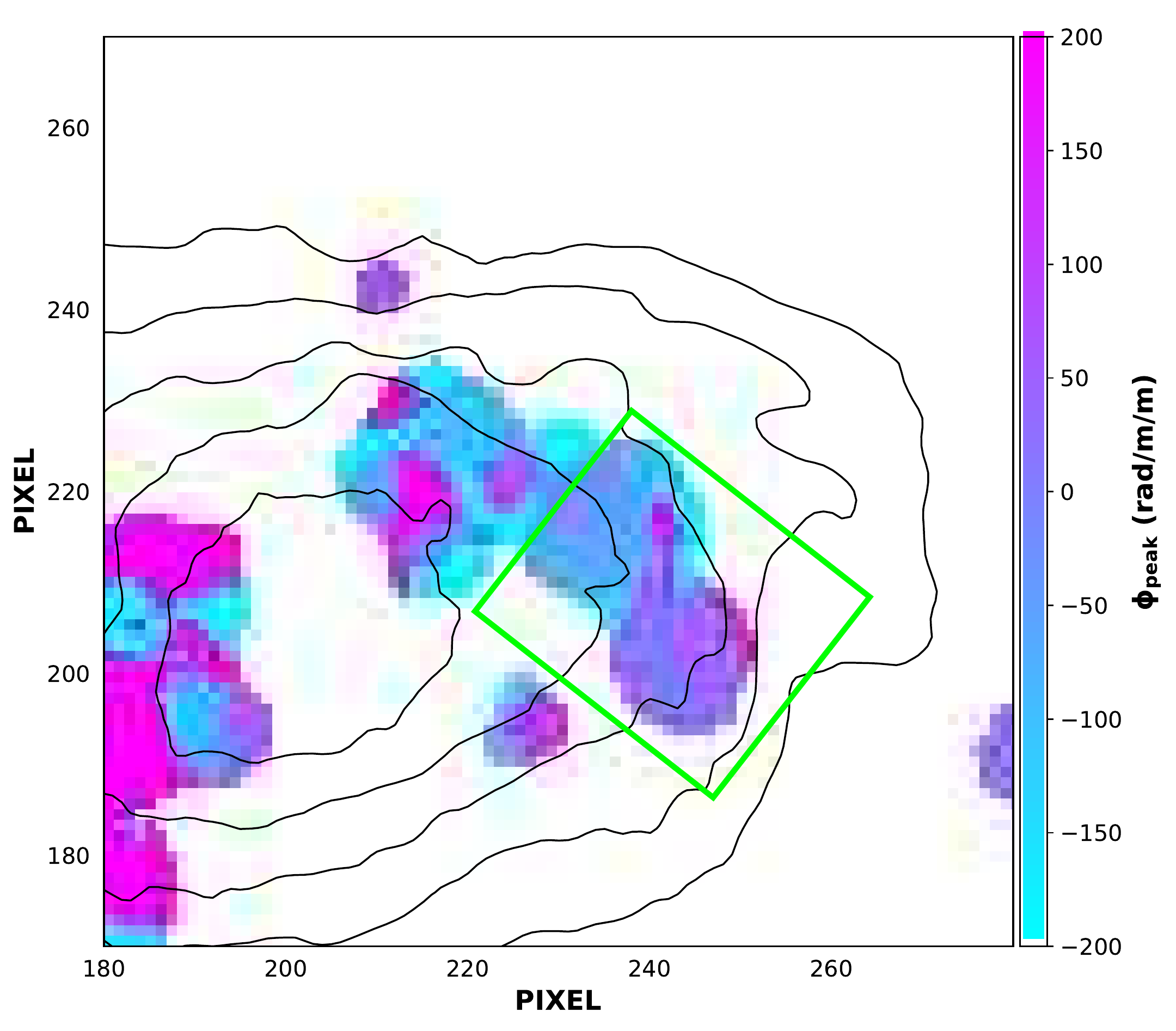}
    \caption{Top panel: Fractional polarization (top panel) and Faraday depth (bottom panel) of the radio diffuse emission in SIM in the 2-4 GHz band. A threshold is imposed to have the same total intensity dynamical range of our observation in the region of the relic.  Black contours are from the X-ray luminosity image in the energy band 0.5-2 keV: they start at 10$^{40}$ erg/s and they are spaced by a factor of two. The green box shows the region where we extracted the average fractional polarization and |$\phi_{\rm peak}$| values}
    \label{fig:pol_70}
\end{figure}

 \begin{figure}
	\includegraphics[width=\columnwidth]{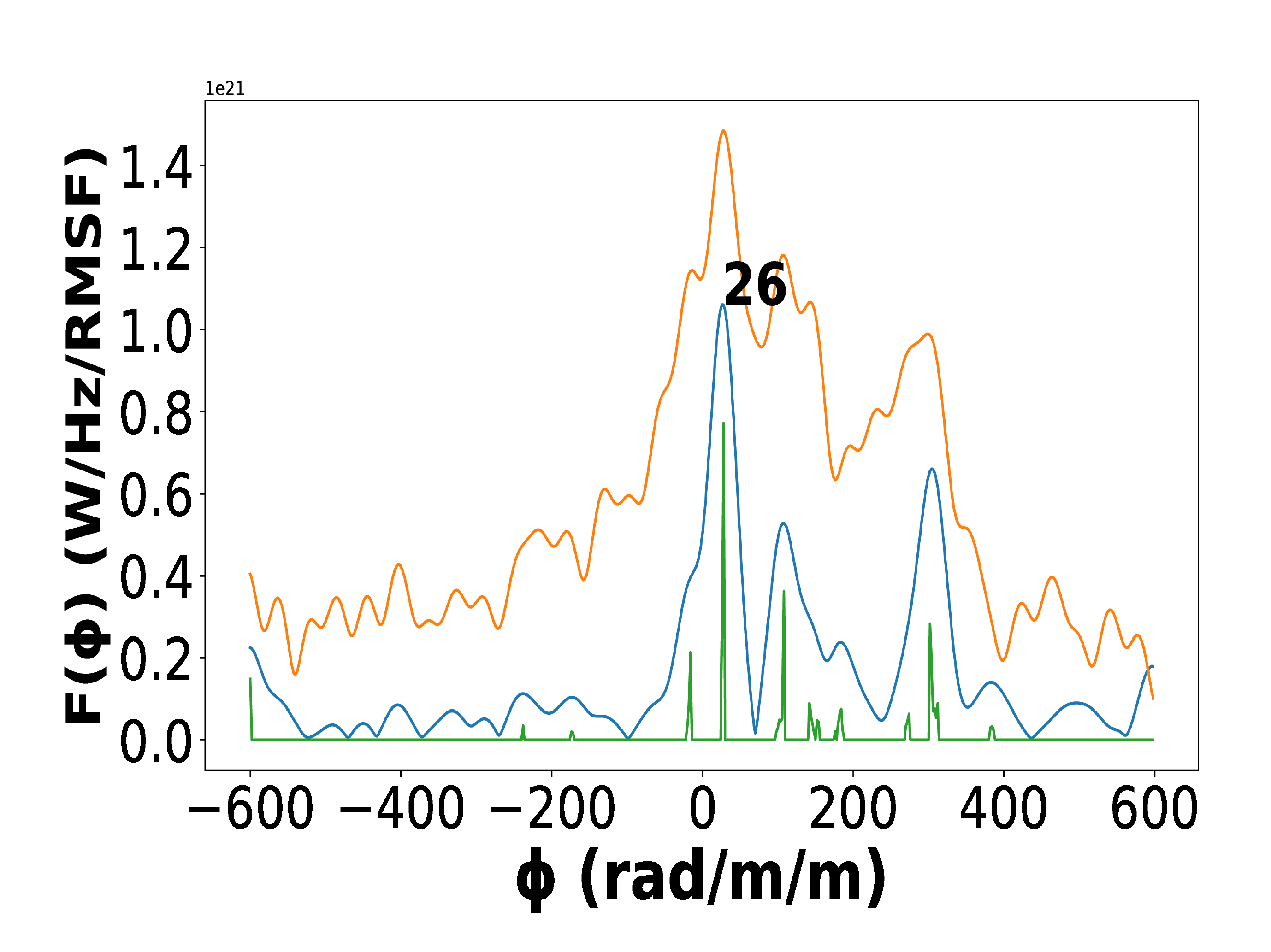}
    \caption{ Same image as the bottom-right panel of Fig.~\ref{fig:pixels} but for the simulated radio relic.}
    \label{fig:high_res}
\end{figure}

\begin{figure}
	\includegraphics[width=\columnwidth]{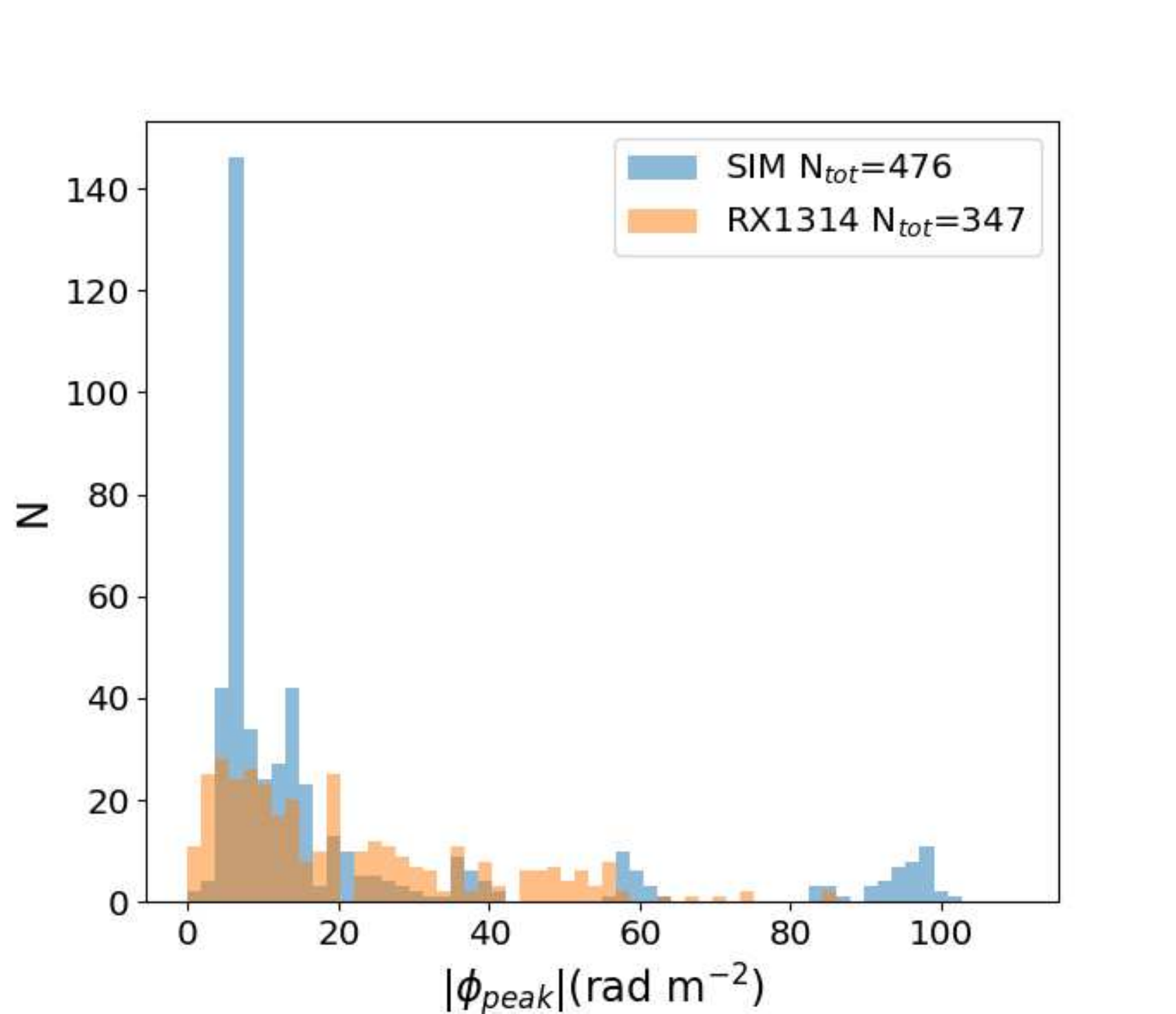}
    \caption{Distribution of Faraday depths. The |$\phi_{\rm peak}$| values of the simulation were reduced by a factor 5 to be compared with the observation, due to the different thermal electron density. For the observation (RX1314), the values are obtained in the northern region of the W relic arc in the S-band measurement. For the simulation (SIM) the values are extracted in the region marked with the green box in Fig.~\ref{fig:pol_70}.}
    \label{fig:RMdistr}
\end{figure}

By comparing the simulated radio power from the relic at $3$ GHz to the observed one of \RX,  we find that an electron acceleration efficiency of $\xi_\mathrm{e} \geq  1000\%$ (referred to the kinetic energy flux of shocked cells) is  needed, meaning that the kinetic energy of this shock would not be enough to explain our $3 \rm ~GHz$ observation of \RX. This value obviously requires something beyond Diffusive Shock Acceleration, and as argued in other works it can be attained only if a pre-existing population of fossil electrons (with $\gamma = 10^2-10^3$) was present upstream of the merger shock which produced the western radio relic in \RX \citep[][]{Markevitch05,Pinzke13,Kang15,Botteon16a,Eckert16a}. Our modeling suggests that this is reasonable, provided that the injection of fossil electrons happened no longer than $\sim ~\rm 1 Gyr$ before the $M \approx 2$ shock  crosses the region where the western radio relic is now visible.

The origin of such fossil population is not clear as it could be for the E relic. It could originate from an AGN that is no longer active or from previous shocks or turbulence generated in the cluster formation process. Considering that the typical time scale for turbulent acceleration larger than 1 Gyr on scales of 500 kpc \citep{Brunetti14} the observed radio halo cannot have supplied already aged electron in the region where the W relic is observed. Fossil electrons have $\gamma = 10^2-10^3$ and to detect them and discover their origin we would need low frequency observation in the MHz regime.

Since the average synchrotron power of a relativistic electron population is $\langle P\rangle\propto\gamma^2B^2$  (if $B \leq 3.2 ~\mu G ~ (1+z)^2$ as in our case here),  in principle either a higher magnetic field, or a larger acceleration efficiency (and/or a higher density of seed fossil electrons) are needed to match the observation. However, as we shall see, the polarization properties of the west relic can be used to constrain the magnetic field amplitude, which in turn sets a lower limit on the amount of necessary particle re-acceleration (or higher efficiency).

We performed RM synthesis on the simulated data. We obtained Stokes $I$, $Q$ and $U$ 2D projections in the same frequency range and sub-bands of our measurements in the S-band. We smoothed all images with a Gaussian kernel of the physical size of the restoring beam of the S-band image (i.e., FWHM $\sim$80 kpc). With the same procedure described in Sec.~\ref{sec:RMsynth} we obtained images of $\phi_{\rm peak}$ and $P=|\widetilde{F}(\phi_{\rm peak})|$ and we obtained the linear degree of polarization map by dividing the polarization and total intensity smoothed images. Fractional polarization and $\phi_{\rm peak}$ images for the 2D projection at $\theta$=70$^{\circ}$ are shown in Fig.~\ref{fig:pol_70}. In all the images we imposed a threshold in order to obtain the same dynamical range of our observations in total intensity.

Also from the simulation we obtained Faraday-complex spectra in some regions of the relic. In order to produce a high resolution Faraday spectrum of the simulated relic, we also performed the RM-synthesis on the simulated data between 1 and 4 GHz reaching a resolution of 37 rad m$^{-2}$ as in the S+L-band measurement. The high resolution spectrum of a representative pixel is shown in Fig.~\ref{fig:high_res}. The Faraday spectrum is shown together with the clean components found during the RM clean procedure and the final cleaned Faraday spectrum. Several filaments are clearly separated after the RM-clean. The filaments in the simulation appear to be spread over $\sim$ 300 rad m$^{-2}$ in Faraday depth and they have different luminosity.
  
We can use as a benchmark the average fractional polarization and the distribution of $\phi_{\rm peak}$ values in the most external region of the relic, marked with the green box in Fig.~\ref{fig:pol_70}. We chose a region where we obtained Faraday-simple spectra in the S-band as in the northern arc of the W relic of \RX, thus we can more easily compare the results of the RM synthesis. The average fractional polarization for $\theta\sim70^{\circ}$ in the smoothed image is $32 \ \%$ and it is consistent with what we obtained from the observation at the same physical resolution (i.e., $31\pm3 \ \%$ in the western relic). This value takes into account the depolarization due to the smoothing of the polarization vectors on physical scales between 15 and 80 kpc since the simulated magnetic field has important sub-structure on such scales. 

In SIM, we found higher values of $\phi_{\rm peak}$ compared to the $\phi_{\rm cl}$ values obtained in \RX in the S-band, but the thermal electron density in the region of the shock is on average $\sim$5 times lower in \RX than in SIM (see Tab.~\ref{tab:simul}). To make a fair comparison on the effect of the magnetic field on the Faraday rotation in the two clusters, we compare in Figure~\ref{fig:RMdistr} the distribution of Faraday depths obtained in the simulation reduced by a factor of 5 with the observed one (foreground subtracted). The average and median values of the two distributions are consistent within the uncertainties. This indicates that the magnetic field in the relic region in \RX is already of the order of the simulated one (i.e., $\sim1\mu$G), and hence that only a higher (re)acceleration efficiency can explain the power radio emission from the W relic. We notice that while the role of re-accelerated fossil electrons has been already proposed as key ingredient to explain observed radio relics \citep[e.g,][]{Pinzke13,Kang15,Kang16}, this is the first time in which we can constrain its importance by simultaneously fixing the uncertainties on the magnetic field in the relic region, thanks to polarization data and simulations.

In spite of similar median values, the two distributions in Fig.~\ref{fig:RMdistr} are very different. For this comparison we used the peak values of S-band spectra in the green region of Fig.~\ref{fig:pol_70}, where Faraday spectra are not resolved and show a single component. The simulated distribution is thus dominated by the emission of the brightest filament at a certain Faraday depth and the narrower distribution compared to the data is due to the fact that the same filament dominates in the sampled spatial region. The broader distribution of RM values found in the northern arc of the radio relic shows that various emitting structures at different Faraday depths are co-spatially located in a region equivalent to the one considered in the simulation.

\subsection{Faraday-complex emission}
\label{sec:discussthick}

As  reported in Sec.~\ref{sec:polarizationW}, we observed Faraday-complex emission in the nose of the W relic. This means that linearly polarized emission originates from different layers of the radio relic which experienced different amounts of Faraday rotation. Faraday complex structures were observed also in the Toothbrush radio relic \citep{vanWeeren12}. In this region, we observed  features in the Faraday dispersion functions spread over a range of 150-250 rad m$^{-2}$.

As an example, in Fig.~\ref{fig:pixels}, the S- and S+L-band Faraday dispersion function (i.e. Faraday spectrum) is shown for two pixels. The first row shows a Faraday-simple source (i.e., a pixel at the position of the source labelled with $C$ in Fig.~\ref{fig:optic}), the second row a Faraday-complex one (i.e., a pixel in the nose region).

The RM clean found a single component in the spectrum of the Faraday-simple source, both with the S- and S+L-band resolution. In the S-band, the Faraday-complex pixel is observed as a single peak but with a FWHM slightly larger than the RMSF resolution (i.e., $\delta\phi=$188 rad m$^{-2}$): some cleaned components are found, within a range of 200 rad m$^{-2}$. Instead, in the S+L-band measurement, where we have higher resolution in Faraday space, the RM clean on the Faraday-complex spectrum found several components, with a similar flux density, and a width of 150 rad m$^{-2}$.

The shape of a complex Faraday spectrum is indicative of the medium generating the emission. A single broad feature in the Faraday-spectrum originates from a regular distribution of magnetic fields and thermal electron density while a series of peaks are expected if filamentary magnetic fields structures are overlaid along the same line of sight. In both cases the Faraday rotation occurs in the emitting region. Our simulation suggested that we should expect a series of peaks in our case since the magnetic fields has filamentary structures with a width in Faraday space lower than 40 rad m$^{-2}$ (see Fig.~\ref{fig:RXSIM} and Fig.~\ref{fig:high_res}).

Although the RM clean procedure helps in reducing secondary lobes when a peak is detected at high signal-to-noise ratio, it is not well suited to detect Faraday-complex sources, since it assumes a thin model for every detected component, and it easily diverges at lower  signal-to-noise values. Furthermore, the performance of RM synthesis in detecting Faraday-complex sources strongly depends on the frequency coverage of the observation. Some techniques have been proposed to overcome such problems and to optimize the reconstruction of Faraday-complex spectra \citep{Frick10,Li11}. All of them assume a model for the synchrotron-emitting and Faraday rotating medium, and nowadays their performances are still tested and debated \citep[see, e.g.,][for the comparison of RM clean with other methods]{Sun15,Miyashita16,Schnitzeler18,Ideguchi18}. 

On the basis of the spectrum shown in the bottom right panel of Fig.~\ref{fig:pixels} it is thus hard to distinguish between a single broad feature or a convolution of several Faraday-simple spectra originating from different filaments. For this reason we also studied the wavelength dependence of the degree of polarization. If filamentary and complex structures of magnetic field originate the complex Faraday spectrum, internal Faraday depolarization is expected \citep{Arshakian11}. The dispersion of RM values that we detected in our unresolved structures should be the same as the turbulent dispersion causing the depolarization, $\sigma_{\rm RM}$. 

\begin{figure}
	\includegraphics[width=1.11\columnwidth]{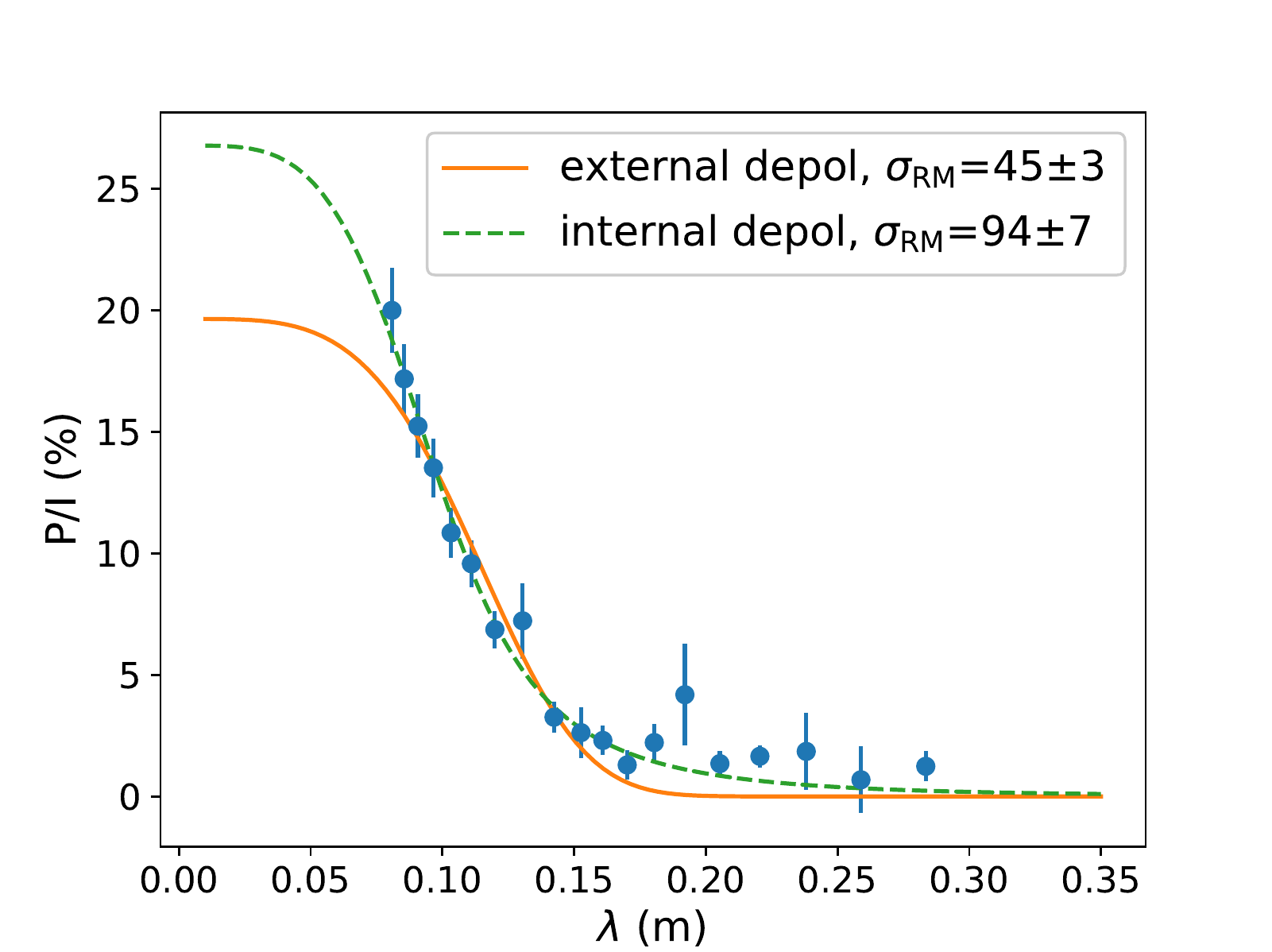}
    \caption{Depolarization profile of the nose of the western relic. Fractional polarization was computed in each sub-band separately. Green dashed line is the fitted internal depolarization model while the orange line is the external depolarization model.}
    \label{fig:depol_nose}
\end{figure}

We  imaged the $Q$ and $U$ data cubes between 1 and 4 GHz into 20 frequency sub-bands of 150 MHz each and we smoothed the data to 30\arcsec resolution to increase the signal-to-noise ratio. We calculated the fractional polarization of each sub-bands separately. The depolarization trend computed in the region of the nose, marked with a green circle in the top panel of Fig.~\ref{fig:polSL}, is shown in Fig.~\ref{fig:depol_nose}. We fitted both internal and external Faraday depolarization as modeled in \citet{Arshakian11}. Both depend on RM dispersion, $\sigma_{\rm RM}$, and wavelength. Internal Faraday depolarization well represents our data with a $\sigma_{\rm RM}$ of 94$\pm$7 rad m$^{-2}$. This shows that the depolarization is occurring in the same emitting volume. We note that the dispersion value is of the same order of magnitude of the range of Faraday depths that we detected with the RM synthesis. Several magnetic field filaments along the line of sight in a turbulent ionized gas are likely the origin of the Faraday-complex structures and of depolarization in the nose region.

We studied the depolarization trend also in the northern region of the E relic, marked with the green circle in the upper panel of Fig.~\ref{fig:polSL}. In this case, the RM dispersion for internal and external Faraday depolarization are 12$\pm$2 rad m$^{-2}$ and 7$\pm$1 rad m$^{-2}$, respectively, and the reduced $\chi^{2}$ is 1.2 for both. Under the simplified assumption that the magnetic field is the same at the position of the two relics, the ratio of $\sigma_{\rm RM}$ is equal to the ratio of the thermal electron densities along the line of sight \citep[see e.g. the definition of $\sigma_{\rm RM}$ given in][]{Arshakian11}. This would imply that the thermal electron density in the region of the E relic is $0.13\pm0.02$ or $0.16\pm0.02$ times lower than in the W relic region, considering only internal or external depolarization, respectively. Overall this is in agreement with the upper limit from the X-rays and with the estimate made in Sec.~\ref{sec:discussAGNrelic}. Anyway, simulations probed that every shock has a peculiar and complex morphology that does not allow a simplified and general description \citep{Hoeft11}.

When the Faraday spectrum is unresolved, the peak value of the spectrum represents the total polarized flux density of the source in Jy/beam, integrated along the Faraday spectrum.  Conversely, in the case of a resolved Faraday-complex emission, the peak value represents only the main component of the spectrum, and it is a fraction of the total emission of the source.  As a result, the polarized flux density of Faraday-complex structures estimated from the main peak is lower than the integrated one, and the fractional polarization that we compute with the RM synthesis should be regarded as a lower limit.

For the purpose of this paper, we decided to use the information that we obtained directly from the RM synthesis technique as it is implemented in \texttt{pyrmsynth}. In the pixels where we observed a Faraday-complex spectrum, we considered as integrated polarization  that one obtained from the peak of the S-band data, where the broad feature is only partially resolved. In this way, we have the best possible estimate for the polarized flux density. Instead, we used S+L-band data to recover the RM value of the brightest component ($\phi_{\rm peak}$) with higher precision. This value should be used with caution for Faraday-complex spectra.
 
 Even if we were able to recover the true Faraday-complex spectrum of a source, its physical interpretation would not be unique. In fact, there are several distributions of magnetic field, thermal electron density and physical depth that can generate the same observed spectrum. Also beam depolarization should be considered since it could introduce spatially varying $\phi$ components that  may contribute to the complexity of the Faraday spectrum. We  plan to further study this aspect applying the RM synthesis technique in simulation of cluster radio diffuse emission, taking into account both external and internal Faraday depolarization. The necessary frequency coverage and appropriate RM synthesis methods for reconstructing such complicated Faraday spectrum could be also investigated in the comparison with synthetic Faraday-complex spectra.
 
\vspace{-25pt}
\section{Conclusions}
\label{sec:conclusions}

In this paper, we performed a detailed study of the radio emission of the merging galaxy cluster \RX in the frequency range 1-4 GHz. We exploited a variety of JVLA observations: full polarization A, B, C and D configurations between 1 and 2 GHz (L-band) and DnC configuration archival data in the 2-4 GHz S-band. We performed both spectral analysis and RM synthesis on the entire cluster field, and in particular on two radio relics and the central radio halo. This study allows us to investigate possible scenarios for the origin of the extended radio emission observed in \RX.

Our results can be summarized as follow:

1. New multi-configurations JVLA data offer unprecedented view of \RX at 1.5 GHz. We detected the two relics and the central radio halo. The western relic shows interesting substructures: two arcs depart from the X-ray detected shock, one of the two being $\sim$900 kpc long and extending toward the outer region of the cluster. The eastern relic  embeds a NAT radio galaxy, member of the cluster.

2. This study revealed a possible connection between the AGN activity and the extended emission of the E radio relic. The spectral index profile supports a re-acceleration scenario in which a seed population of mildly relativistic particles, pre-accelerated by AGN jets, are  subsequently re-accelerated by merger driven shock waves. Polarization properties of the entire eastern system suggest that the shock waves were able to align the magnetic field lines along the shock plane, but we did not find Faraday rotation caused by the cluster magnetic field in this region. Only very deep X-ray observations could proof the presence of a shock underlying the eastern radio relic.

3. Spectral index profiles and polarization properties, allowed us to disentangle the  contributions of the relic and the halo to the western diffuse emission. The relic shows a typical spectral steepening downstream of the X-ray detected shock, while a different acceleration process should be able to re-energize particles in the radio halo region. The polarization properties support this scenario: only the radio relic is detected in polarized intensity at 3 GHz.

4. We discovered Faraday-complex emission in the northern region of the W relic. This indicates the presence of thermal gas and complex magnetic field morphology within the relic volume. The fractional polarization in the region is low (i.e., on average $8.5\pm0.6 \ \%$ at 3 GHz) due to internal Faraday depolarization. We obtained very different RM dispersion from the depolarization trend of the two relics. However, deriving constraints on underlying magnetic field would require a more detailed knowledge of the gas thermal density, projection effects, and shock-wave morphology.

5. We studied the X-ray shock coincident with the inner arc of the western radio relic. We derived a Mach number consistent with that found by \citet{Mazzotta11} (i.e., $M=1.7^{+0.4}_{-0.2}$ from the X-ray surface brightness jump and $M=2.4^{+1.1}_{-0.8}$ from the temperature jump). The Mach number derived from the radio spectral index assuming DSA mechanism is consistent with the one derived from the X-ray analysis.

6. By comparing our observation with a cosmological simulation, we tested a possible viewing angle of $\sim70^{\circ}$. The consistency check pointed out that the simulation is able to reproduce the observed fractional polarization and the RM of the western relic with a magnetic field of $\sim1\mu$G in the relic region and important sub-structures on scales between 15 and 80 kpc. The required electron acceleration efficiency to match the observed radio power provides the need of significant re-acceleration of fossil electrons. 

In the future, to resolve and interpret Faraday-complex structures, we could exploit the SKA-MID\footnote{https://www.skatelescope.org/}, which will reach a resolution in Faraday space of $\sim5$ rad m$^{-2}$, spanning the frequency range between 350 MHz and 14 GHz. A deeper understanding of the physical interpretation of such structures should be also reached thanks to the comparison with simulation \citep{Loi18}. The combination of radio polarimetric studies, high-resolution spectral index imaging, total intensity radio observation at 50-350 MHz and deep X-ray observations of merger shocks, promises to shed new light on particle acceleration processes occurring in the ICM, on the role played by the fossil plasma and on the properties of the magnetic field at merger shocks.

\vspace{-20pt}
\section*{Acknowledgements}
C.S. and A.Bon. acknowledge support from the ERC-StG DRANOEL, n. 714245. A.Bon. acknowledges support from the MIUR grant FARE SMS. C.S. and A.Bon. acknowledge T. Venturi for useful discussions and for sharing GMRT images. D.W. and F.V. acknowledge financial support from the European Union's Horizon 2020 program under the ERC-StG MAGCOW, n. 714196. The cosmological simulations were performed with the ENZO code (\url{http://enzo-project.org}), which is the product of a collaborative effort of scientists at many universities and national laboratories. The authors gratefully acknowledge the Gauss Centre for Supercomputing e.V. (\url{www.gauss-centre.eu}) for supporting this project by providing computing time through the John von Neumann Institute for Computing (NIC) on the GCS Supercomputer JUWELS at J\"ulich Supercomputing Centre (JSC), under projects n. 11823, 10755 and 9016 as well as hhh42 (P.I. F.V.) and hhh44 (P.I. D.W.). A.Bot. thanks D. Eckert for helpful suggestions regarding the XMM-Newton data analysis. Part of this work was performed under the auspices of the U.S. Department of Energy by Lawrence Livermore National Laboratory under Contract DE-AC52-07NA27344. R.J.vW. acknowledges support of the VIDI research program with project n. 639.042.729, which is financed by the Netherlands Organisation for Scientific Research (NWO). We thank the anonymous referee for the useful comments.

%%%%%%%%%%%%%%%%%%%%%%%%%%%%%%%%%%%%%%%%%%%%%%%%%%

%%%%%%%%%%%%%%%%%%%% REFERENCES %%%%%%%%%%%%%%%%%%

% The best way to enter references is to use BibTeX:
\vspace{-20pt}
\bibliographystyle{mnras}
\bibliography{my_bib} % if your bibtex file is called example.bib

%%%%%%%%%%%%%%%%%%%%%%%%%%%%%%%%%%%%%%%%%%%%%%%%%%

%ent}

%%%%%%%%%%%%%%%%% APPENDICES %%%%%%%%%%%%%%%%%%%%%

%\appendix

%\section{Some extra material}

%If you want to present additional material which would interrupt the flow of the main paper,
%it can be placed in an Appendix which appears after the list of references.

%%%%%%%%%%%%%%%%%%%%%%%%%%%%%%%%%%%%%%%%%%%%%%%%%%

%\end{comment}

% Don't change these lines
\bsp	% typesetting comment
\label{lastpage}
\end{document}